\documentclass[preprint,3p,12pt]{elsarticle}

\usepackage{graphicx}
\usepackage{amsmath}
\usepackage{appendix}
\usepackage[hang]{subfigure}
\usepackage{bm}
\usepackage{float}
\usepackage{mathrsfs}
\usepackage{amssymb}
\usepackage{multirow}
\usepackage{bbding}
\usepackage[misc]{ifsym}

\graphicspath{{./},{draw/}}

\newcommand{\bx}{\mathbf{x}}

\newcommand{\md}{\mathrm{d}}

\journal{Elsevier}
\begin{document}

\begin{frontmatter}



\title{Parallel energy-stable phase field crystal simulations based on domain decomposition methods}

\author[iscas]{Ying Wei}
\author[iscas,sklcs]{Chao Yang\corref{cor1}}
\ead{yangchao@iscas.ac.cn}
\author[amss]{Jizu Huang}
\address[iscas]{Institute of Software, Chinese Academy of Sciences, Beijing,  China}
\address[sklcs]{State Key Laboratory of Computer Science, Chinese Academy of Sciences, Beijing, China}
\address[amss]{Academy of Mathematics and Systems Science, Chinese Academy of Sciences, 
Beijing, China}
\cortext[cor1]{Corresponding author. }


\begin{abstract}
In this paper, we present a parallel numerical algorithm for solving the phase field crystal equation.
In the algorithm, a semi-implicit finite difference scheme is derived based on the discrete variational 
derivative method. Theoretical analysis is provided to show that the scheme is unconditionally energy stable and can achieve second-order accuracy in both space and time. 
An adaptive time step strategy is adopted such that the time step size 
can be flexibly controlled  based on the dynamical evolution of the problem. 
At each time step, a nonlinear algebraic system is constructed from the discretization of the 
phase field crystal equation and solved by a domain decomposition based,
parallel Newton--Krylov--Schwarz method 
with improved boundary conditions for subdomain problems. 
Numerical experiments with several two and three dimensional test cases
show that the proposed algorithm is second-order accurate in both space and time, 
energy stable with large time steps, and  highly scalable to over ten thousands  
processor cores on the Sunway TaihuLight supercomputer.
%Several two and three dimensional test cases are used to validate the 
%accuracy and examine the efficiency of the proposed algorithm. 
%Numerical results show that the proposed algorithm is highly scalable to over ten thousands  
%processor cores on the Sunway TaihuLight supercomputer. 
\end{abstract}


\end{frontmatter}

% \linenumbers


\section{Introduction}
\label{intro}
The phase field crystal (PFC) equation is a popular model for simulating microstructural evolution
in material sciences. As an atomic description of crystalline materials on the diffusive time scale, 
the PFC equation was originally proposed to model the dynamics of crystal
growth by Elder et. al. \cite{PhysRevLett.88.245701, PhysRevE.70.051605}. 
Since then, during the past decade, it has been applied 
with significant successes for the simulation of phenomena found in various solid-liquid systems such as the 
crystal growth in a supercooled liquid \cite{zhang2013, Yang_ascalable, Cheng_anefficient}, 
the crack propagation in a ductile material \cite{Gomez201252, PhysRevE.70.051605}, the dendritic 
and eutectic solidification \cite{ElderPRB, Gomez201252}, and the epitaxial growth \cite{ElderPRB, WuPRB}.
The PFC equation is usually derived from the free-energy functional that originates from the 
more advanced density 
functional theory of Hohenberg and Kohn \cite{prb136.864}, resulting in 
a six-order nonlinear partial differential equation of the atomic density function. 
Due to the existence of the nonlinearity and the high-order terms, the PFC equation requires 
high-fidelity simulations with advanced numerical methods.

Together with the introduction of the PFC model, an explicit Euler method was employed 
to solve the PFC equation successfully by 
Elder et. al. \cite{PhysRevE.70.051605, ElderPRB}, in which the time step size is in proportion to 
the sixth order of grid sizes. The computational cost of the explicit method is high due 
to the small time step for maintaining the stability. 
To relax the restriction of the small time step, some implicit methods were proposed, see, e.g., 
\cite{Nucleation, FASOO, Cheng_anefficient, precondition}.
Although relatively large time step size can be employed in these implicit methods, 
no energy stability analyses were presented. 
Recently, many researches were done to design energy stable time-stepping algorithms 
that are fully free from the time step constraint due to the stability condition; 
examples include the energy stable finite difference methods \cite{Wise2009, wise20092, Elsey2012, zhang2013, 
Yang_ascalable} and the energy stable finite element methods \cite{Gomez201252, vignal2015, guo2016}. 
But because of the lack of the supportive theory, 
%systematic derivation, 
the design of these methods are often problem-dependent and not easy to generalize.
In this paper, we design a second-order discretization scheme for the PFC equation based on 
the discrete variational derivative (DVD) method \cite{DVDM} to  naturally achieve the energy stability.
And by exploiting the DVD method, the numerical scheme can be designed in a general way, which is 
suitable for the PFC equation with different types of the mobility and boundary conditions.
Further more, we employ an adaptive time stepping 
as a companion of the unconditionally energy stable method so as to adjust the time step size without 
losing the accuracy.

When an implicit discretization of the PFC equation is applied, a sparse linear or nonlinear algebraic system 
arises at each time step. And the solution of the discretized system could be time consuming especially 
when a fine mesh is used. It is therefore of great importance to study highly efficient solvers to accelerate 
the simulation at large scale. Despite the fact that some effective approaches, such as the multigrid 
method, have been applied in solving the discretized PFC equations \cite{guo2016, precondition, wise20092}, 
dedicated studies on efficient parallel solution algorithms are less to be seen. In this paper, we propose a highly scalable parallel 
solver based on the Newton--Krylov--Schwarz (NKS) algorithm \cite{NKS} with modified subdomain boundary 
conditions for solving the nonlinear algebraic system arising at each implicit time step. Several key parameters 
in the solver, including the type of the Schwarz preconditioner, the size of the overlap, and the solver for 
subdomain problems, are discussed and tested to achieve the optimal performance. We show by experiments 
that the proposed solver can scale well to over ten thousands processor cores.

The remainder of this paper is organized as follows. 
The PFC equation is introduced in Sec. 2. In Sec. 3, we employ the DVD method to 
obtain an unconditionally stable scheme for 
the PFC equation. In Sec. 4, we introduce the NKS algorithm  to solve the nonlinear system at each time step. 
Several numerical simulations are reported in Sec. 5 and concluding remarks are given in Sec. 6.

\section{Phase Field Crystal Equation}
\label{sec:1}
The free-energy functional in the PFC model takes the following 
dimensionless form \cite{PhysRevLett.88.245701, PhysRevA.15.319}:
\begin{equation}
 F(\phi)=\int_{\Omega}\left\{ \frac{1-\gamma}{2}\phi^2 + \frac{1}{4}\phi^4 -\lvert \nabla \phi \rvert^2 
 +\frac{1}{2} (\Delta \phi)^2\right\}\md\bx,
 \label{GE}
\end{equation}
where  $\gamma>0$ is the quench depth for supercooling the material and 
$\phi: \Omega \rightarrow \mathbb{R}$ is the density distribution function to approximate the 
number density of atoms. In the PFC model, the quench depth is proportional to the deviation 
of the temperature from the melting temperature and the density distribution function 
is conserved during the non-equilibrium process. 
For simplicity, we focus the discussion on a two dimensional rectangle domain 
$\Omega=[0,L_x] \times [0,L_y]$. The three dimesional cases will be studied in the numerical experiments.
We impose the doubly periodic boundary conditions or Neumann-type boundary conditions 
$$\mathbf{n} \cdot \nabla \phi |_{\partial \Omega}=\mathbf{n} \cdot \nabla (\Delta \phi) |_{\partial \Omega}=
\mathbf{n} \cdot \nabla (\Delta^2 \phi) |_{\partial \Omega} =0$$
for $\phi$, with $\mathbf{n}$ being the outward normal of $\partial \Omega$.

Based on \eqref{GE}, a local energy functional  $G(\phi)$ can be defined as
\begin{equation}
G(\phi) = \frac{1-\gamma}{2}\phi^2 + \frac{1}{4}\phi^4 -\lvert \nabla \phi \rvert^2 +\frac{1}{2} (\Delta \phi)^2.
\label{localenergy}
\end{equation}   
Then the PFC equation takes the following form \cite{DVDM}
 \begin{equation}
 \phi_t = \nabla \cdot \left(M(\phi) \nabla \frac{\delta G}{\delta \phi}\right),
 \label{PFCeq}
 \end{equation}
 where $M(\phi) \geq 0$ is the mobility, and
  \begin{equation}
  \begin{aligned}
 \frac{\delta G}{\delta \phi} &= \frac{\partial G}{\partial \phi} 
 - \nabla \cdot \left(\frac{\partial G}{\partial \phi_x}, \frac{\partial G}{\partial \phi_y}
\right)^T +\left(\frac{\partial^2}{\partial x^2},\frac{\partial^2}{\partial y^2}\right)^T\cdot \left(\frac{\partial G}{\partial \phi_{xx}}, \frac{\partial G}{\partial \phi_{yy}}\right)^T \\
&= (1-\gamma)\phi + \phi^3 +2\Delta \phi + \Delta^2 \phi.
\end{aligned}
 \end{equation}
 is the variational derivative of 
 the local energy $G(\phi)$.
For the PFC equation, it can be proved that the time derivative of the free-energy $F(\phi)$ satisfies 
\begin{equation}
\label{DIS1}
\begin{aligned}
 & \frac{d}{dt}F = \frac{d}{dt}\int_\Omega G \md\bx \\
  &= \int_\Omega \left(\frac{\partial G}{\partial \phi}\frac{\partial \phi}{\partial t}+
  \frac{\partial G}{\partial \phi_x}\frac{\partial \phi_x}{\partial t}+\frac{\partial G}{\partial \phi_{y}}\frac{\partial \phi_{y}}{\partial t}
  +\frac{\partial G}{\partial \phi_{xx}}\frac{\partial \phi_{xx}}{\partial t}+\frac{\partial G}{\partial \phi_{yy}}\frac{\partial \phi_{yy}}{\partial t}
  \right)\md\bx \\
  &= \int_\Omega\frac{\delta G}{\delta \phi}\phi_t \md\bx + B_1 =  \int_\Omega\frac{\delta G}{\delta \phi}\nabla \cdot \left(M(\phi) \nabla \frac{\delta G}{\delta \phi}\right) \md\bx + B_1\\ 
    &= - \int_\Omega M(\phi) \nabla\frac{\delta G}{\delta \phi} \cdot \nabla\frac{\delta G}{\delta \phi} \md\bx + B_1 + B_2 \leq 0,
\end{aligned}
\end{equation}
which demonstrates the energy dissipative of the system. Here $B_1$ and $B_2$ are the boundary terms coming from the integration-by-parts formula. In particular, we have $B_1=B_2=0$ for the periodic and Neumann-type boundary conditions.

\section{The discretization of the PFC equation}
\label{sec:2}
In the section, we use the DVD method to construct a numerical scheme of the PFC equation. 
By denoting $G_\delta$ as $G(\phi+\delta \phi,...,\phi_{yy}+\delta\phi_{yy})$, 
the following integral relationship between the variational derivative and the local energy holds
\begin{equation}
\label{condG}
\begin{aligned}
&~\int_\Omega\left(G_\delta
-G(\phi,\phi_x,\phi_{y}, \phi_{xx},\phi_{yy})\right)\md\bx\\ 
&=\int_\Omega \left(\frac{\partial G}{\partial \phi}\delta \phi + \frac{\partial G}{\partial \phi_x}\delta \phi_x 
+ \frac{\partial G}{\partial \phi_y}\delta \phi_y + \frac{\partial G}{\partial \phi_{xx}}\delta \phi_{xx} 
+ \frac{\partial G}{\partial \phi_{yy}}\delta \phi_{yy}\right)\md\bx\\
&~~~~+O\left((\delta \phi)^2\right)\\
&=\int_\Omega \frac{\delta G}{\delta \phi}\delta \phi \md\bx 
+ B_3+ O\left((\delta \phi)^2\right)\\ 
\end{aligned}
\end{equation}
The first equality comes from the Taylor expansion formula. The second equality comes from 
the integration-by-parts formula, and the boundary term $B_3=0$
under the given boundary conditions. Eq.~\eqref{condG} plays an important role in the DVD method 
due to the fact that it shows 
the connection between the discrete variational derivative and the discrete local energy. 􏰙

We use a uniform mesh of $N_x\times N_y$ elements with mesh sizes 
$\Delta x = L_x/N_x$ and $\Delta y = L_y/N_y$ to cover the computational domain $\Omega$. The solution $\phi$ is approximated as $\phi_{i,j}\approx \phi(x_i,y_j)$, in which 
$(x_i, y_j) = \left((i-\frac{1}{2})\Delta x, (j-\frac{1}{2})\Delta y\right)$, $1\leq i \leq N_x, 1\leq j \leq N_y$. 
$\phi_{i,j}^{(n)}$ denotes the numerical solution at $n$-th time step corresponding to time  $t_n$. 
We introduce some useful notations as following 
$$D_x^+\phi_{i,j}=\frac{\phi_{i+1,j}-\phi_{i,j}}{\Delta x},\ \ D_x^-\phi_{i,j}=\frac{\phi_{i,j}-\phi_{i-1,j}}{\Delta x},$$
 $$D_x \phi_{i,j}=\frac{\phi_{i+\frac{1}{2},j}-\phi_{i-\frac{1}{2},j}}{\Delta x}, \ \ D_x^{\langle 2 \rangle}\phi_{i,j} = \frac{\phi_{i+1,j}-2\phi_{i,j} +\phi_{i-1,j}}{(\Delta x)^2},$$
 and $D_y^+, D_y^-, D_y, D_y^{\langle 2 \rangle}$ are defined similarly. 
In this article, we use the subscript $d$ to 
indicate the corresponding discrete forms of operators and functions. And we denote 
$$\nabla_d=(D_x, D_y)$$
as the discrete gradient operator. The Laplacian operator $\Delta$ 
is discretized by
$$\Delta_d=\nabla_d \cdot \nabla_d = D_x^{\langle 2 \rangle} + D_y^{\langle 2 \rangle},$$
and operator $\nabla \cdot M(\phi) \nabla$ is discretized by
$$(\nabla \cdot M(\phi) \nabla)_d = \nabla_d \cdot M(\phi_{i,j}) \nabla_d = D_x M(\phi_{i,j}) D_x 
+ D_y M(\phi_{i,j}) D_y,$$
where the value of $M$ on the midpoint of an edge is approximated by the averaged value of $M$ on the two adjacent nodes of the edge; 
for instance, $M(\phi_{i+\frac{1}{2},j}) \approx \frac{M(\phi_{i+1,j}) + M(\phi_{i,j})}{2}$.

To compute the discrete variational derivative, we first define the discrete local energy and discrete free-energy for 
the PFC equation. The discrete approximation of the local free-energy $G$ in Eq. \eqref{localenergy} is
\begin{equation}
\label{relations}
\begin{aligned}
  G_d(\phi_{i,j}^{(n)})&= \frac{1-\gamma}{2}(\phi_{i,j}^{(n)})^2 +\frac{1}{4}(\phi_{i,j}^{(n)})^4
  +\frac{1}{2}\left(D_x^{\langle 2\rangle}\phi_{i,j}^{(n)}+D_y^{\langle 2\rangle}\phi_{i,j}^{(n)}\right)^2\\ 
          & - \left(\frac{(D_x^+\phi_{i,j}^{(n)})^2+(D_x^-\phi_{i,j}^{(n)})^2}{2}+
          \frac{(D_y^+\phi_{i,j}^{(n)})^2+(D_y^-\phi_{i,j}^{(n)})^2}{2}\right).
   \end{aligned}
\end{equation}
We define the discrete free-energy $F_d$ at time $t_n$ as 
\begin{equation}
\label{freeenergy}
\begin{aligned}
F_d^{(n)} = \sum_{i=1}^{N_x}\sum_{j=1}^{N_y}G_d(\phi_{i,j}^{(n)})\Delta x \Delta y. 
\end{aligned}
\end{equation}
Taking $\delta \phi := \phi^{(n+1)}_{i,j} - \phi^{(n)}_{i,j}$,  
we can obtain  
\begin{equation}
G_\delta-G(\phi,\phi_x,\phi_y,\phi_{xx},\phi_{yy}) :=  G_d(\phi_{i,j}^{(n+1)})-G_d(\phi_{i,j}^{(n)}).
\end{equation}
Then the discretization of Eq.~\eqref{condG} is derived as a summation formula 
\begin{equation}
\label{DVD}
\begin{aligned}
&~~~~\sum_{i,j} \left(G_{d}(\phi_{i,j}^{(n+1)})-G_{d}(\phi_{i,j}^{(n)})\right)\Delta x\Delta y\\
&=\sum_{i,j} \frac{\delta G_d}{\delta (\phi^{(n+1)},\phi^{(n)})_{i,j}}\left(\phi_{i,j}^{(n+1)}-\phi_{i,j}^{(n)}\right)\Delta x \Delta y,
\end{aligned}
\end{equation}
where $\frac{\delta G_d}{\delta (\phi^{(n+1)},\phi^{(n)})_{i,j}}$ expresses the discrete variational derivative 
obtained by solving Eq. \eqref{DVD}.
According to Eq.~\eqref{relations} and \eqref{DVD}, we can obtain the discrete variational derivative as \begin{equation}
\label{discretedG}
\begin{aligned}
 & \frac{\delta G_d}{\delta(\phi^{(n+1)},\phi^{(n)})_{i,j}}= \left[(1-\gamma)+2\Delta_d +\Delta_d^2 \right]\left(\frac{\phi_{i,j}^{(n+1)}+\phi_{i,j}^{(n)}}{2}\right)\\ 
  &~~~~~~~~~~~~~~~~~+\frac{(\phi_{i,j}^{(n+1)})^3+(\phi_{i,j}^{(n+1)})^2\phi_{i,j}^{(n)}+\phi_{i,j}^{(n+1)}(\phi_{i,j}^{(n)})^2+(\phi_{i,j}^{(n)})^3}{4}.
   \end{aligned}
\end{equation}

When a first-order finite difference formula is applied, we may obtain a semi-implicit scheme for Eq.~\eqref{PFCeq} as
\begin{equation}
\frac{\phi_{i,j}^{(n+1)}-\phi_{i,j}^{(n)}}{\Delta t_n} = \nabla_d \cdot M_d \nabla_d 
 \frac{\delta G_d}{\delta(\phi^{(n+1)},\phi^{(n)})_{i,j}}.
\label{NSOP}
\end{equation} 
Here, $\Delta t_n = t_{n+1}-t_n$ is the time step size, and $M_d = M\left(\frac{\phi_{i,j}^{(n+1)}+\phi_{i,j}^{(n)}}{2}\right)$. 
The scheme can be viewed as the standard Crank--Nicolson scheme, 
which is trivial for the linear terms, but is non-trivial when dealing with the nonlinear terms. 
Numerical simulations presented later will show that the scheme~\eqref{NSOP} has numerical second-order accuracy 
in time and space \cite{NMFP}. 
%Numerical second-order or Proved second-order 
It's worth mentioning that the discrete variational derivative in Eq.~\eqref{discretedG} derived by DVD method happens 
to have a similar form with the numerical scheme for PFC equation in \cite{zhang2013}. However, the numerical scheme presented in \cite{zhang2013} employ a first-order forward difference instead of a second-order centered one 
to discrete the gradient operator, therefore the scheme is formally first-order 
accurate in space in order to maintain the energy stability.

The scheme~(\ref{NSOP}) has a crucial feature that 
the dissipation property is kept no matter what time step size is adopted. The proof can be 
derived analogously to the continuous case. In addition, it can be easily proved that the scheme~\eqref{NSOP} 
keeps the conservation property of the density distribution function $\phi$ during the whole time evolution process.

\vspace{0.2cm}
\noindent\textbf{Proposition.}
Under the periodic or Neumann-type boundary conditions, the numerical scheme~\eqref{NSOP} is 
unconditionally energy 
stable. More precisely, for any time step size $\Delta t > 0$, the solution of the scheme~\eqref{NSOP}
satisfies the energy dissipation
\begin{equation}
F_d(\phi^{(n+1)})\leq F_d(\phi^{(n)}).
\end{equation}
\textbf{Proof.}
With the periodic boundary conditions, we present two vital formulas that will be used in the proof
\begin{equation}
\label{summation}
\begin{aligned}
&~\sum_{i=1}^{N_x}g_{i,j}(D_x M_{i,j}D_x)g_{i,j}= -\sum_{i=1}^{N_x}\frac{M_{i,j}}{2}\left[(D_x^+ g_{i,j})^2 +(D_x^- g_{i,j})^2\right],\\
&~\sum_{j=1}^{N_y}g_{i,j}(D_y M_{i,j}D_y)g_{i,j}= -\sum_{j=1}^{N_y}\frac{M_{i,j}}{2}\left[(D_y^+ g_{i,j})^2 +(D_y^- g_{i,j})^2\right].
   \end{aligned}
\end{equation}
The Eq.~\eqref{summation} can be regarded as a discrete analogue of the integration-by-parts formula and their demonstration is omitted here. According to Eq. \eqref{summation}, the dissipation property of the discrete free-energy $F_d$ is given by 
\begin{equation}
\begin{aligned}
  &~ \frac{F_d(\phi^{(n+1)})-F_d(\phi^{(n)})}{\Delta t_n}\\ 
   &= \sum_{i=1}^{N_x}\sum_{j=1}^{N_y}\frac{\delta G_d}{\delta (\phi^{(n+1)},\phi^{(n)})_{i,j}}\left(\frac{\phi_{i,j}^{(n+1)}-\phi_{i,j}^{(n)}}{\Delta t_n}\right)\Delta x\Delta y\\ 
  &=\sum_{i=1}^{N_x}\sum_{j=1}^{N_y}\left(\frac{\delta G_d}{\delta (\phi^{(n+1)},\phi^{(n)})_{i,j}}\right)\nabla_d \cdot M_{i,j} \nabla_d\left(\frac{\delta G_d}{\delta (\phi^{(n+1)},\phi^{(n)})_{i,j}}\right)\Delta x\Delta y \\ 
  &= -\sum_{i=1}^{N_x}\sum_{j=1}^{N_y}\frac{M_{i,j}}{2} \left[(D_x^+ \mathcal{A})^2 +(D_x^- \mathcal{A})^2 + (D_y^+ \mathcal{A})^2 + (D_y^- \mathcal{A})^2\right]\Delta x\Delta y \leq 0.
   \end{aligned}
\end{equation}
In the third equality, we simplify the notation $\frac{\delta G_d}{\delta (\phi^{(n+1)},\phi^{(n)})_{i,j}}$ 
as $\mathcal{A}$ for brevity. The demonstration can be derived analogously for the Neumann-type 
boundary conditions.

The unconditional stability of the semi-implicit scheme \eqref{NSOP} is obtained from the dissipation 
property. But a blind increase of the time step size is adverse for keeping the computational accuracy. 
Numerical experiments show that using a large constant time step may produce nonphysical solutions 
\cite{zhang2013}. This is because the PFC equation, similar to other phase-field equations, contains 
multiple time scales that may vary in orders of magnitude during the coarsening and phase separation 
processes. Therefore an adaptive control of the time step size is necessary in the numerical simulation, 
in which the time step size is selected based on the desired solution accuracy and the dynamic 
features of the system. 

%In the semi-implicit scheme \eqref{NSOP}, a nonlinear system as follows 
%\begin{equation}
%{\cal F}(\mathbf{\Phi},\mathbf{\Phi}^n;\Delta t_n):=\frac{\mathbf{\Phi}-\mathbf{\Phi}^{(n)}}{\Delta t_n}+{\cal E}(\mathbf{\Phi},\mathbf{\Phi}^{(n)})=0
%\label{implicit-s}
%\end{equation}
%needs to be solved at each time step with the time step size $\Delta t_n$. Here, $\mathbf{\Phi}^{(n)}$ 
%represents the numerical solution at $t_n$ and $\mathbf{\Phi}$ represents the unknown numerical solution 
%at $t_{n+1}$, ${\cal E}$ depends on the spatial discretization scheme.

In this paper, we exploit the adaptive time step strategies described in \cite{zhang2013}, 
the time step size is computed by
\begin{equation}
\label{sencondt}
\Delta t_n = \max \left(\Delta t_{min},\frac{\Delta t_{max}}{\sqrt{1+\eta \lvert F'(t)\rvert^2}} \right),
\end{equation}
where $\lvert F'(t)\rvert =|\frac{F_d^{(n)}-F_d^{(n-1)}}{\Delta t_{n-1}}|$ corresponds to the change rate of free-energy on the two previous time steps, 
and $\eta$ is chosen to adjust the level of adaptivity in which large (small) $\eta$ indicates strengthening (easing) 
the restriction to the time step size. $\Delta t_{min}$ and $\Delta t_{max}$ are defined as the lower and upper bounds of the time step size, namely $\Delta t_{min} \leq \Delta t_n \leq \Delta t_{max}$.
In the PFC model, the free-energy $F$ decays quickly at the early stage of dynamics 
because of the nonlinear interaction, and then decays rather slowly until it reaches a steady state. 
By using the adaptive time step strategy in Eq. \eqref{sencondt}, the time step size is adjusted promptly based on the change rate of the free-energy $F$, in which the large $\lvert F'(t)\rvert$ leads to the small time step size, and the small $\lvert F'(t)\rvert$ yields the large time step size. 

\section{Newton--Krylov--Schwarz Algorithm}
In the semi-implicit scheme \eqref{NSOP}, the PFC equation is discretized into a nonlinear system
\begin{equation}
{\cal F}(\mathbf{\Phi})=0
\label{NLS}
\end{equation}
at each time step. In the paper, we solve the nonlinear system~\eqref{NLS} on a parallel supercomputers by adopting 
an NKS type algorithm \cite{NKS}. The NKS algorithm consists of three important components: (1) an inexact Newton method as the outer iteration; (2) a Krylov method as an inner iteration for the linear Jacobian system at each 
Newton iteration; and (3) a Schwarz preconditioner to improve the Krylov method.

At each time step, the nonlinear system \eqref{NLS} is solved by an inexact Newton method. The solution of the previous time step is set to be the initial guess which has a great impact on the convergence of the iteration. 
At the $(m+1)$-th step of the inexact Newton iteration, the new approximate solution $\mathbf{\Phi}_{m+1}$ is obtained from the current approximate solution $\mathbf{\Phi}_{m}$ through
\begin{equation}
\label{newton}
\mathbf{\Phi}_{m+1}=\mathbf{\Phi}_{m} +\lambda_m\mathbf{S}_m,\ \ \ m=0,1,\cdots .
\end{equation}
 Here $\lambda_m$ is the step length determined by a line search procedure \cite{NMFU}, and 
$\mathbf{S}_m$ is the search direction obtained by approximately solving the following linear Jacobian system 
\begin{equation}
J_m\mathbf{S}_m = -{\cal F}(\mathbf{\Phi}_m),
\label{Jacobi}
\end{equation}
where $J_m=\frac{\partial {\cal F}(\mathbf{\Phi}_m)}{\partial \mathbf{\Phi}_m}$ is the Jacobian matrix. 
The stopping condition for the Newton iteration \eqref{newton} is 
\begin{equation}
\|{\cal F}(\mathbf{\Phi}_{m+1})\| \leq \max\{\varepsilon_r
\|{\cal F}(\mathbf{\Phi}_0)\|, \varepsilon_a\},
\end{equation}
where $\varepsilon_r, \varepsilon_a \geq 0$ are the relative and absolute tolerances for the 
nonlinear iterations, respectively. 
Compared to the classical Newton method, the inexact method is superior especially when the number 
of unknowns is large 
(e.g., of the order of millions or larger) due to the reason that the linear Jacobian system 
is solved approximately instead of exactly, leading to a substantial reduction of the computational cost. 
%Instead of solving the , a right-preconditioned linear system is solved by gmres until ......

To accelerate the convergence of the linear solver, a right-preconditioned linear system
\begin{equation}
\label{hjacobi}
J_mH_m^{-1}(H_m\mathbf{S}_m) = -{\cal F}(\mathbf{\Phi}_m),
\end{equation}
is solved instead of the original Jacobian system \eqref{Jacobi}.
In our study, the Generalized Minimal Residual (GMRES) method \cite{gmres} is applied to approximately 
solve the right-preconditioned linear system \eqref{hjacobi}
until the linear residual $\mathbf{r}_m = J_m\mathbf{S}_m + {\cal F}(\mathbf{\Phi}_m)$ satisfies 
the stopping condition
$$\|\mathbf{r}_m\| \leq  \textnormal{max} \{\xi_r\|{\cal F}(\mathbf{\Phi}_m)\|, \xi_a\},$$
where $\xi_r, \xi_\alpha \geq 0$ are the relative and absolute tolerances for the linear iterations, respectively. 
In the GMRES method, the additive Schwarz type preconditioner $H_m^{-1}$ is the key to 
the success of the linear solver. To define the additive Schwarz type preconditioner 
$H_m^{-1}$ , we first partition the computational domain $\Omega$ into $np$ non-overlapping 
subdomains $\Omega_k,\ (k = 1, 2, \cdots, np)$, then extend each subdomain by $3\delta$ mesh 
layers to form an overlapping decomposition $\Omega =\cup_{k=1}^{np}\Omega_k^{\delta}$, in which 
$3$ is the stencil width of the finite difference scheme for the PFC equation. 
The classical additive Schwarz preconditioner  \cite{DDA}
is defined as
\begin{equation}
H_m^{-1}(\delta\delta)=\sum_{k=1}^{np}(R_k^{\delta})^T \textnormal{inv}(A_k)R_k^{\delta}.
\label{invM}
\end{equation}
Here $R_k^{\delta}$ serves as a restriction operator that maps a vector to a new one that is defined in the 
subdomain $\Omega_k^{\delta}$, by discarding the components outside $\Omega_k^{\delta}$; 
$(R_k^{\delta})^T$ represents an extension operator that maps a vector defined in the 
subdomain $\Omega_k^{\delta}$ to a new one that is defined in the whole domain, by putting zeros at the components outside $\Omega_k^{\delta}$. 

There are two modified versions of the AS preconditioner that may have some
potential advantages. The first one is the left restricted 
additive Schwarz (left-RAS, \cite{cai99sisc}) preconditioner 
that reads
\begin{equation}\label{eq:ras}
H_m^{-1}(0\delta) =  \sum_{k=1}^{np} (R_{k}^0)^T \textnormal{inv}(A_{k}) R_{k}^{\delta}.
\end{equation}
The only difference between the left-RAS preconditioner and the AS preconditioner
is the extension operator.
Instead of $(R_{k}^{\delta})^T$, the left-RAS preconditioner uses $(R_{k}^0)^T$ 
which puts zeros not only 
outside $\Omega_k^{\delta}$ but also outside $\Omega_k$.
The other modification to the AS preconditioner is the
right restricted  additive Schwarz (right-RAS, \cite{cai03sinum_ash}) preconditioner 
that is given by
\begin{equation}\label{eq:ash}
H_m^{-1}(\delta 0) =  \sum_{k=1}^{np} (R_{k}^{\delta})^T \textnormal{inv}(A_{k}) R_{k}^0.
\end{equation}
The only difference between the right-RAS preconditioner and the AS preconditioner
is the restriction operator.
Instead of $R_{k}^{\delta}$, the right-RAS preconditioner uses $R_{k}^0$ 
which ignores the entries outside $\Omega_k$ when doing the extension.

In Eq.~\eqref{invM}-\eqref{eq:ash}, the subdomain matrix $A_k$ can be directly generated as
\begin{equation}
A_k=R_k^{\delta}J_m(R_k^{\delta})^T.
\label{Ak}
\end{equation}
But people usually do not employ the method in Eq. \eqref{Ak} due to the high computational cost caused by 
the usage of the global matrix $J_m$ in the formula. 
Likewise, we use the discretization of the subdomain problem to generate $A_k$. Therefore, we need to 
consider what boundary conditions should be imposed on the subdomain boundaries. 
%Choosing different interface conditions can lead to very different convergence results. 
In our approach, we employ the boundary conditions as follows
\begin{equation}
\phi=0,\ \ \ \ \Omega_k^{\delta+1} \backslash \Omega_k^{\delta},
\end{equation}
where $\Omega_k^{\delta+1}$ is the $3$ mesh layers expansion region of $\Omega_k^{\delta}$ that 
ensures all mesh points in $\Omega_k^{\delta}$ can acquire 
sufficient information to perform the stencil calculations when solving the subdomain problem.
After defining suitable interface conditions for the subdomain problems, we then solve them 
either directly by using a sparse LU factorization or approximately by using a sparse incomplete 
LU (ILU) factorization.

A great advantage of the additive Schwarz preconditioners is that communication only occurs 
between neighboring subdomains during the restriction and extension processes. The major cost 
of the additive Schwarz preconditioners is the subdomain solves which are done sequentially 
without any inter-process communication. Therefore the additive Schwarz preconditioners 
is naturally suitable to parallel computing as long as the number of iterations is kept low. We further 
remark that compared to the classical AS preconditioner, the communication in the two restricted 
versions is reduced approximately by half because only the restriction or the extension step requires communication. And experiment results show that the restricted Schwarz preconditioners also 
have similar levels of convergence rate to the classic case, if not superior to. This may further improve 
the performance of the preconditioner.

\section{Numerical experiments}
In this section, we investigate the numerical behavior and parallel performance of the proposed algorithm. 
We begin with several numerical tests for the PFC equation to validate the discretization of the proposed algorithm. 
In addition, we investigate different performance-related 
parameters in the NKS algorithm to obtain the best performance. We consider four test cases, including three two 
dimensional (2D) test cases and a three dimensional (3D) test case. 
%in which the mobility $M(\phi)$ is fixed to be $1$ for convenience. 
We mainly focus on (1) the verification of the numerical accuracy of the semi-implicit method, (2) 
the parallel performance of the semi-implicit method for various parameters, including the subdomain solvers and preconditioners, and (3) the parallel scalability of the proposed algorithm.

We perform our numerical experiments on the Sunway TaihuLight supercomputer, 
which tops the TOP--500 list as of June, 2016, with a peak performance greater than 100 PFlops.
The computing power of TaihuLight is provided by a homegrown many--core SW26010 CPU \cite{Fu2016}, 
in which we only enable one core per CPU socket for the current study. The algorithm for the PFC equation 
is implemented on top of the Portable, Extensible Toolkits for Scientific computations (PETSc, \cite{petsc}) 
library. The stopping conditions for the nonlinear and linear iterations are as follows. 
\begin{itemize}
\setlength{\itemsep}{0pt}
\item The relative tolerance for the nonlinear iteration: $\varepsilon_r = 1\times10^{-8}$.
\item The absolute tolerance for the nonlinear iteration: $\varepsilon_a = 1\times10^{-10}$.
\item The relative tolerance for the linear iteration: $\xi_r = 1\times10^{-3}$.
\item The absolute tolerance for the linear iteration: $\xi_a = 1\times10^{-11}$.
\end{itemize}

To test the accuracy of our scheme, we define the relative $l_2$ error as follows:
\begin{equation}
l_2 = \left(\frac{\sum_{i=1}^{N_x}\sum_{j=1}^{N_y}(\phi_{i,j}-\tilde{\phi}_{i,j})^2}{\sum_{i=1}^{N_x}\sum_{j=1}^{N_y}\tilde{\phi}^2_{i,j}}\right)^{\frac{1}{2}},
\end{equation}
where $\phi_{i,j}$ is the numerical solution, $\tilde{\phi}_{i,j}$ is the analytical solution.

\subsection{Validation of the proposed method}

\noindent A. Crystal growth in a 2D supercooled liquid
\vspace{0.2cm}

First, we use the PFC equation to model the crystal growth in a 2D supercooled homogeneous liquid. 
The simulation is conducted on a periodic square domain $\Omega = [0,128]^2$ with a random 
initial value $\phi_{i,j}^{(0)}=\bar{\phi} + \sigma_{i,j}$ in which $\bar{\phi} = 0.07$ is the average density of the liquid-crystal 
system and $\sigma_{i,j}\in[-0.07,0.07]$ is chosen randomly. The positive parameter $\gamma$ is set to be $0.025$, 
and the mobility $M(\phi)$ is set to be $1$ for convenience.
\begin{figure}[!b]
\begin{center}
\qquad\scriptsize{(a) $t = 1$}\qquad\qquad\qquad\qquad
\qquad\qquad\qquad\scriptsize{(b) $t = 500$}\\
{\includegraphics[width=0.45\textwidth]{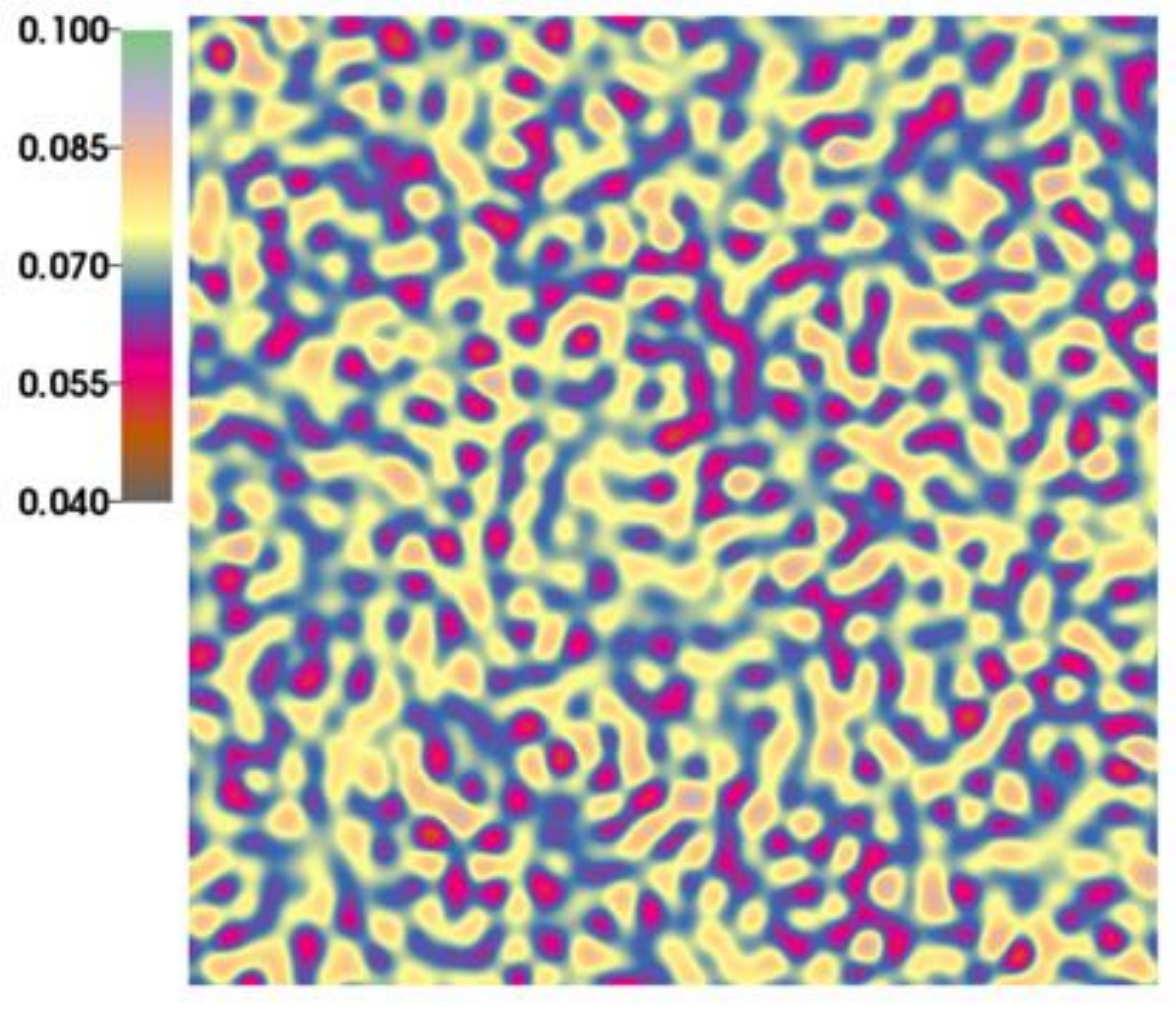}}
{\includegraphics[width=0.45\textwidth]{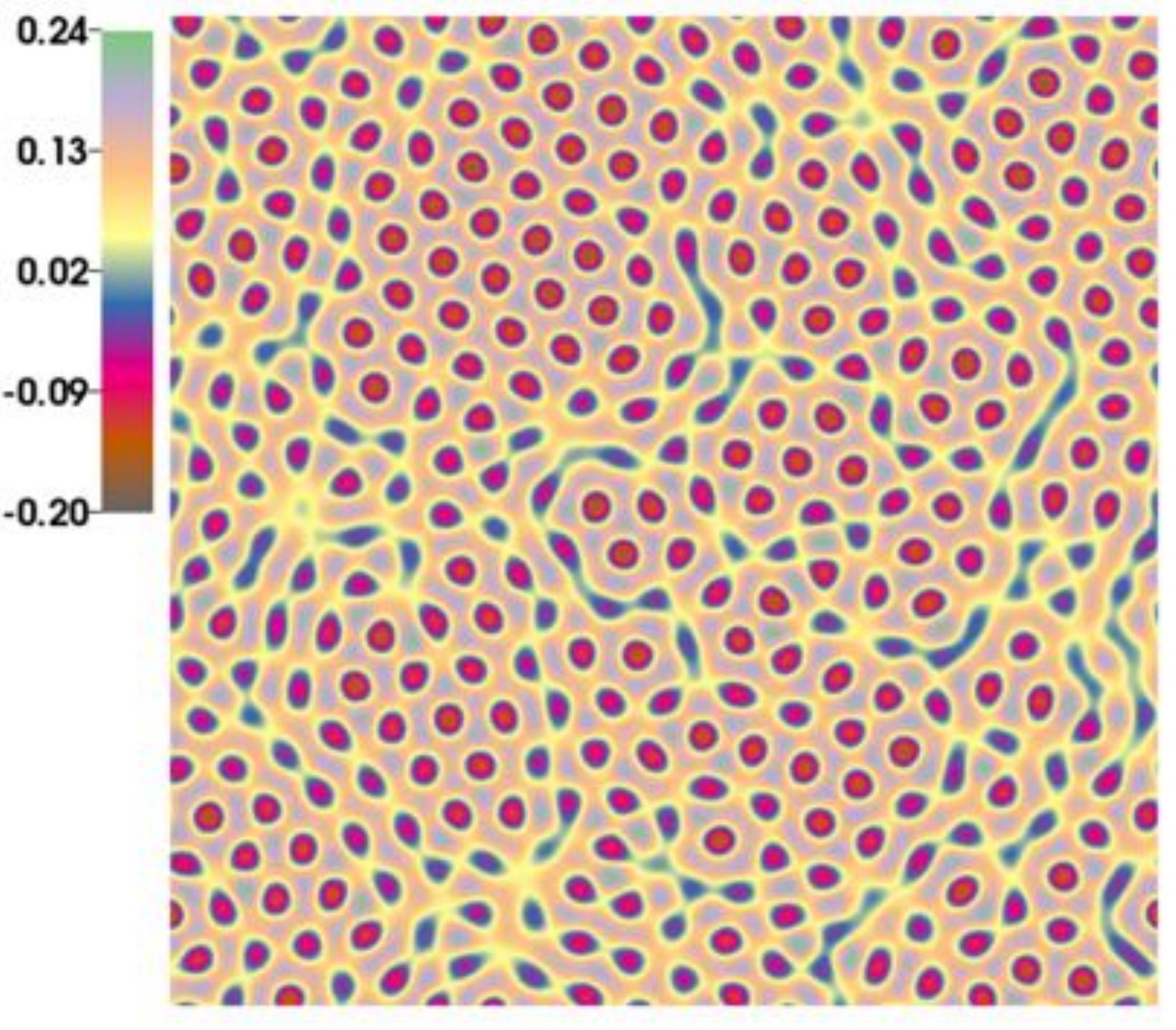}}
\\~~\\
\qquad\scriptsize{(c) $t = 1,200$}\qquad\qquad\qquad\qquad
\qquad\qquad\qquad\scriptsize{(d) $t = 3,200$}\\
{\includegraphics[width=0.45\textwidth]{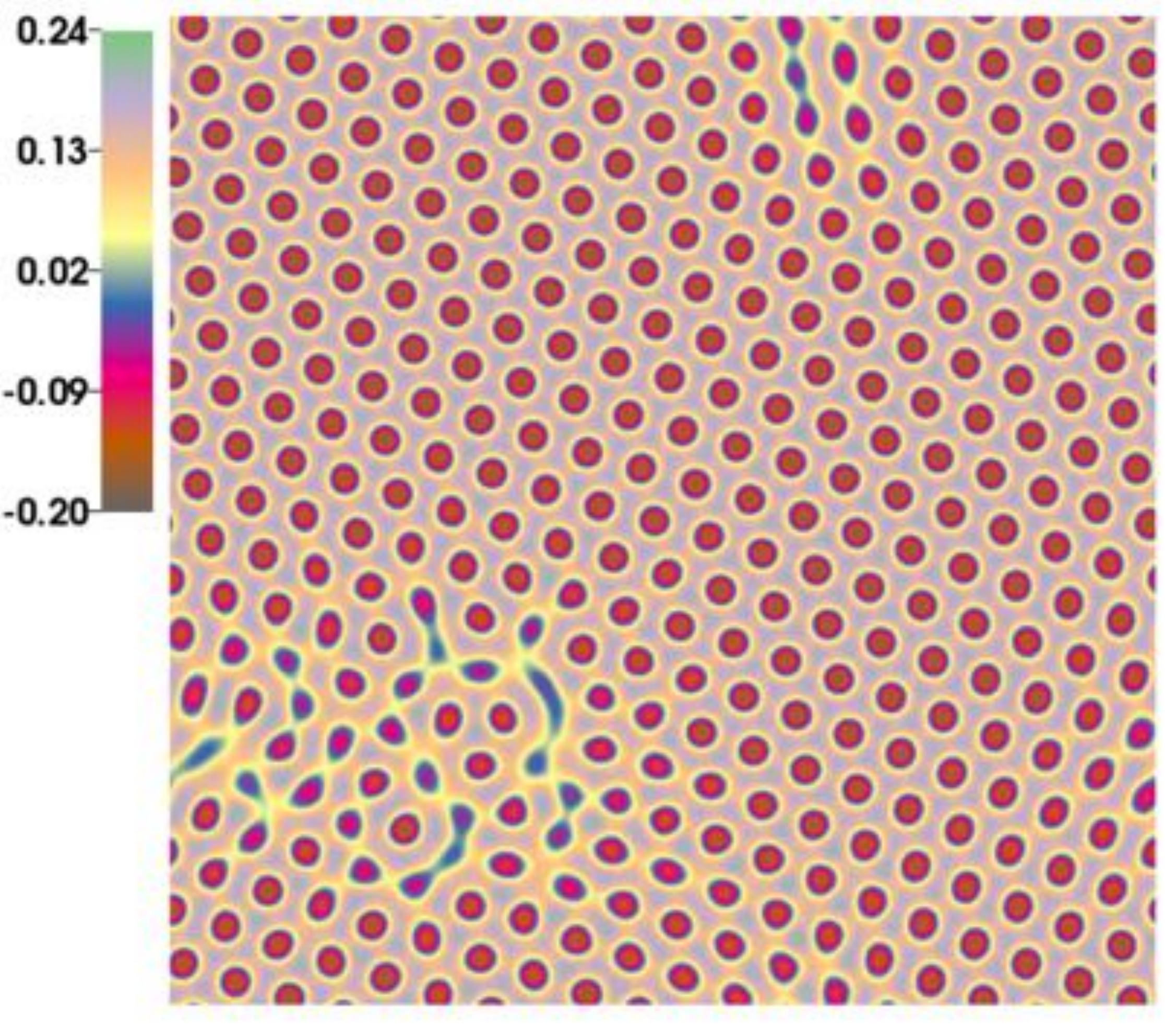}}
{\includegraphics[width=0.45\textwidth]{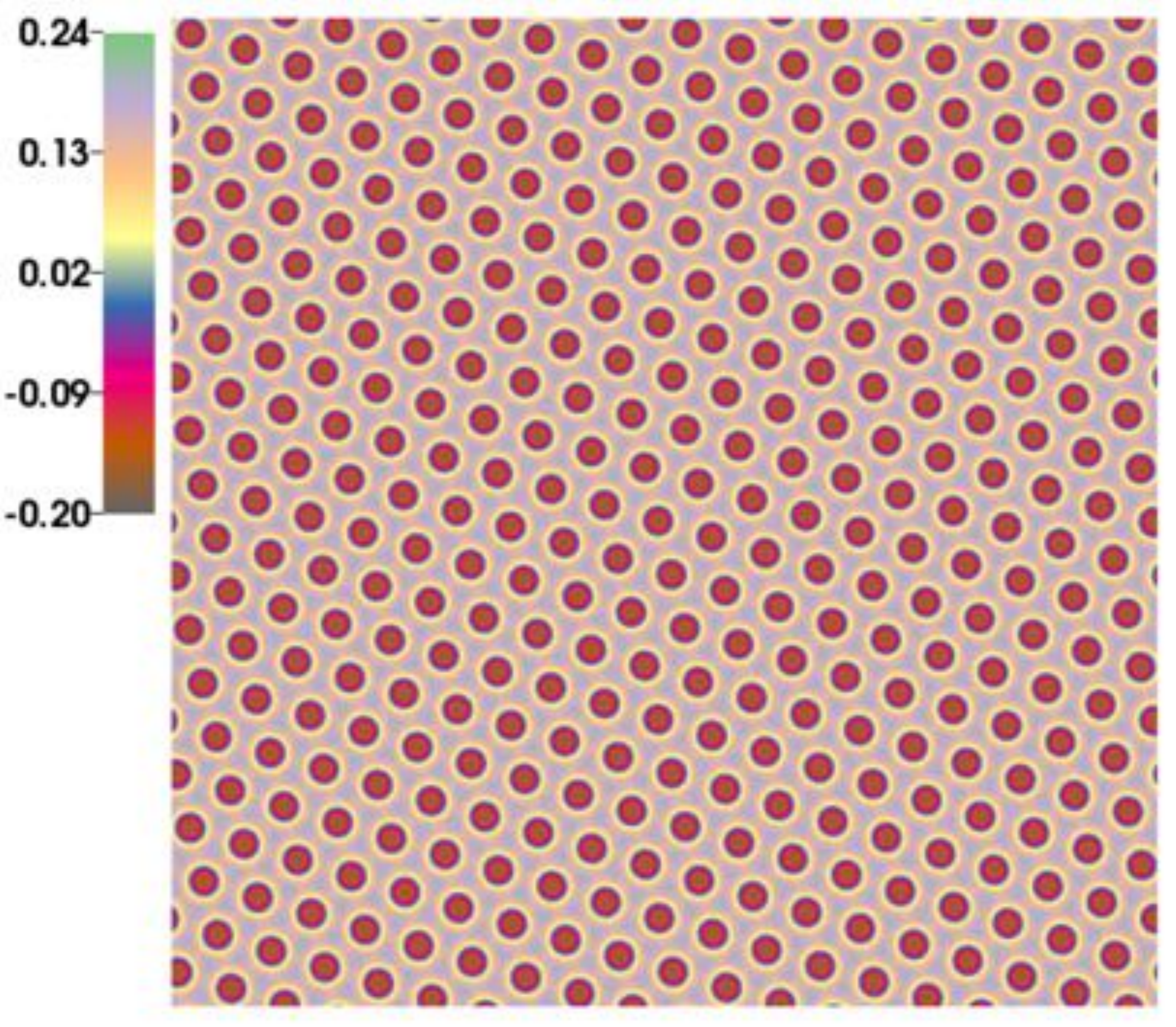}}
\end{center}
\caption{The density distribution of crystal growth in a 2D supercooled liquid.}
\label{ex1:sol}    
\end{figure}
Similar simulations were reported in \cite{Cheng_anefficient, wise20092, Yang_ascalable}.
We perform the simulation on a $256\times 256$ uniform mesh.
The time step size is adaptively controlled by using the adaptive time step strategy with $\Delta t_{min} = 0.01, 
\Delta t_{max}= 20$ and the parameter $\eta = 400,000$. 

\begin{figure}[!h]
\begin{center}
\scriptsize{(a)}\qquad\qquad\qquad\qquad\qquad\qquad\qquad
\qquad\qquad\qquad\scriptsize{(b)}\qquad\qquad\\
{\includegraphics[width=0.48\textwidth]{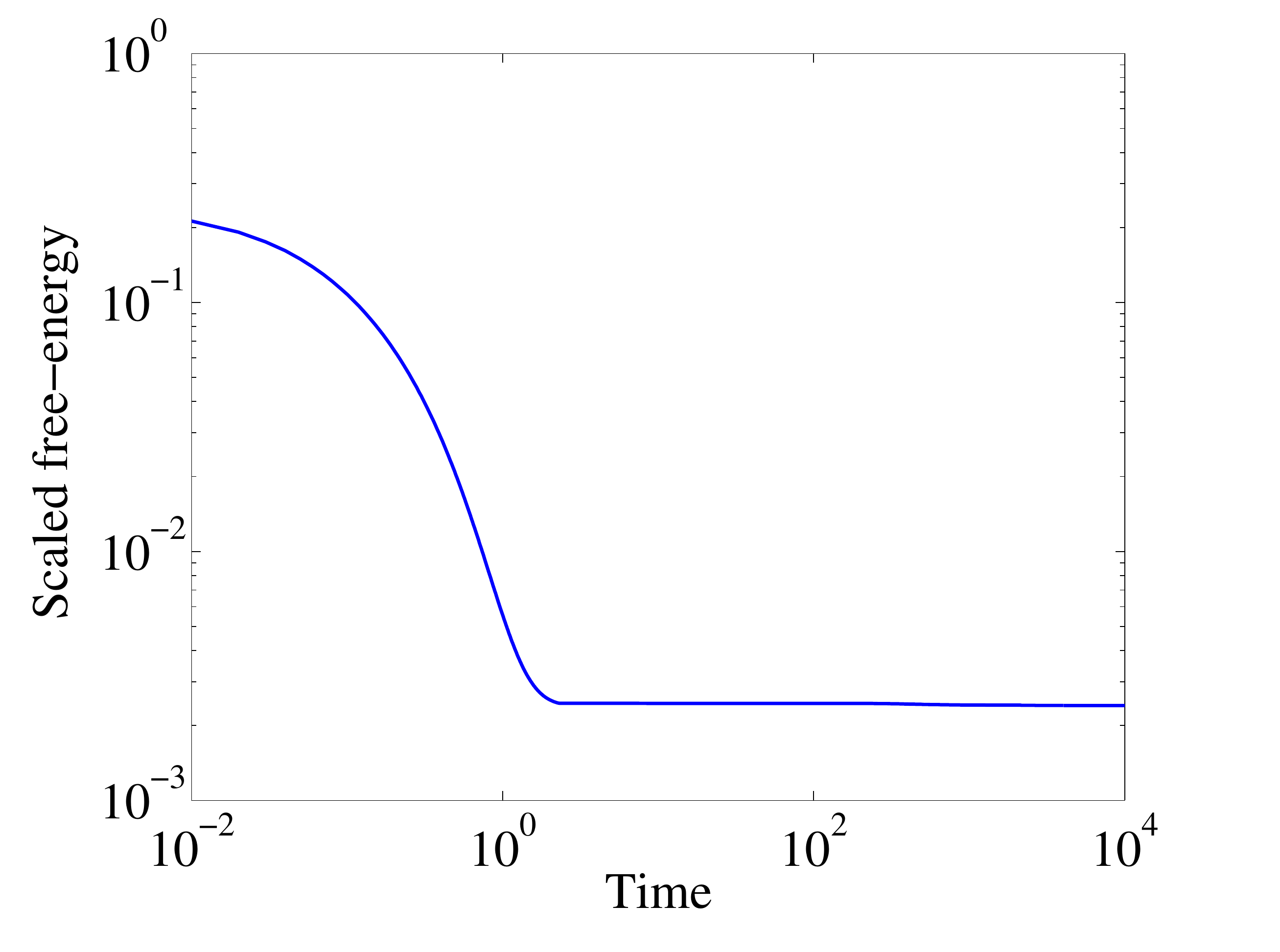}}
{\includegraphics[width=0.48\textwidth]{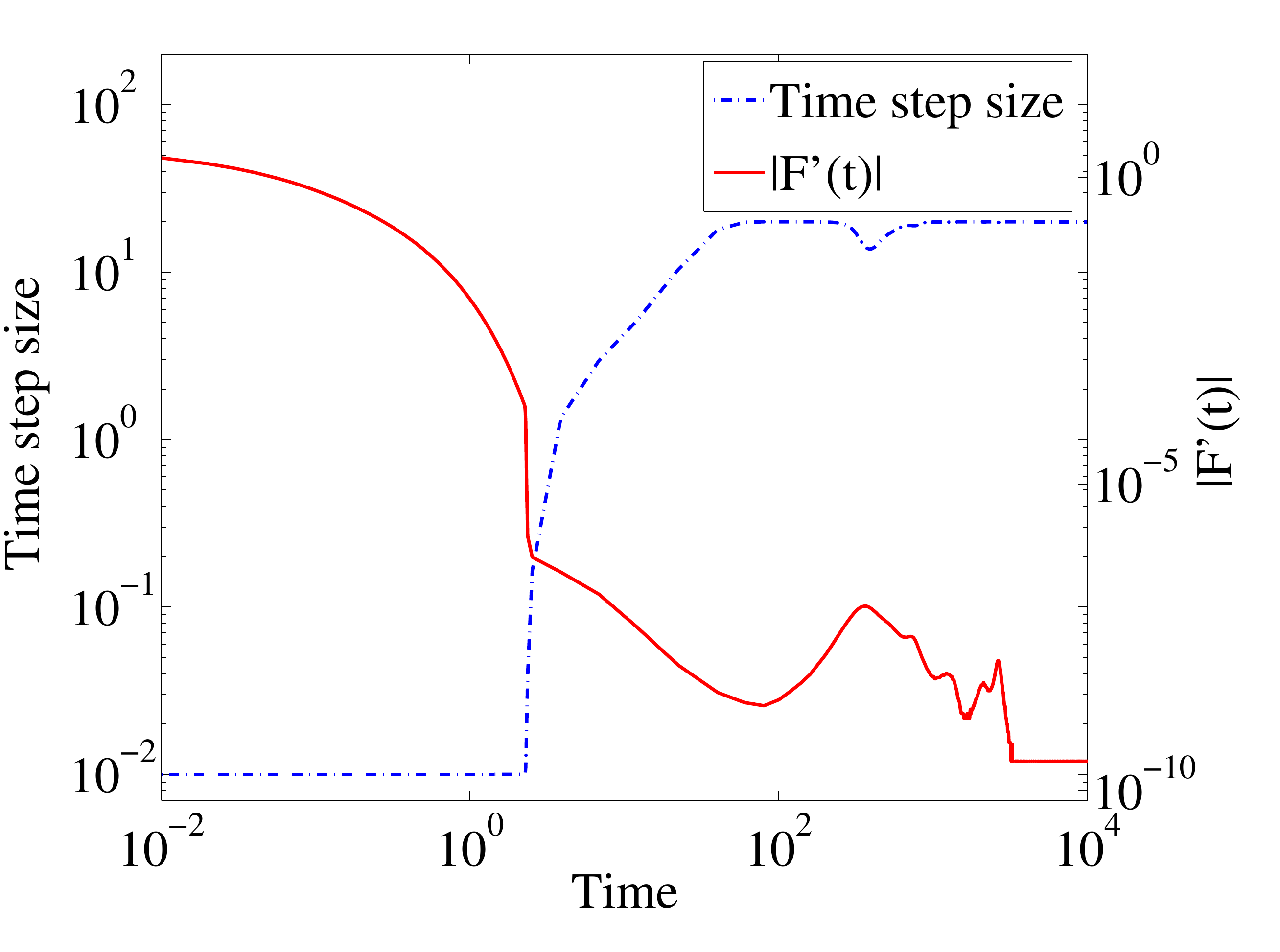}}
\end{center}
\caption{
The scaled total free-energy (a) and the history of the time step size (b). }
\label{ex1:energy} 
\end{figure}

Fig.~\ref{ex1:sol} shows the pseudocolor plots of the density distribution at times $t=1$, $500$, $1,200$ and $3,200$, 
from which we observe: (1) the fluid quickly crystallizes under the supercooling before $t = 1,200$; (2) after $t = 1,200$ 
the crystallized material gradually stabilizes as a solid lattice with periodic hexagonal pattern; (3) the domain is 
filled perfectly with periodic regular hexagonal pattern at $t = 3,200$. 
We present the scaled total free-energy $F/(L_xL_y)$ and the history of the time step size in 
Fig.~\ref{ex1:energy}, respectively. As shown in Fig.~\ref{ex1:energy} (a), the free-energy decreases monotonically 
to the minimal as the solution evolves to the steady state. 
We observe that the free-energy decays quickly at the early stage 
and then decays rather slowly.  
From Fig.~\ref{ex1:energy} (b), we observe that 
the time step size is rapidly adjusted from $\Delta t_{min}$ to 
$\Delta t_{max}$ due to the change of the free-energy. The time step is controlled by the variation of the free-energy 
on the two previous time step, in which the small time step means that the free-energy varies quickly and the 
large time step indicates that the free-energy varies slowly, which is 
consistent with the variation of the free-energy in Fig.~\ref{ex1:energy} (a).  
Compared to the other publications of the PFC equation \cite{wise20092, zhang2013, guo2016}, 
we obtain the same solution that ascertains the rightness of the semi-implicit scheme.

\begin{figure}[!h]
\begin{center}
\qquad\scriptsize{(a)}\qquad\qquad\qquad\qquad\qquad\qquad\qquad
\qquad\qquad\qquad\scriptsize{(b)}\\
{\includegraphics[width=0.48\textwidth]{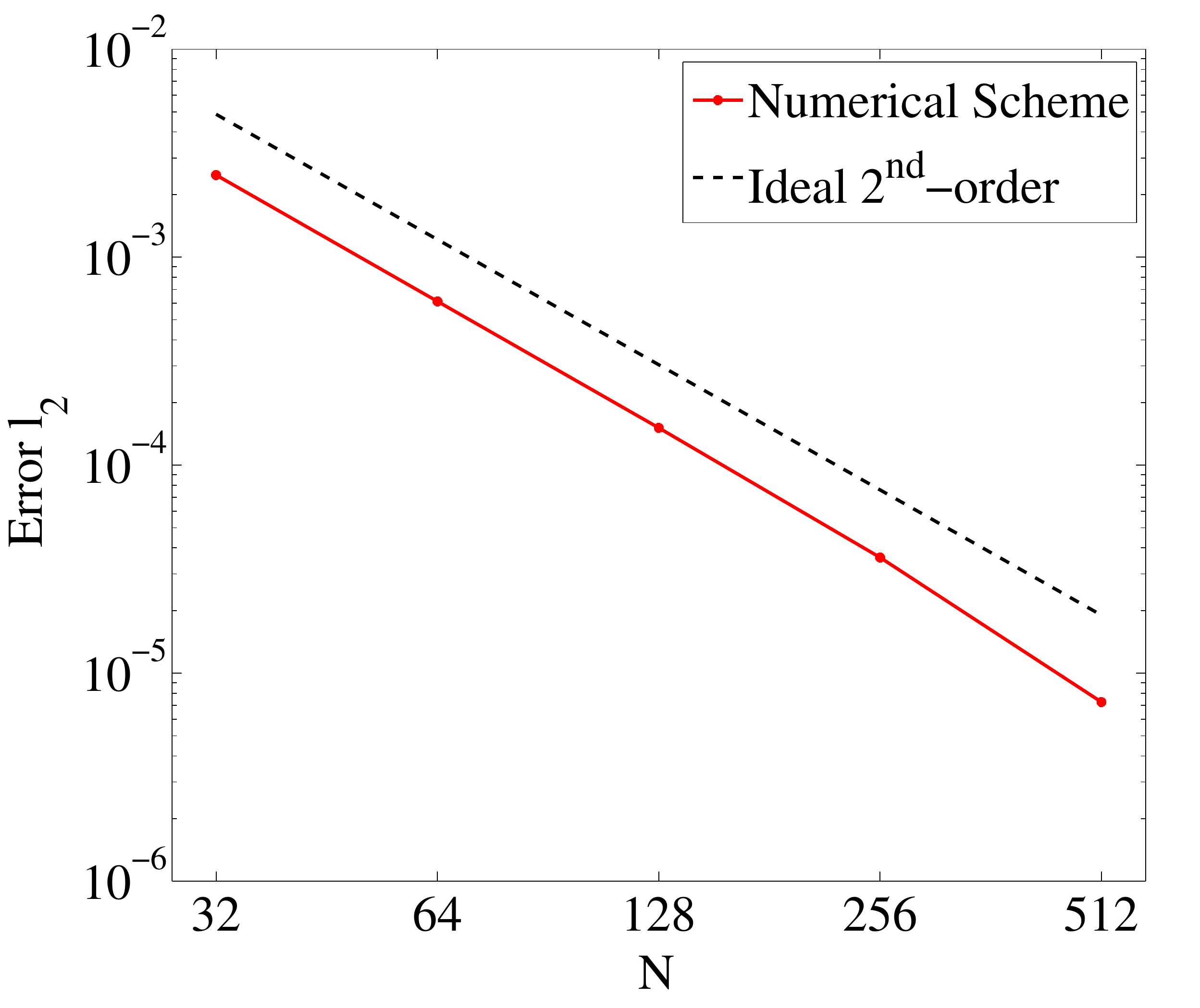}}
{\includegraphics[width=0.48\textwidth]{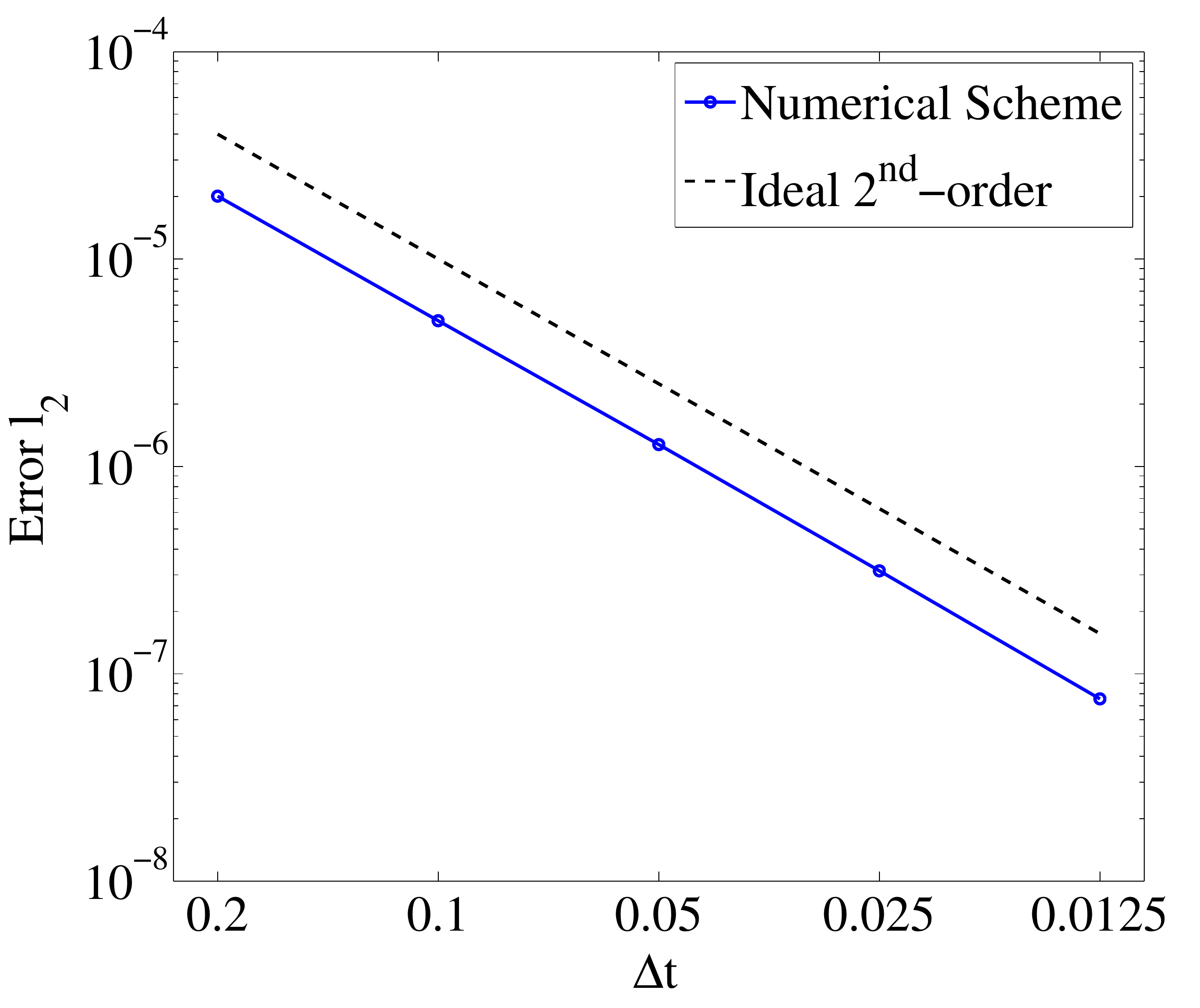}}
\end{center}
\caption{The $l_2$ error of the proposed scheme, (a) the spatial $l_2$ errors, (b) the temporal $l_2$ error.}
\label{ex1:accuracy} 
\end{figure}

Next, we study the accuracy of the proposed scheme. Because the random initial value is inappropriate 
for the check of the grid convergence, we consider the following smooth initial value
 \begin{equation}
 \phi^{(0)}(x,y) = 0.5\sin\left(\frac{2\pi x}{32}\right)\sin\left(\frac{2\pi y}{32}\right),\ \ (x,y)\in\Omega,
 \end{equation}
where $\Omega=[0,32]^2$ is periodic. The positive parameter $\gamma$ and mobility $M(\phi)$ is 
unchanged. 
%Since the exact density distribution function $\phi$ is unknown, we take the numerical solution obtained by 
%the presented scheme with a very fine mesh and very small fixed time step size as the analytical solution 
%$\tilde{\phi}_{i,j}$. 
To understand the accuracy of the spatial discretization, we run the test on gradually refined meshes. 
Since the exact density distribution function $phi$ is unknown, the numerical solution on a very fine 
mesh $1,024 \times 1,024$ and small time step size $\Delta t = 0.01$ is taken as the analytical solution 
$\tilde{\phi}_{i,j}$. The $l_2$ errors at time $t = 5$ for several mesh sizes ($N_x=N_y=N$) are plotted 
in Fig.~\ref{ex1:accuracy} (a) which clearly shows the pesented semi-implicit scheme has a second-order 
accuracy in space. 
Next we fix the spatial mesh as $512\times512$ to study the accuracy of the semi-implicit scheme in time. 
The numerical solution at a fixed time step size $\Delta t=0.0005$ is regarded as the analytical solution. 
The $l_2$ errors at time $t = 5$ are illustrated in Fig.~\ref{ex1:accuracy} (b), which validates the 
second-order accuracy of the semi-implicit scheme in time.

%\vspace{1.0cm}

\vspace{0.2cm}
\noindent B. Polycrystalline growth in a 2D supercooled liquid
\vspace{0.2cm}

The second test case is the polycrystalline growth in a 2D supercooled liquid in which the growth of different 
orientated crystallites is studied. In the simulation, three initial crystallites with hexagonal pattern oriented in 
different directions are seeded in the liquid. Similar numerical experiments were reported in 
\cite{PhysRevLett.88.245701, wise20092, Gomez201252, guo2016}. The simulation is conducted on a square 
domain $\Omega=[0,400]^2$, in which a uniform mesh with $1,024\times 1,024$ elements is applied. 
The positive parameter $\gamma$ is set to be $0.25$.
The time step size is adaptively controlled with $\eta = 5,000, \Delta t_{min}=0.02$, and  
$\Delta t_{max} = 10$.
To define the initial value, we first set the density function $\bar{\phi}$ to be a constant $0.285$ in 
the computational domain, and then replace three 
hexagonal lattice crystallites in three $25\times25$ square patches of the domain. 
\begin{figure}[!b]
\begin{center}
\qquad\scriptsize{(a1) $t = 0$}\qquad\qquad\qquad\scriptsize{(a2) $t = 200$}
\qquad\qquad\scriptsize{(a3) $t = 430$}\qquad\qquad\qquad\scriptsize{(a4) $t = 10000$}\\
{\includegraphics[width=0.24\textwidth]{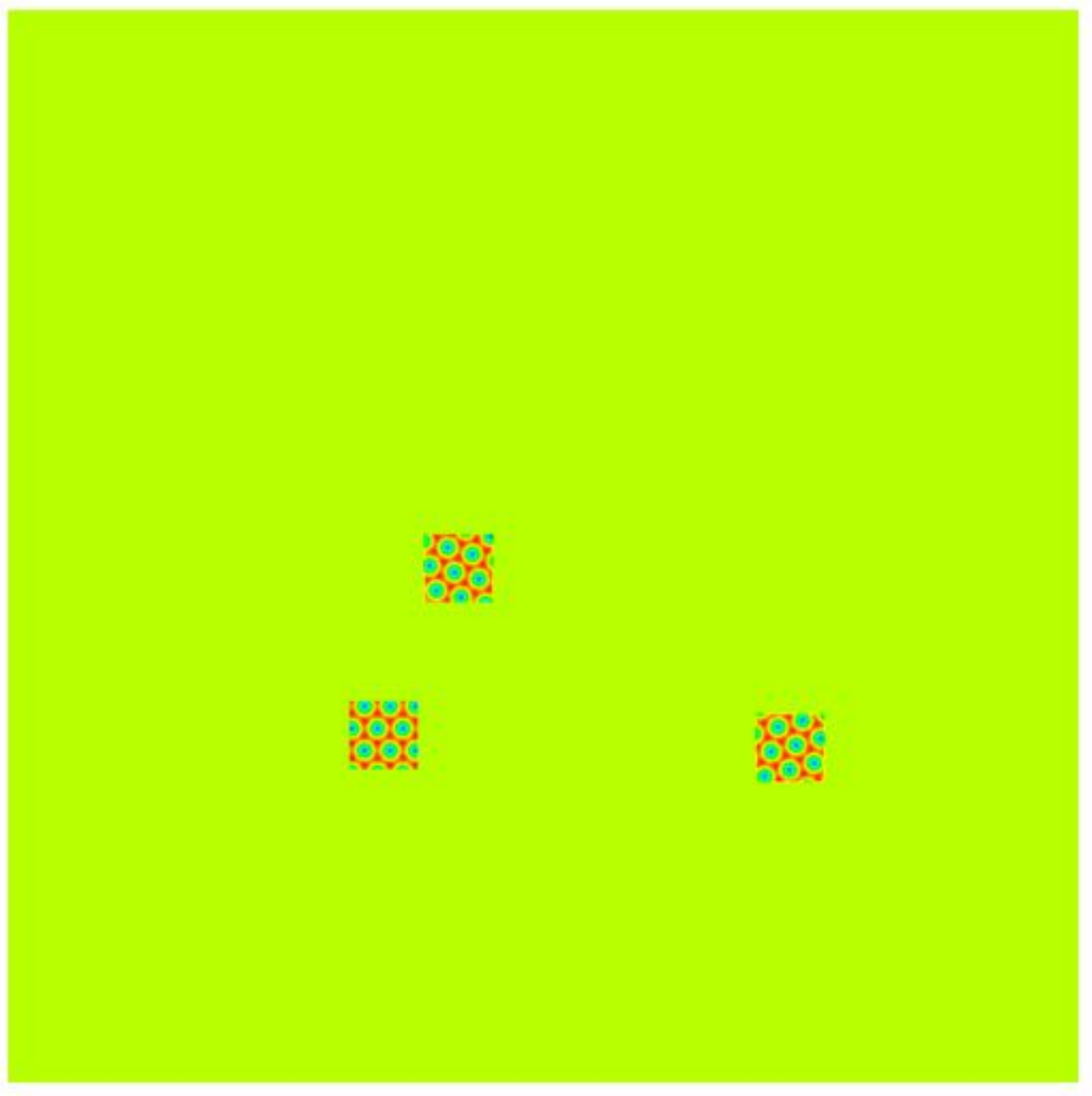}}
{\includegraphics[width=0.24\textwidth]{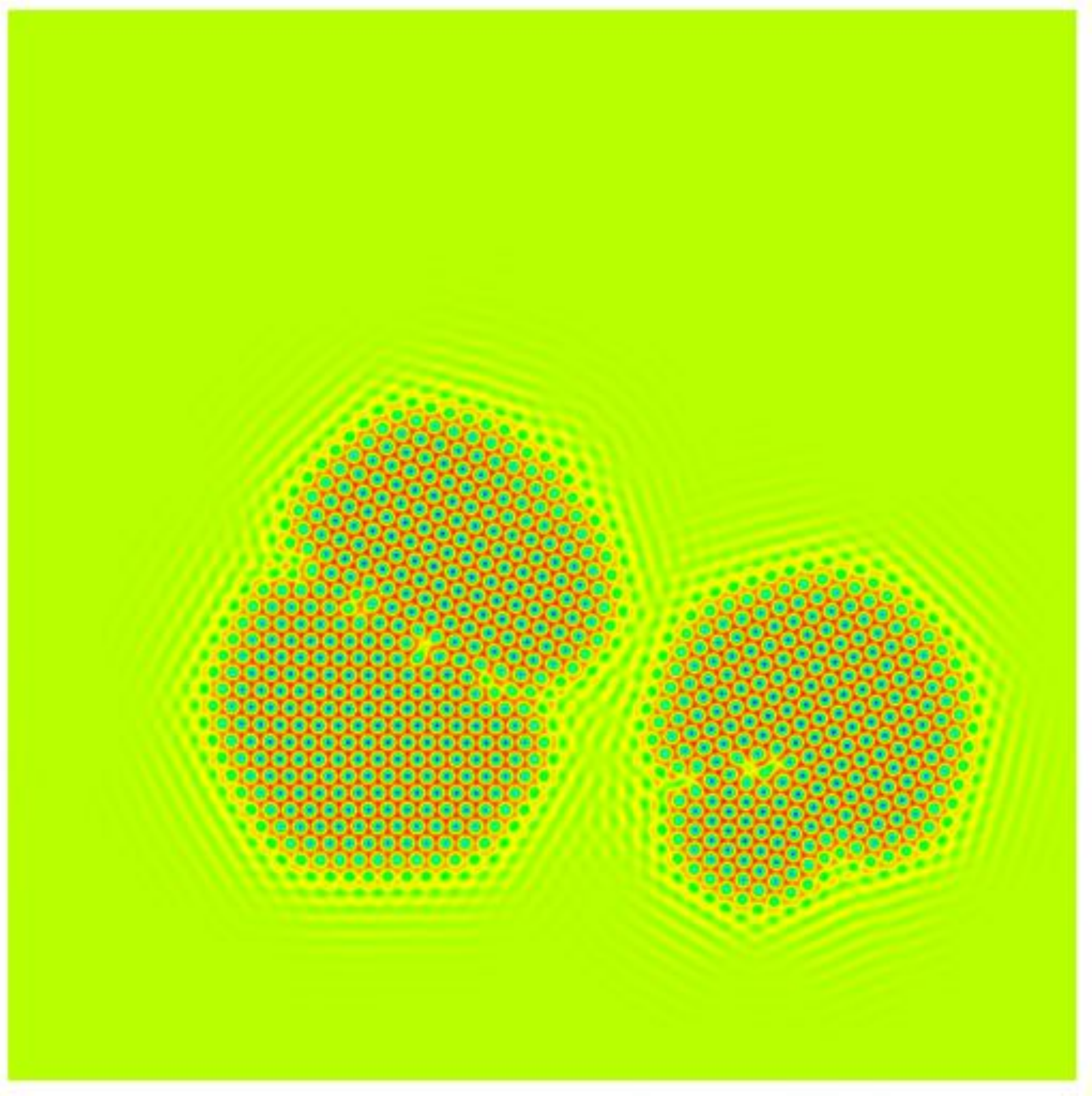}}
{\includegraphics[width=0.24\textwidth]{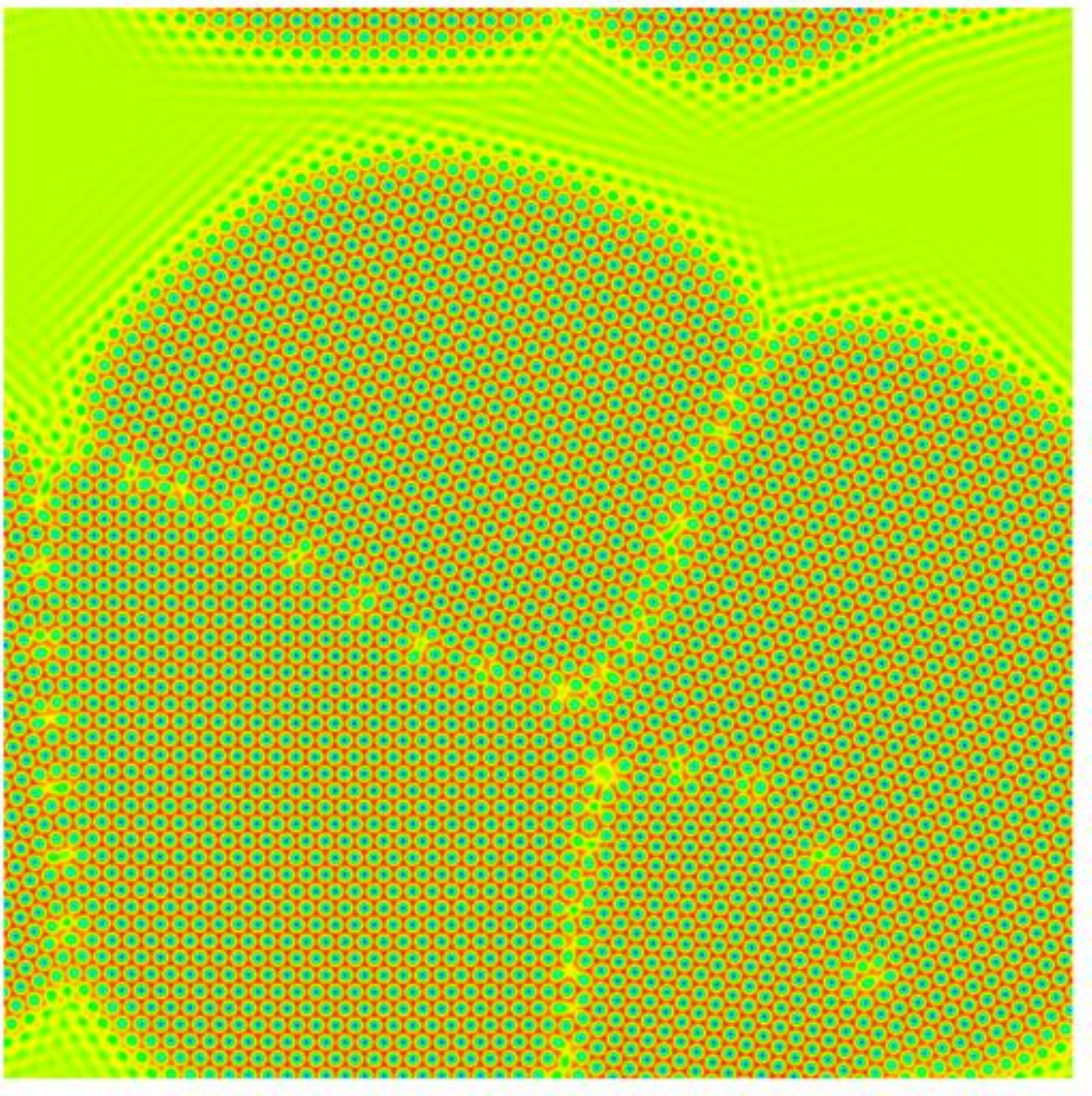}}
{\includegraphics[width=0.24\textwidth]{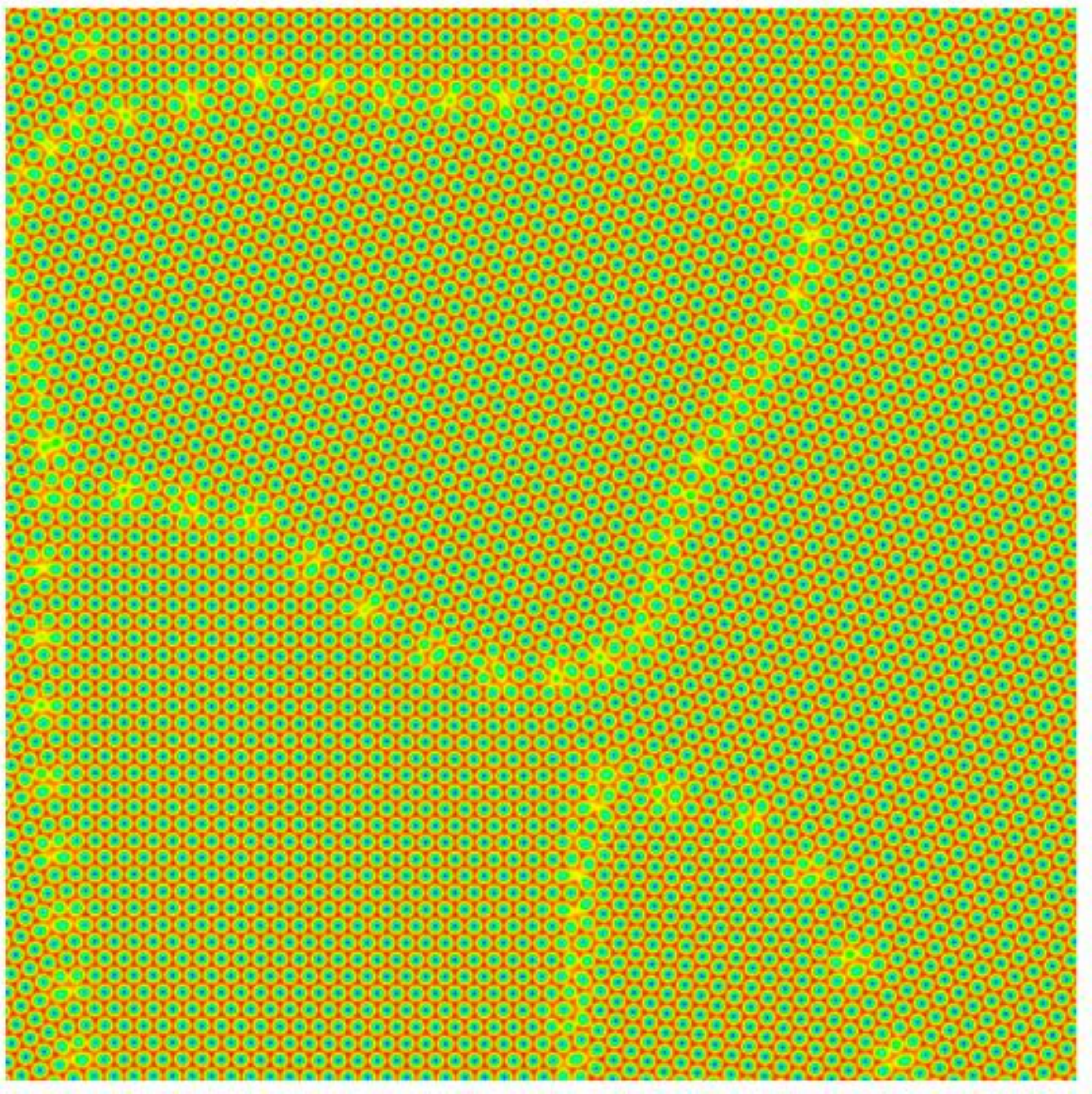}}
\\~~\\
\qquad\scriptsize{(b1) $t = 0$}\qquad\qquad\qquad\scriptsize{(b2) $t = 200$}
\qquad\qquad\scriptsize{(b3) $t = 430$}\qquad\qquad\qquad\scriptsize{(b4) $t = 10000$}\\
{\includegraphics[width=0.24\textwidth]{t0}}
{\includegraphics[width=0.24\textwidth]{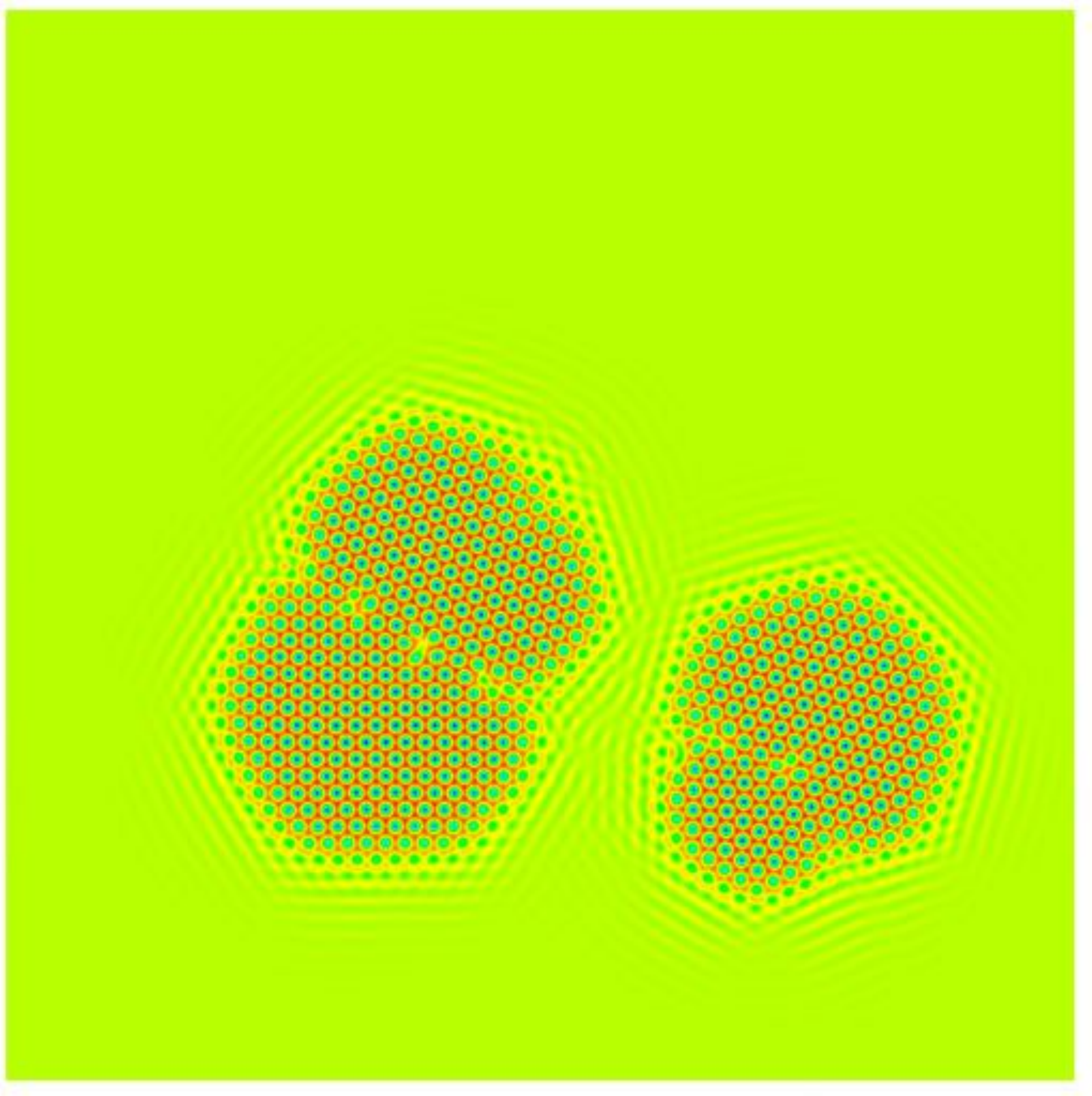}}
{\includegraphics[width=0.24\textwidth]{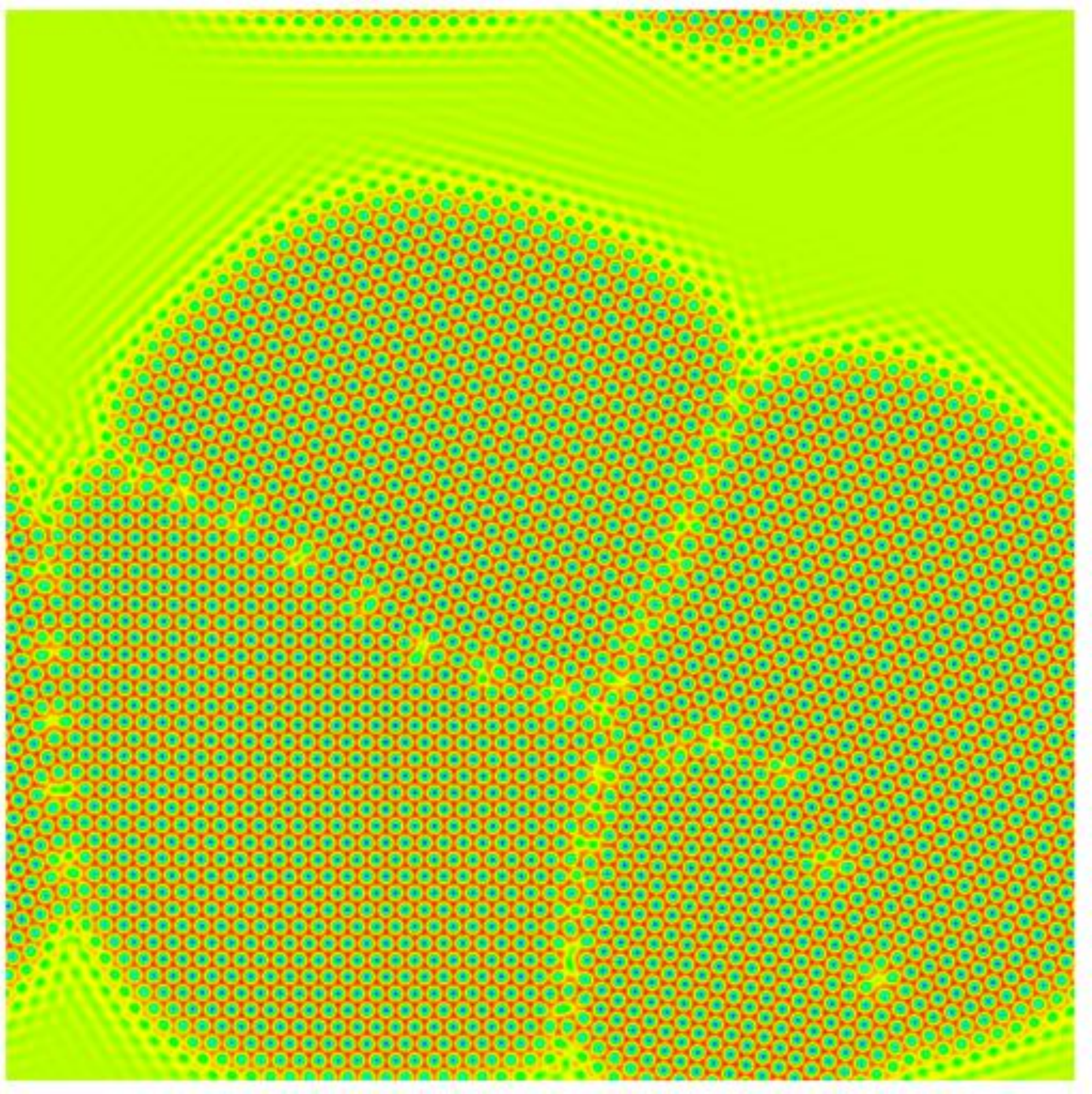}}
{\includegraphics[width=0.24\textwidth]{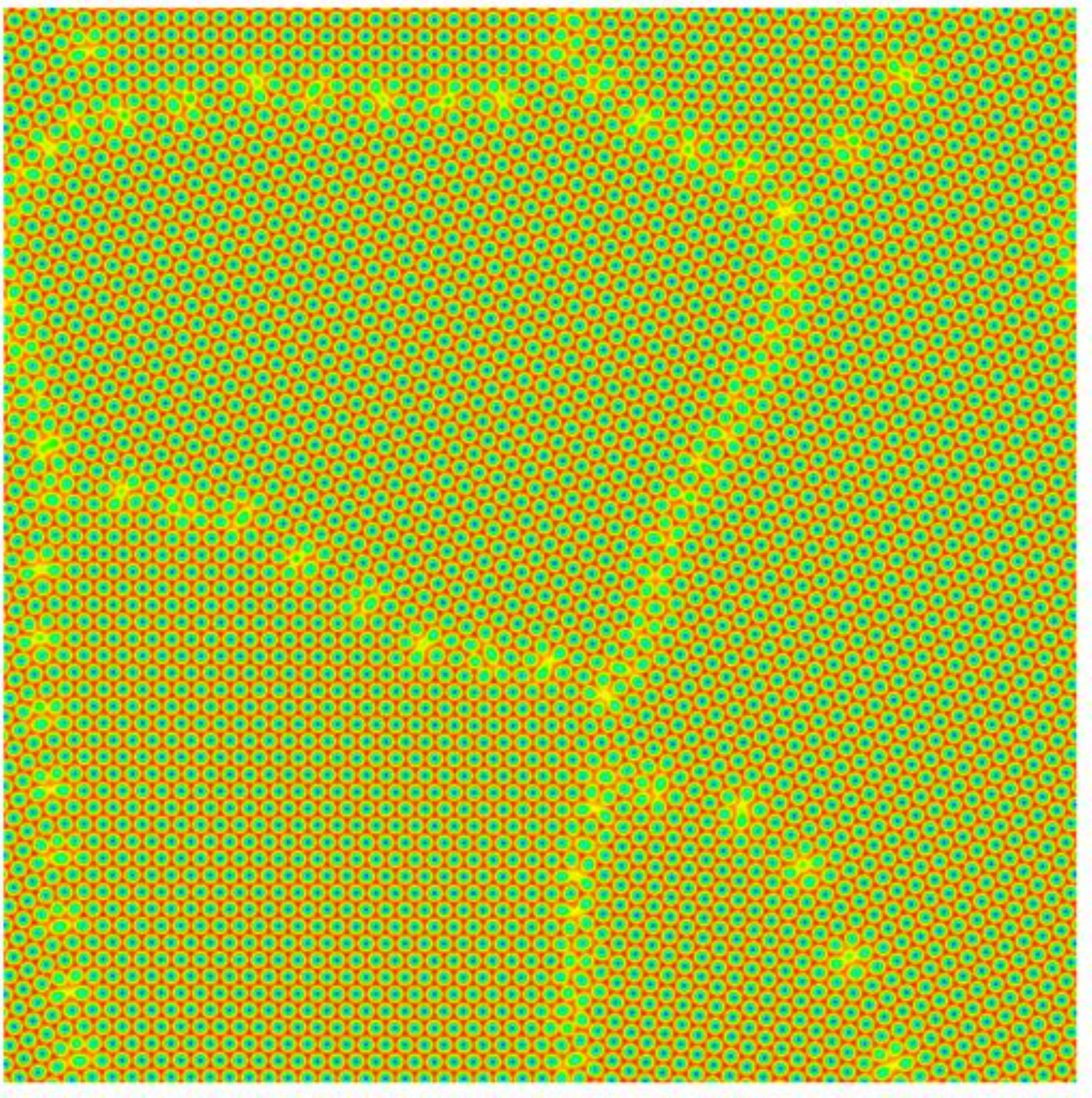}}
\\~~\\
\qquad\scriptsize{(c1) $t = 0$}\qquad\qquad\qquad\scriptsize{(c2) $t = 200$}
\qquad\qquad\scriptsize{(c3) $t = 430$}\qquad\qquad\qquad\scriptsize{(c4) $t = 10000$}\\
{\includegraphics[width=0.24\textwidth]{t0}}
{\includegraphics[width=0.24\textwidth]{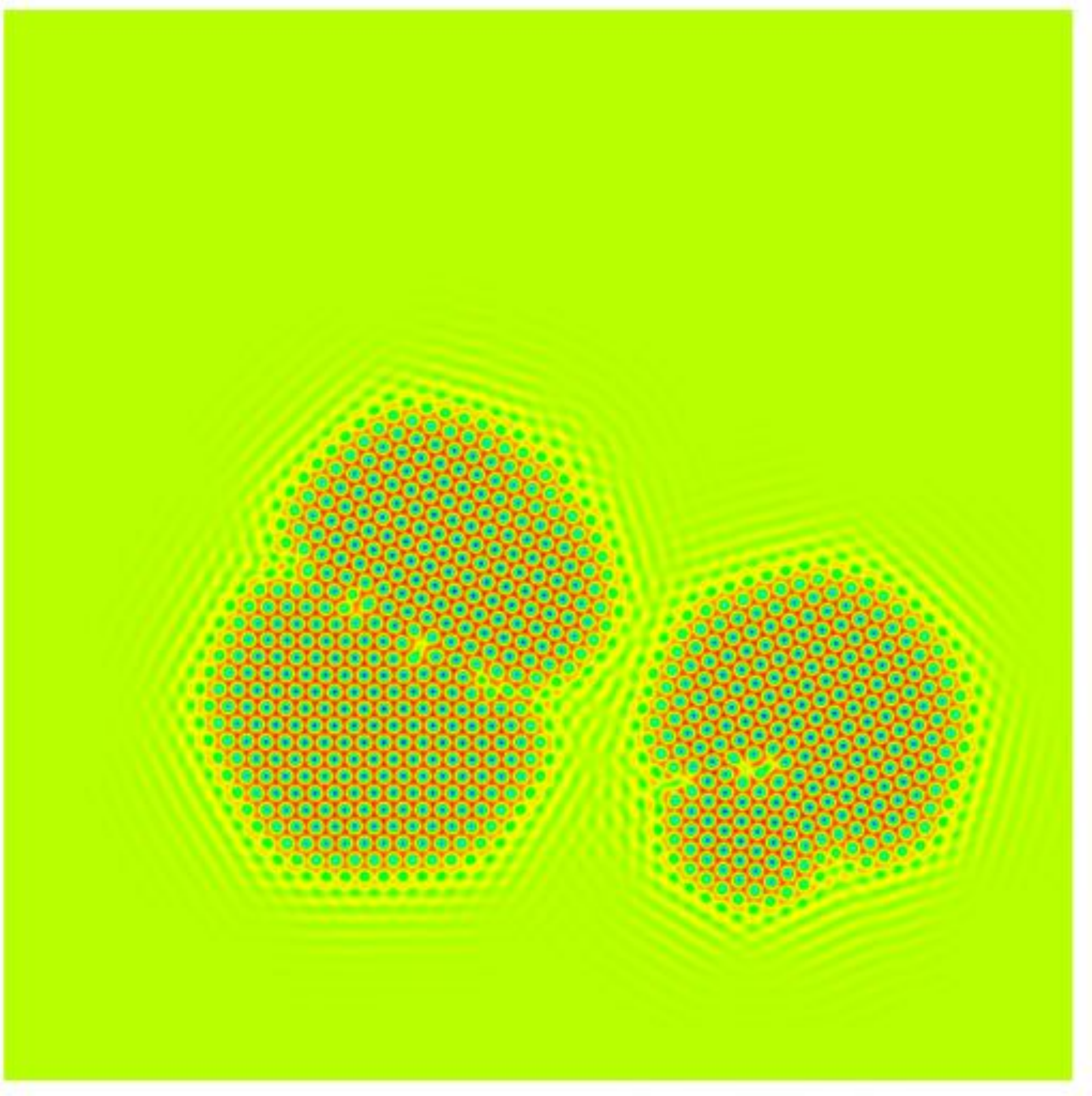}}
{\includegraphics[width=0.24\textwidth]{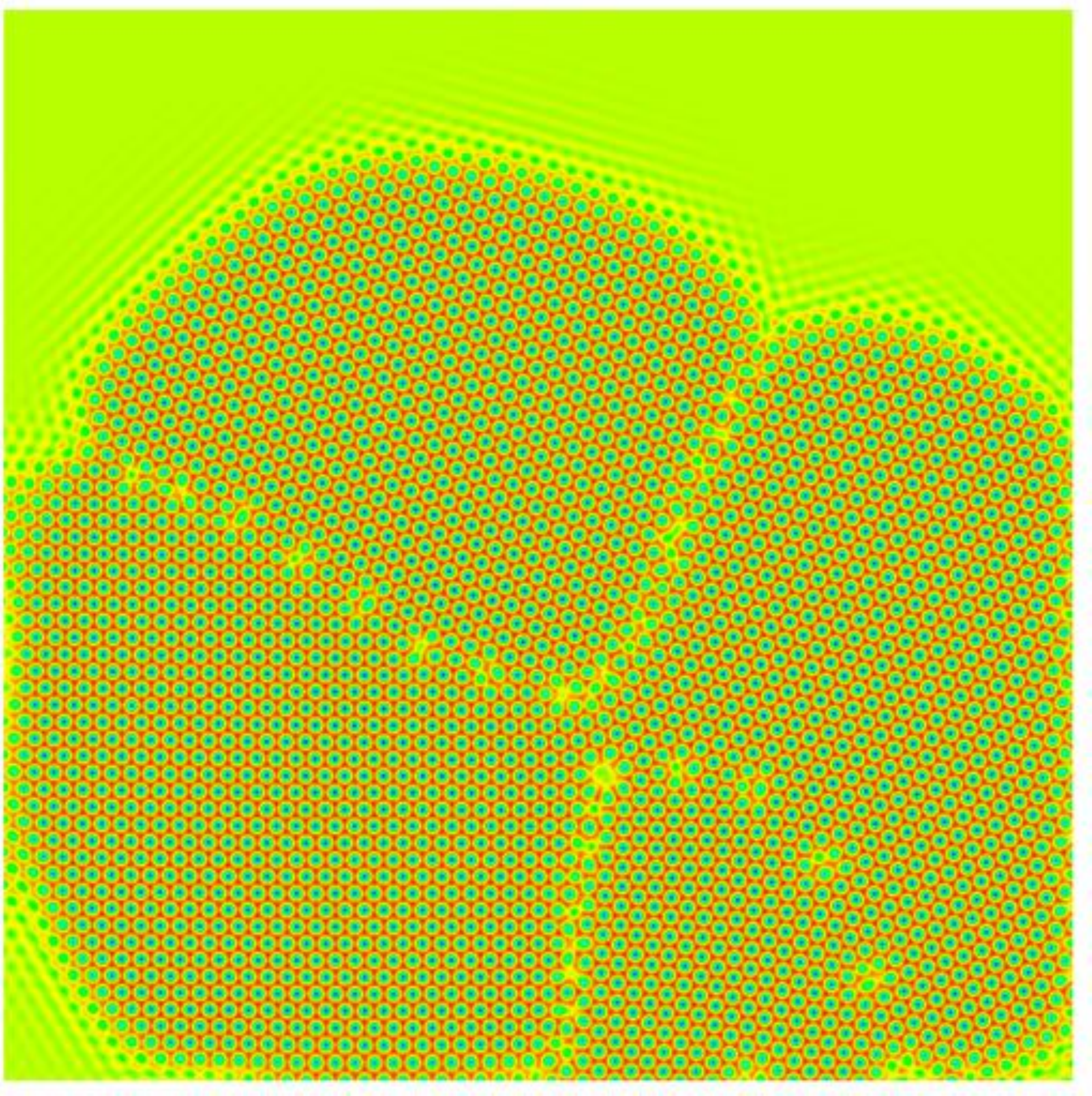}}
{\includegraphics[width=0.24\textwidth]{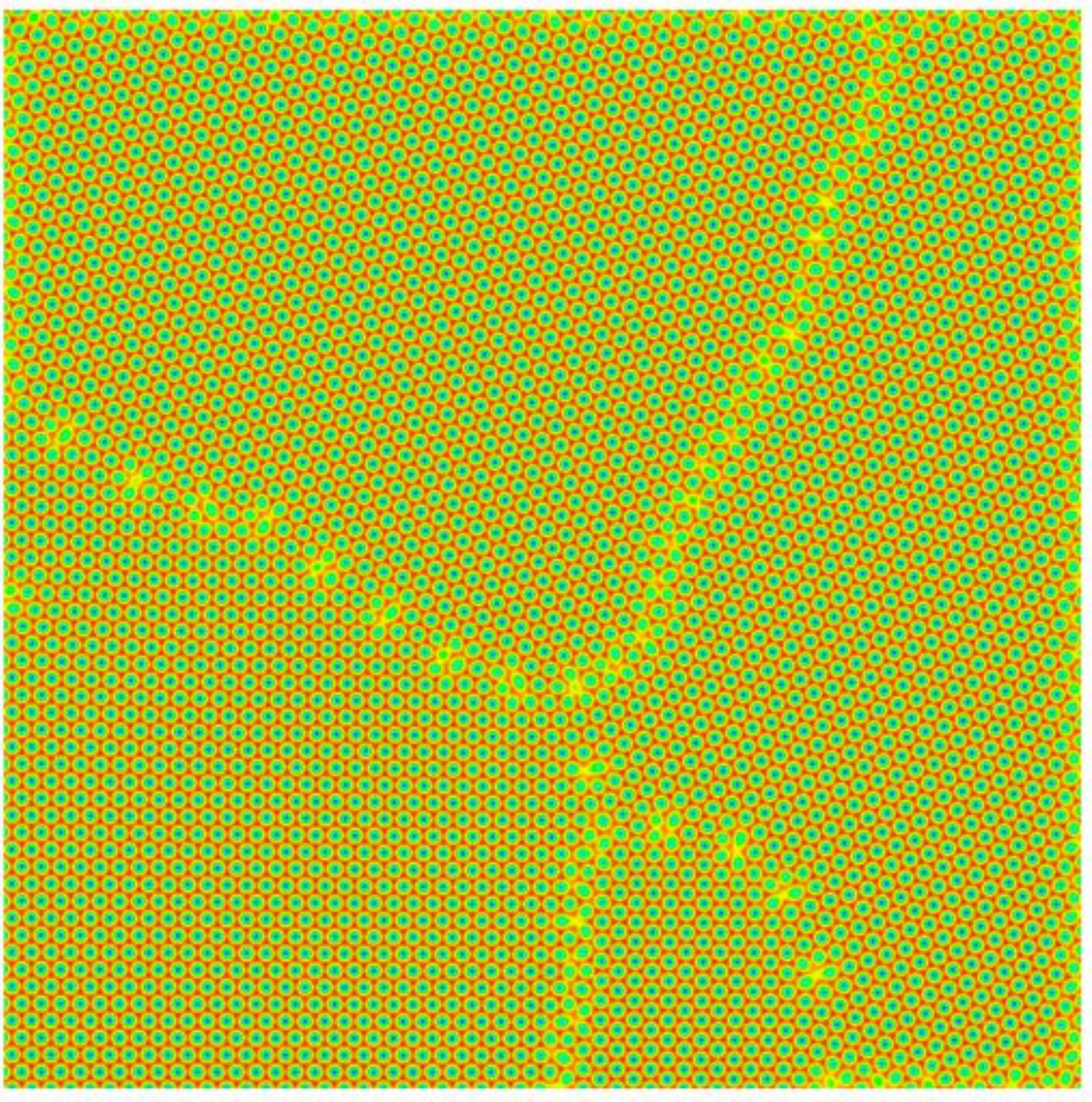}}
\\~~\\
\qquad\scriptsize{(d1) $t = 0$}\qquad\qquad\qquad\scriptsize{(d2) $t = 200$}
\qquad\qquad\scriptsize{(d3) $t = 430$}\qquad\qquad\qquad\scriptsize{(d4) $t = 10000$}\\
{\includegraphics[width=0.24\textwidth]{t0}}
{\includegraphics[width=0.24\textwidth]{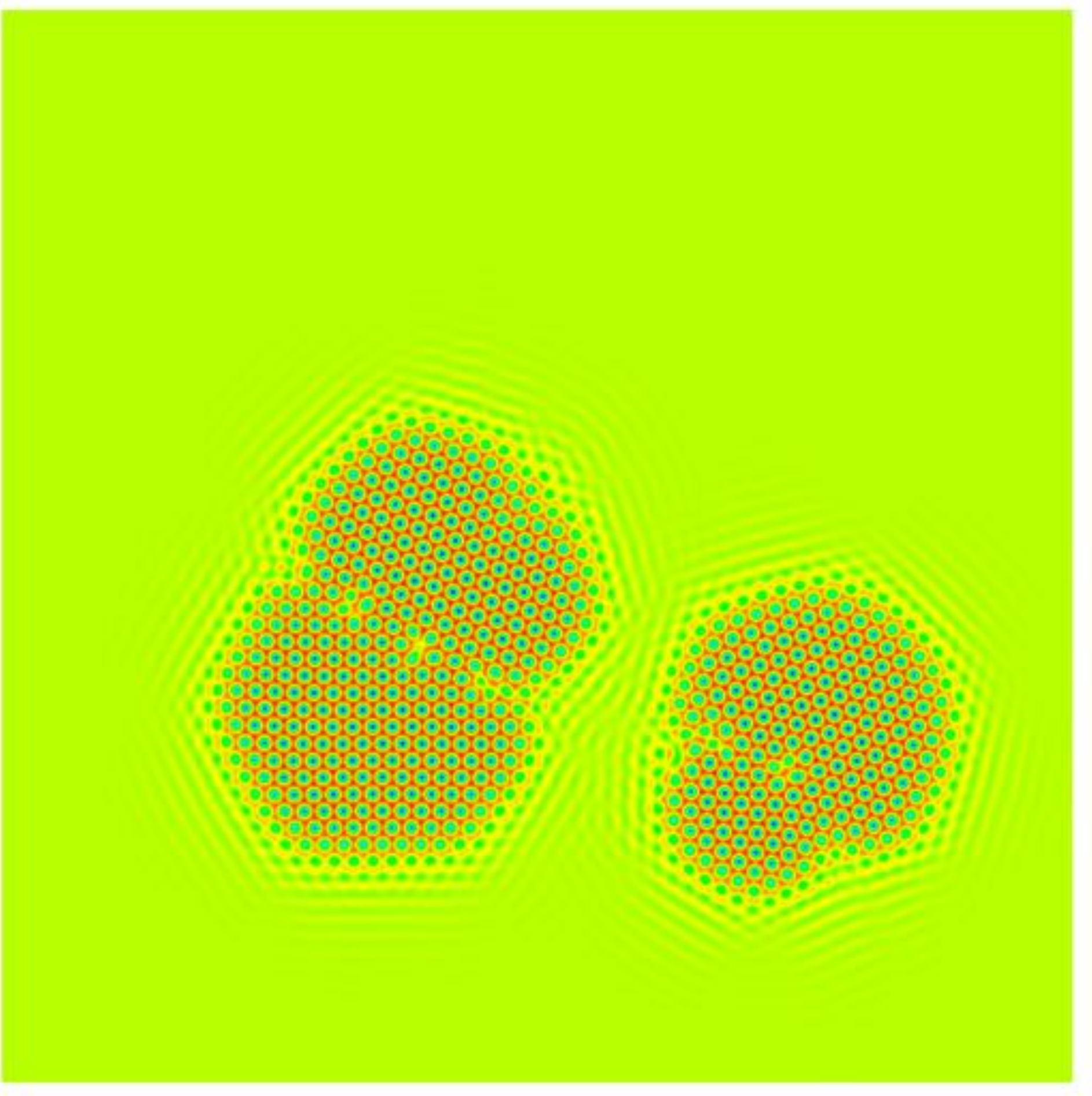}}
{\includegraphics[width=0.24\textwidth]{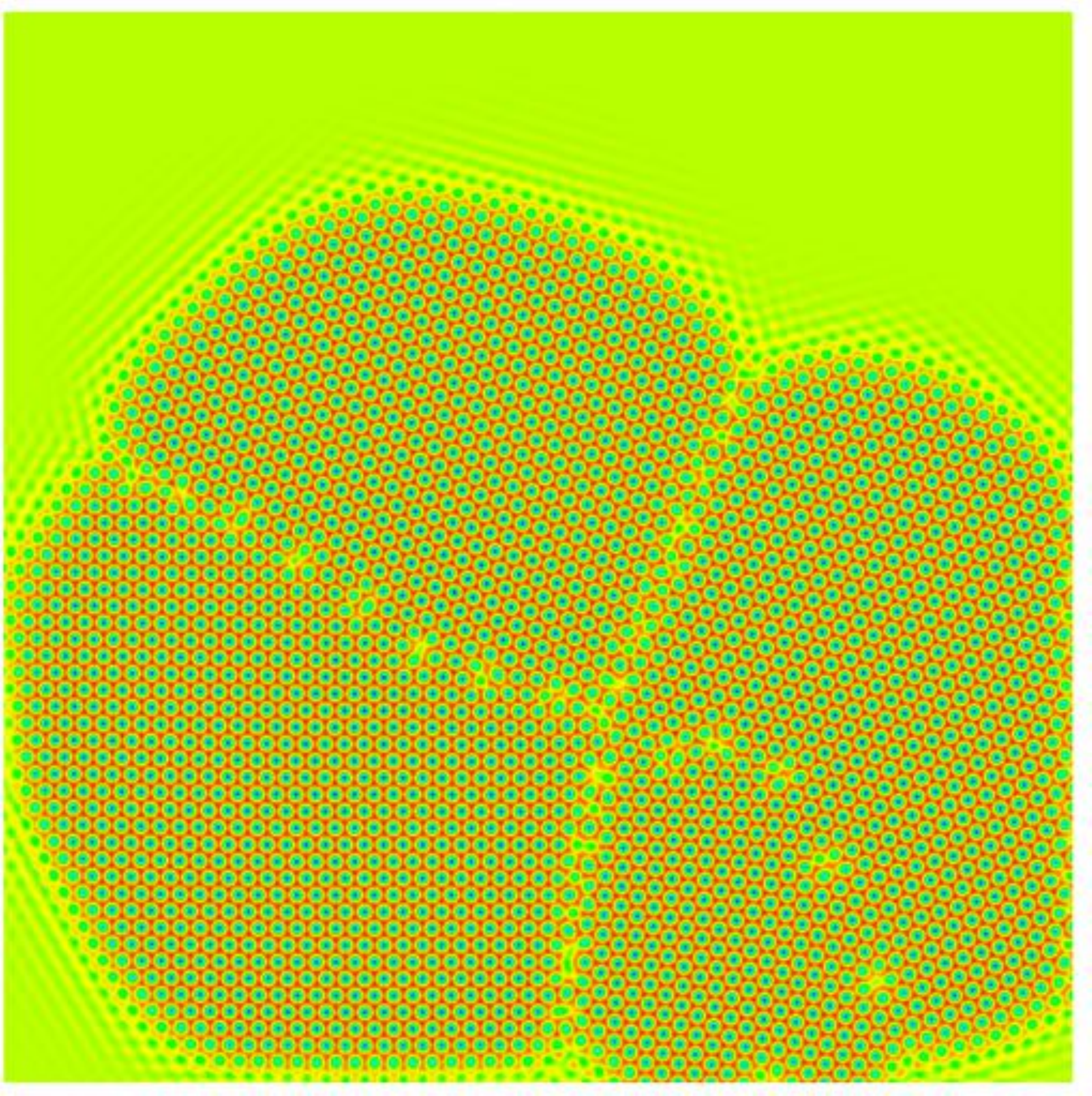}}
{\includegraphics[width=0.24\textwidth]{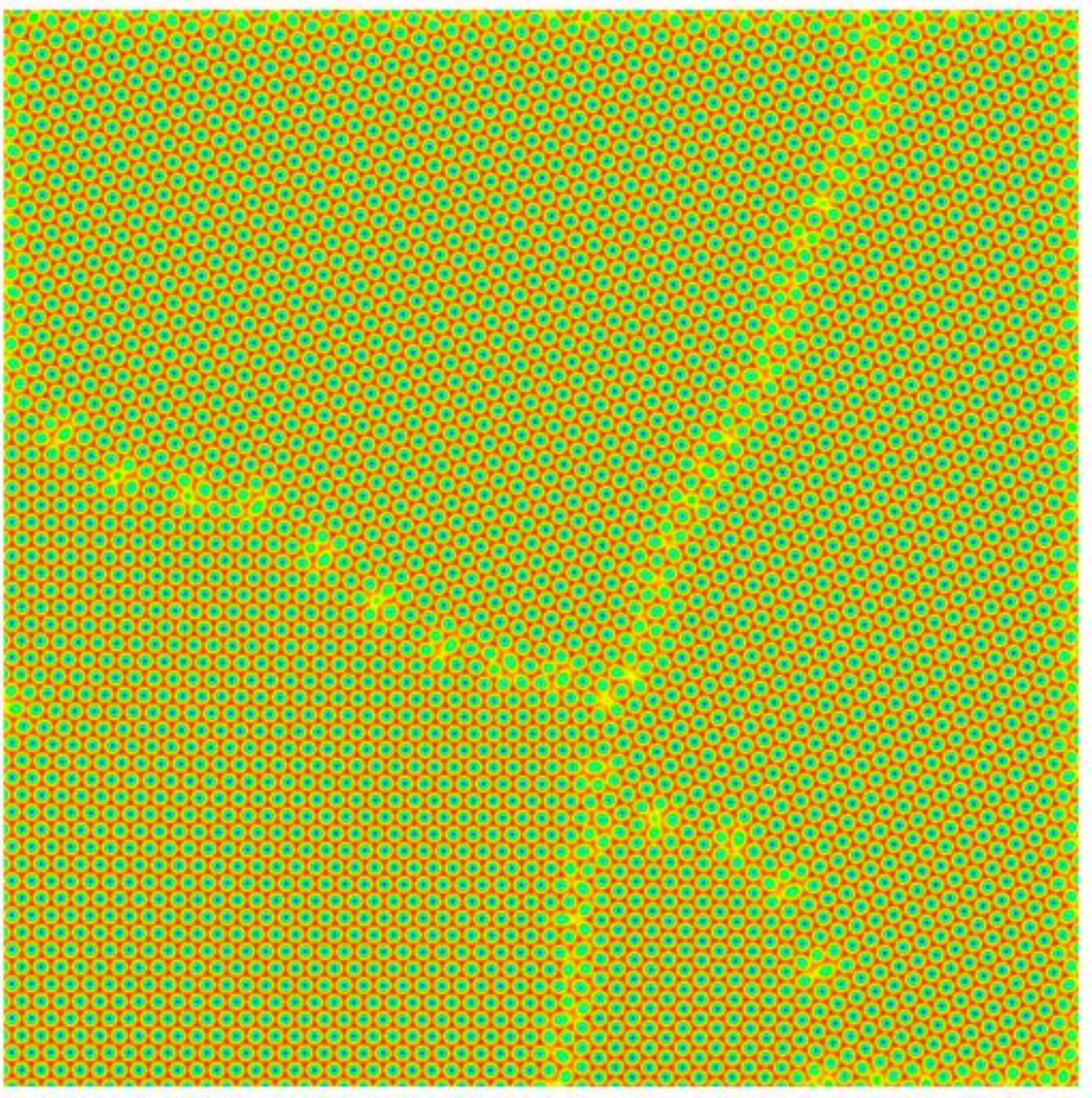}}
\end{center}
\caption{The density distribution of polycrystalline growth in a 2D supercooled liquid, 
(a1-a4): periodic boundary conditions with $M(\phi)=1$; 
(b1-b4): periodic boundary conditions with $M(\phi)=1-\phi^2$; 
(c1-c4): Neumann-type boundary conditions with $M(\phi)=1$; 
(d1-d4): Neumann-type boundary conditions with $M(\phi)=1-\phi^2$.}
\label{ex2:sol}
\end{figure}
We use the following periodic function to describe the two-dimensional hexagonal lattice structure
\begin{equation}
\phi(\mathbf{x})=\bar{\phi}+A\left[\cos\left(\frac{q_y}{\sqrt{3}}y\right)\cos\left(q_xx\right) - 0.5\cos\left(\frac{2q_y}{\sqrt{3}}y\right)\right],
\end{equation}
where $A = 0.446$ is a constant which represents an
amplitude of the fluctuations in density, and $q_x = q_y = 0.66$ are the period-related parameters. 
To obtain the orientated crystallites, we 
define a local Cartesian coordinate system $\tilde{\mathbf{x}}$ that is obtained by rotating the original 
Cartesian coordinate $\bx=(x, y)$ with a certain angle $\omega$. Here $\tilde{\mathbf{x}}$ is defined as
\begin{equation}
 \tilde{\mathbf{x}} = 
\left(\begin{array}{c}
\tilde{x} \\
\tilde{y} \\
\end{array}
\right)=
\left(\begin{array}{c}
\cos(\omega)x - \sin(\omega)y\\
\sin(\omega)x + \cos(\omega)y\\
\end{array}
\right).
\end{equation}
We imply rotation angles  $\omega = -\pi/4, 0, \pi/4$
on three crystal patches, respectively.  

%In this test case, we compare the periodic and Neumann-type boundary conditions on the computational domain.  
%respectively, with the constant mobility $M(\phi)=1$ and $M(\phi)=1-\phi^2$ are both used. 
We consider four test cases, which are (a) periodic boundary conditions with $M(\phi)=1$; 
(b) Neumann-type boundary conditions with $M(\phi)=1$; 
(c) periodic boundary conditions with $M(\phi)=1-\phi^2$; 
and (d) Neumann-type boundary conditions with $M(\phi)=1-\phi^2$.
Fig.~\ref{ex2:sol} shows the distributions of the density function under these situations. 
The solutions $\phi$ are displayed at times $t=0,\, 200,\, 430,\,10,000$, from which 
we observe the growth of the crystallines. We observe a common phenomena that the 
three hexagonal lattice crystallites grow separately and then fuse with each other as time goes on.  
The defects and the dislocations caused by the different alignment of the crystallites can be clearly 
observed in Fig.~\ref{ex2:sol}. 
In the cases with Neumann-type boundary conditions, the ``X'' shape defects and the dislocations 
finally appear in the domain. In the cases with periodic boundary conditions, the hexagonal lattice 
crystallites can go through the boundaries resulting to the appearance of defects and dislocations near the 
boundaries. It is worth to note that the results of $M(\phi)=1$ and $M(\phi)=1-\phi^2$ are 
almost the same due to $\phi<1$, which agree well with the results reported in \cite{guo2016}.

\begin{figure}[!h]
\begin{center}
\qquad\scriptsize{(a)}\qquad\qquad\qquad\qquad\qquad\qquad\qquad
\qquad\qquad\qquad\scriptsize{(b)}\\
{\includegraphics[width=0.48\textwidth]{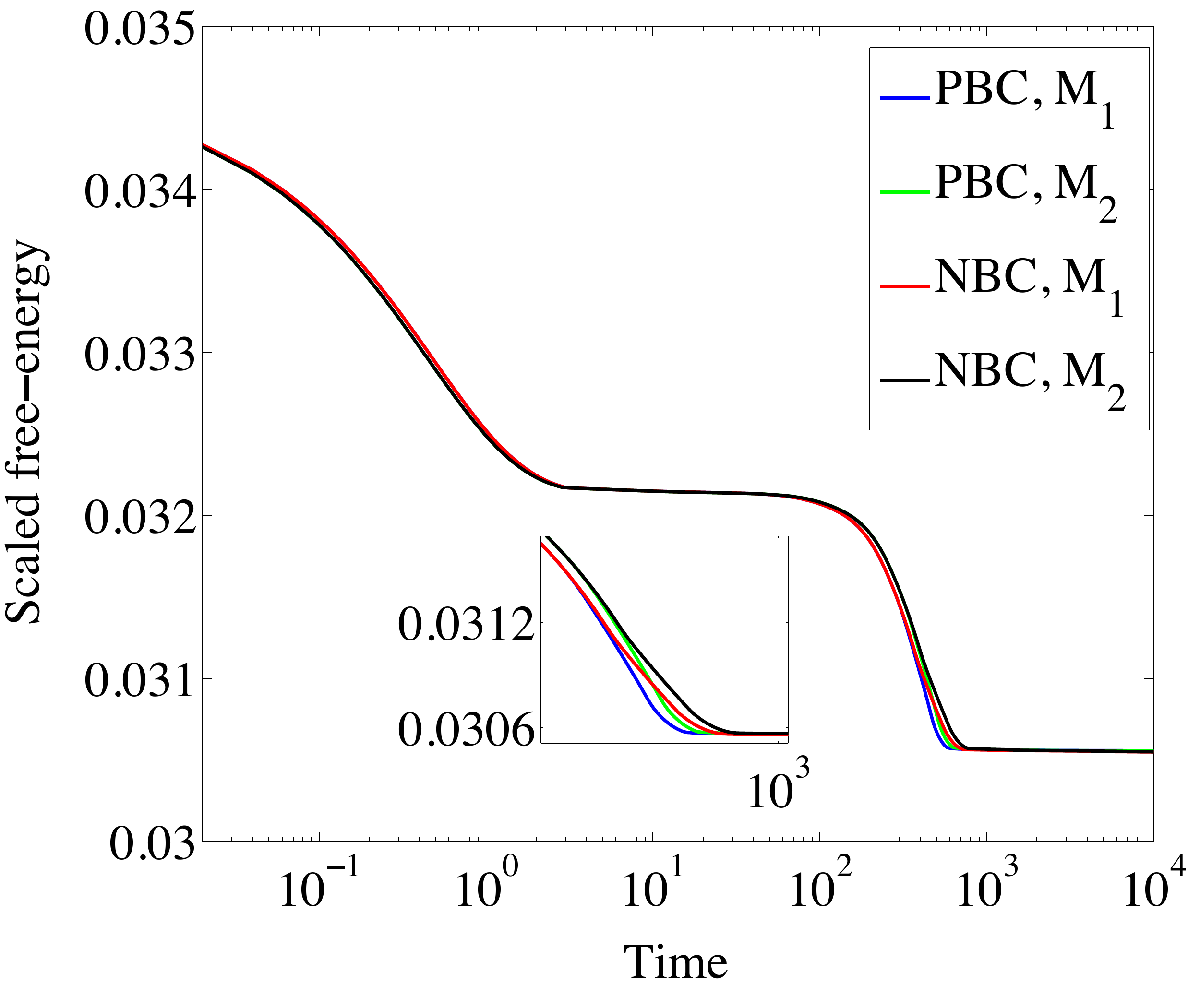}}
{\includegraphics[width=0.48\textwidth]{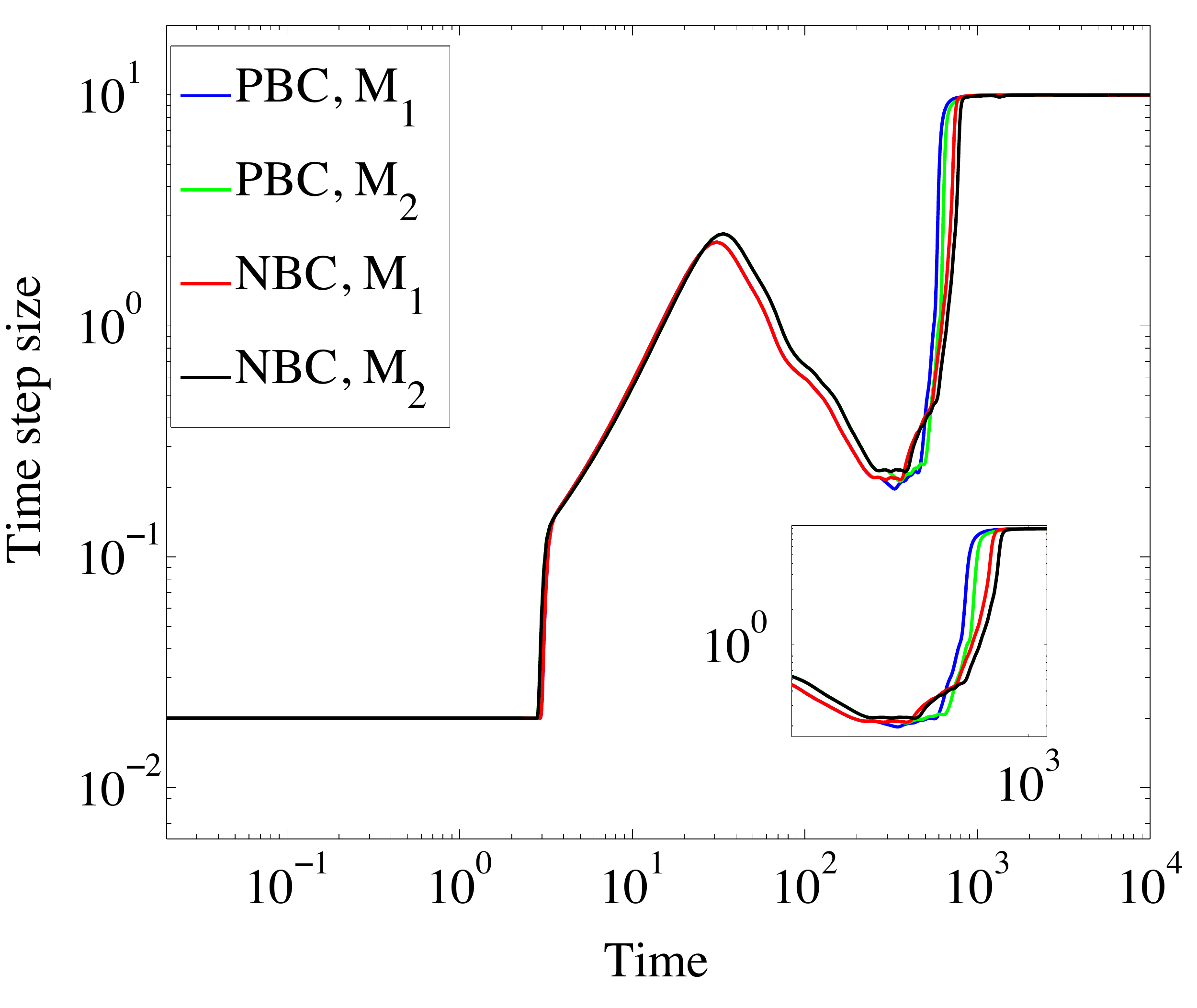}}
\end{center}
\caption{
The scaled total free-energy (a) and the history of the time step size (b), here PBC means the periodic boundary 
conditions, NBC means the Neumann-type boundary conditions and $M_1=1, M_2=1-\phi^2$ are the mobilities.}
\label{ex2:energy}
\end{figure}

We show the scaled total free-energy and the history of the time step size of the tests in 
Fig.~\ref{ex2:energy}. For the four tested scenarios, the total free-energy decreases monotonically 
at a similar pathway. 
Meanwhile, the time step sizes are successfully adjusted from $\Delta t_{min}$ to $\Delta t_{max}$, and have almost same line graph. In this sense, we summarize that these two kinds of boundary conditions and mobilities have little 
impact on the free-energy in test case B.

\vspace{0.2cm}
\noindent C. Crack propagation in a 2D ductile material
\vspace{0.2cm}

In the test case, we employ the PFC equation to model the crack propagation in a periodic rectangle domain with 
ductile material. Similar simulations can be found in \cite{PhysRevE.70.051605, Gomez201252}.
In order to define the initial value, we first set a crystal lattice given by the expression
\begin{figure}[!h]
\begin{center}
\scriptsize{(a) $t = 0$}\\
{\includegraphics[width=0.82\textwidth]{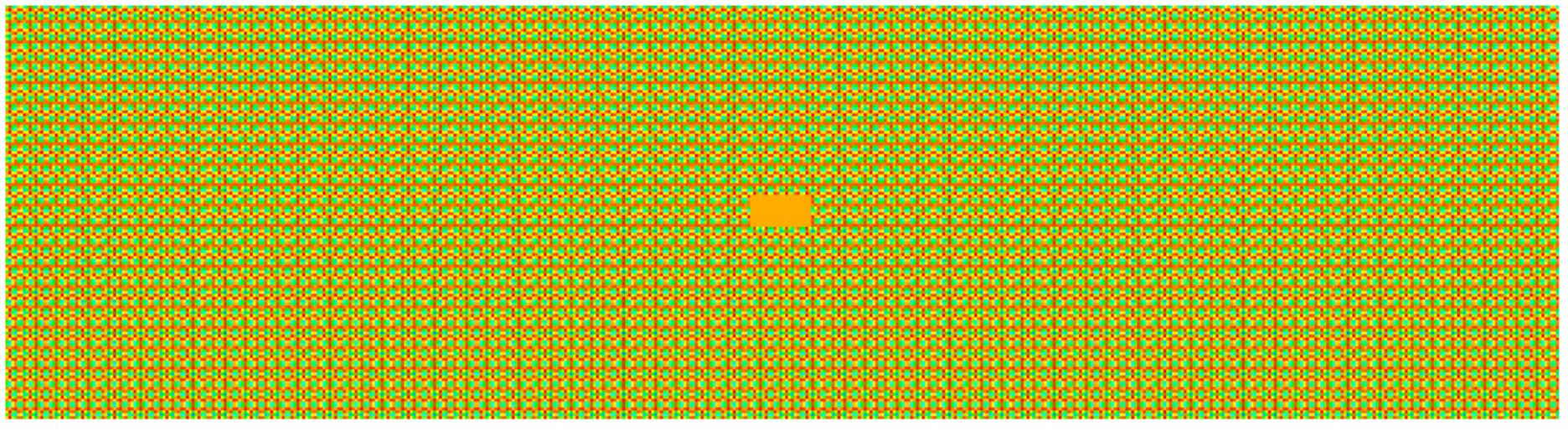}}
\\~~\\
\scriptsize{(a) $t = 8,000$}\\
{\includegraphics[width=0.82\textwidth]{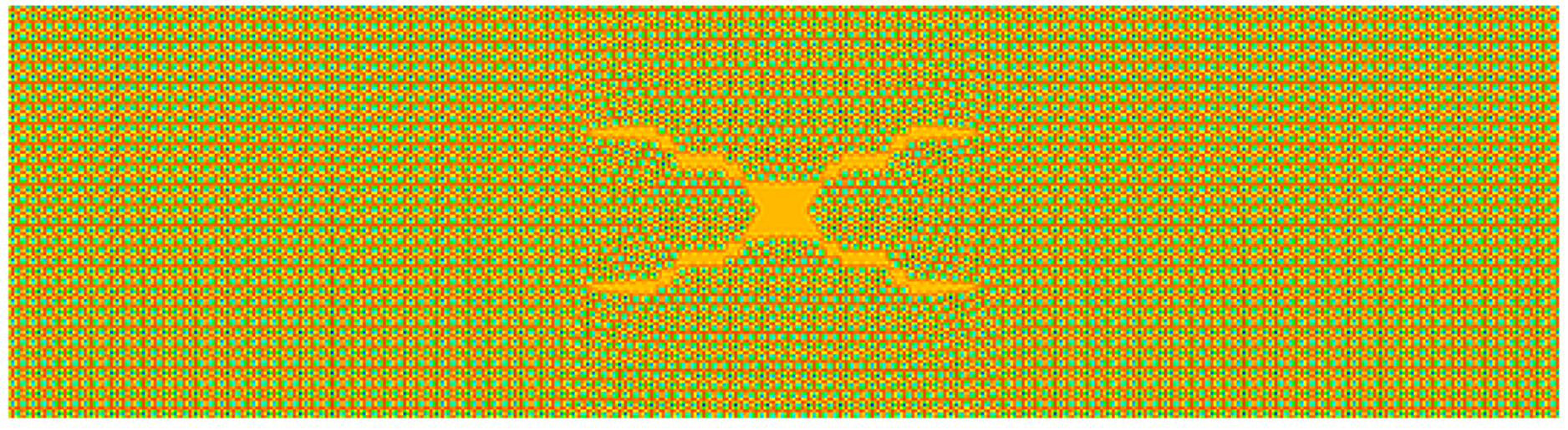}}
\\~~\\
\scriptsize{(b) $t = 16,000$}\\
{\includegraphics[width=0.82\textwidth]{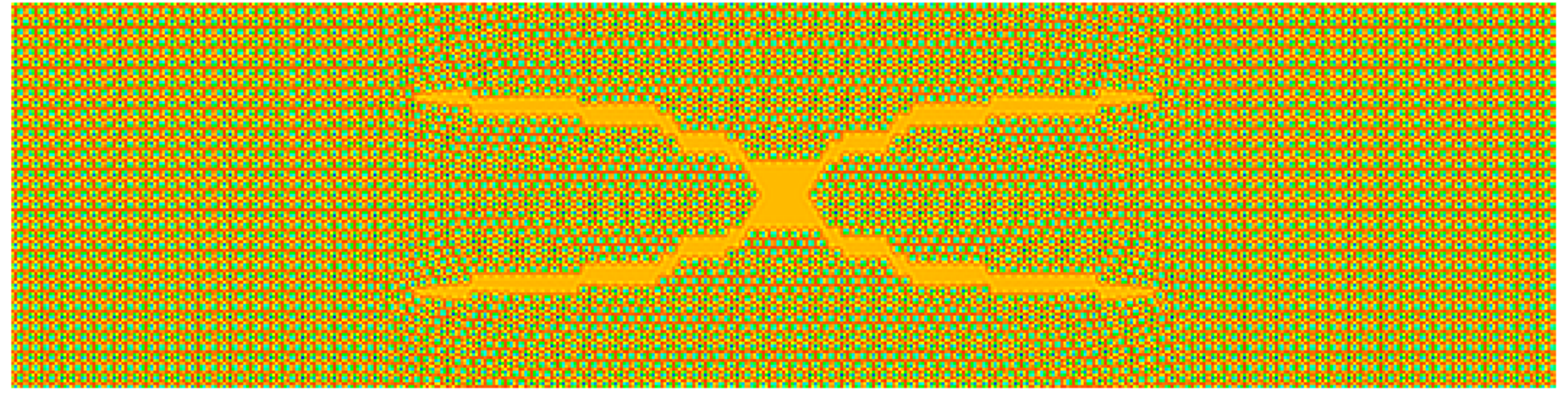}}
\end{center}
\caption{The density distribution of crack propagation in a 2D ductile material. In the three pictures, we only show a center part of the computational domain with size $1,024\Delta x\times 256\Delta y$. }
%colorbar -1.25--1.2
\label{ex3:sol}
\end{figure}
\begin{equation}
\phi(\mathbf{x})=0.49+\cos\left(\frac{q_y}{\sqrt{3}}y\right)\cos\left(q_xx\right) - 0.5\cos\left(\frac{2q_y}{\sqrt{3}}y\right),
\end{equation}
in the computational domain. Here, $q_x$ and $q_y$ are the parameters which determine the crystal period. 
We take $q_x=\sqrt{3}/2$, and $q_y = 0.9q_x$, which means that the initial crystal  
has no stretching in the $x$ direction and a 1/9 stretching in the $y$ direction. 
\begin{figure}[!h]
\begin{center}
\scriptsize{(a)}\qquad\qquad\qquad\qquad\qquad\qquad
\qquad\qquad\qquad\scriptsize{(b)}\\
{\includegraphics[width=0.48\textwidth]{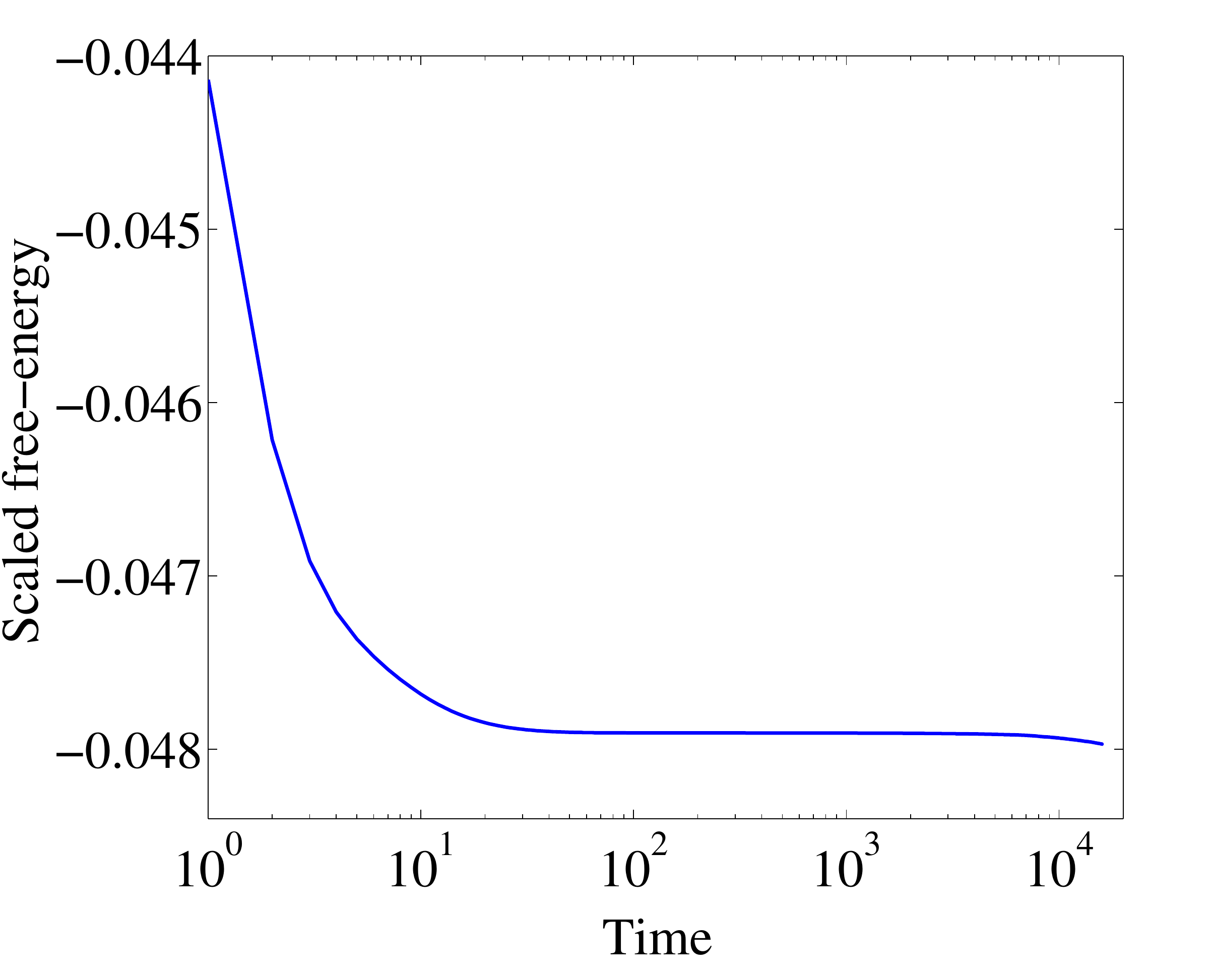}}
{\includegraphics[width=0.48\textwidth]{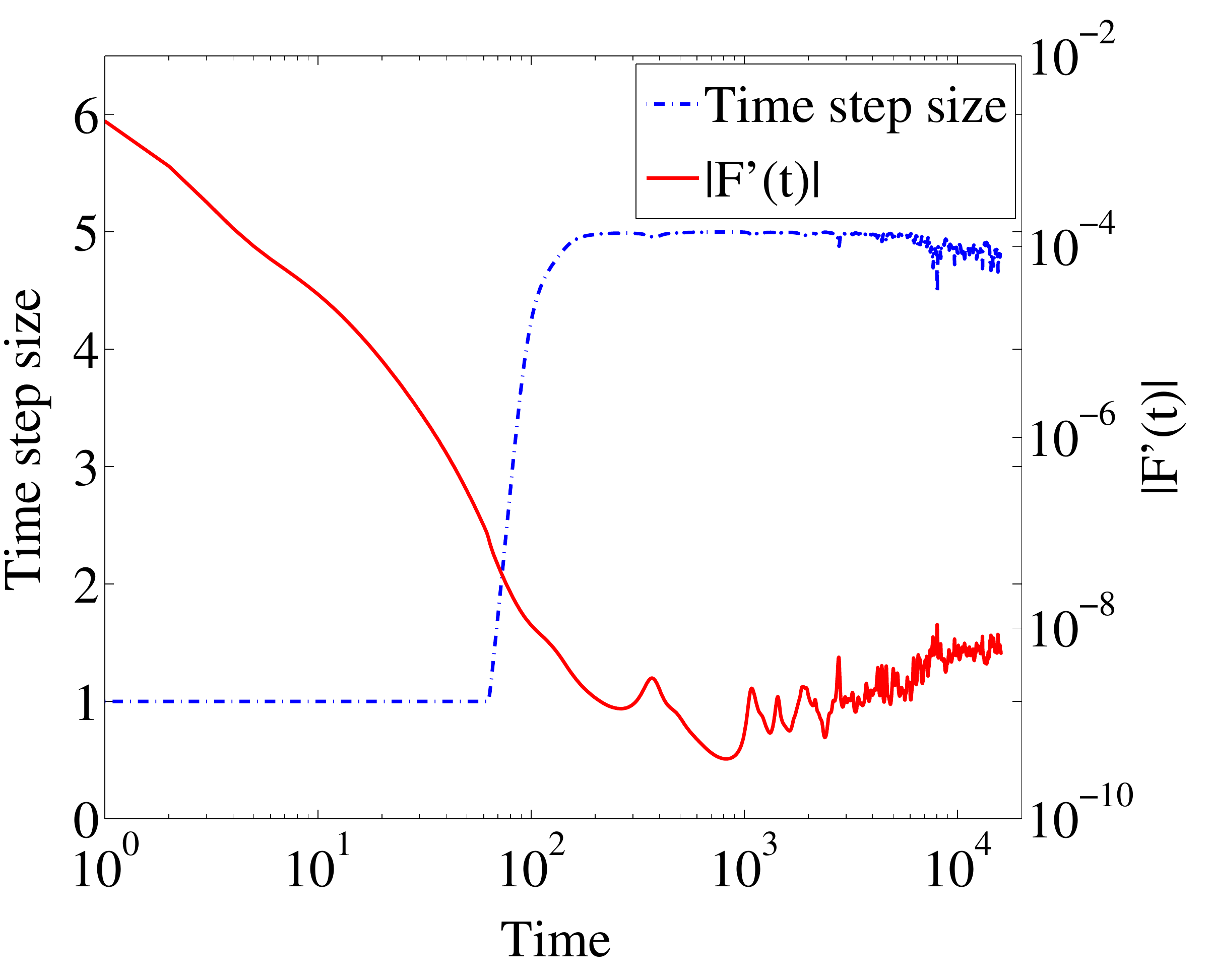}}
\end{center}
\caption{
The scaled total free-energy (a)
and the history of the time step size (b). }
\label{ex3:energy} 
\end{figure}
The numerical simulation is conducted on a maximum periodic system $591\frac{2\pi}{q_x}\times 
76\frac{2\pi\sqrt{3}}{q_y}$ contained in domain $4,096\pi/3 \times 1,024\pi/3$. 
The computational mesh is composed of $4,096\times 1,024$ elements.
In the centre of computational domain, a notch of size $40􏰰\Delta x \times􏰯 20\Delta y$ is cut out 
and replaced with a coexisting liquid ($\phi = 0.79$) \cite{PhysRevE.70.051605}􏱍. The notch provides a nucleating cite for a crack to start propagating. In this simulation, the parameter $\gamma$, $M(\phi)$ equal $1$, and the time step size is adaptively controlled with $\Delta t_{min} = 1, \Delta t_{max} = 5$ and $\eta = 100$.

Fig.~\ref{ex3:sol} depicts the pseudocolor plots of the density distribution at three times, $t=0$ (the initial shape), 
and $t=8,000,\,16,000$. We observe that the crack grows from a little rectangle and keeps growing 
outward like a tree on the endings. 
We show the scaled total free-energy and the history of the time step size in  Fig.~\ref{ex3:energy}.
It can be seen that there is no increase in free-energy, and the time step size increases from 
$\Delta t_{min}$ to $\Delta t_{max}$.

\vspace{0.2cm}
\noindent D. Polycrystalline growth in a 3D supercooled liquid
\vspace{0.2cm}

The experiment is conducted to simulate the polycrystalline growth in three dimensional space, which can be 
regarded as the 3D version of test B. Similar simulation was reported in \cite{vignal2015}.
The computational domain is 􏰕$[0,80\pi]^3$, and periodic boundary conditions are imposed in all directions. 
An uniform mesh comprised of $256\times256\times256$ elements is used, 
and the time step size is controlled adaptively with $\Delta t_{min} = 0.5, \Delta t_{max} = 10$, 
and $\eta = 500$. 
Three initial crystallite spheres with BCC 
configuration oriented in different directions are placed in the domain. We can predict that the grain boundaries 
emerge eventually when the crystallites meet due to the inconformity of orientations. 
\begin{figure}[!b]
\begin{center}
\scriptsize{(a) $t = 0$}\\
{\includegraphics[width=0.4\textwidth]{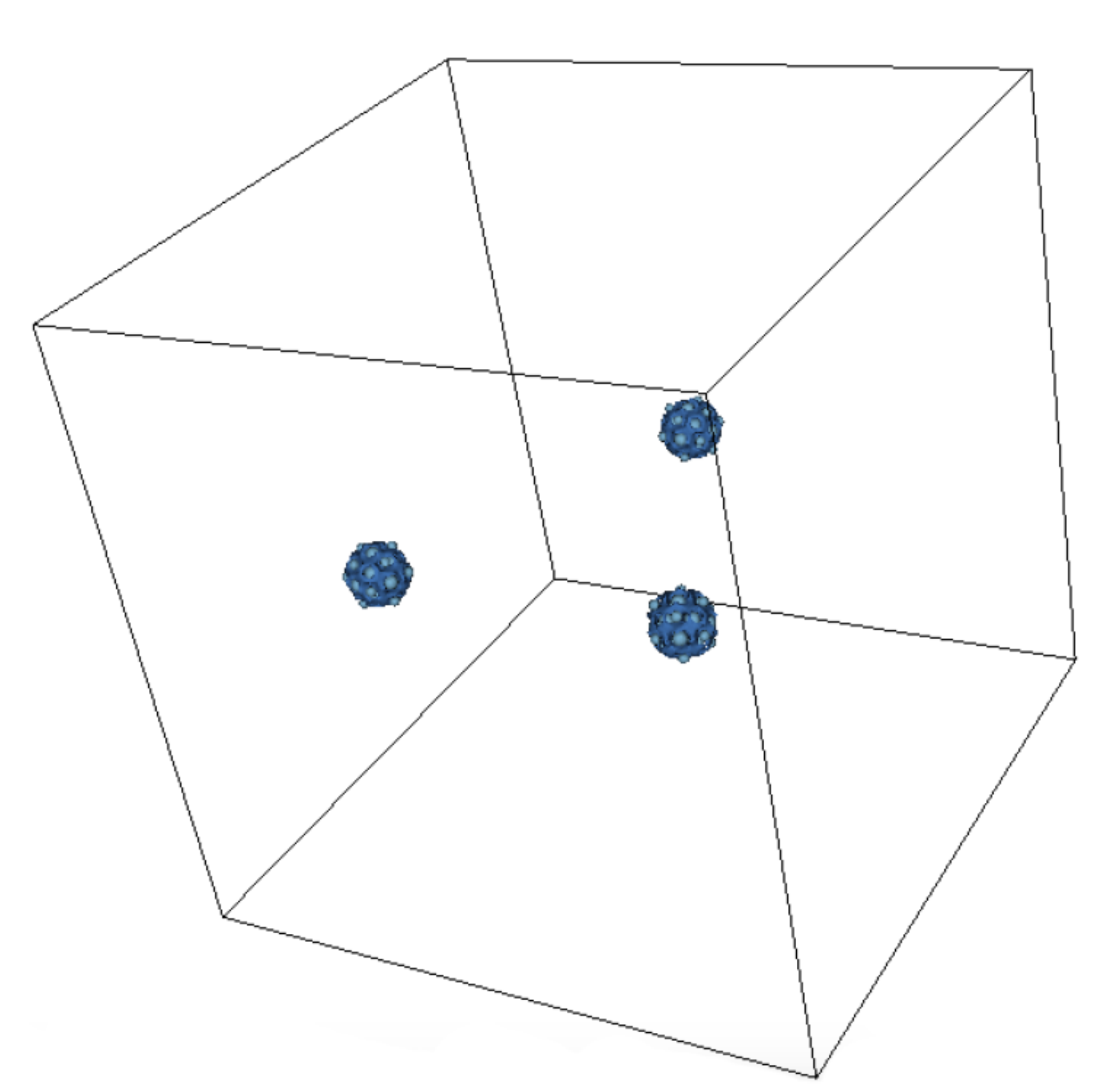}}\qquad
{\includegraphics[width=0.4\textwidth]{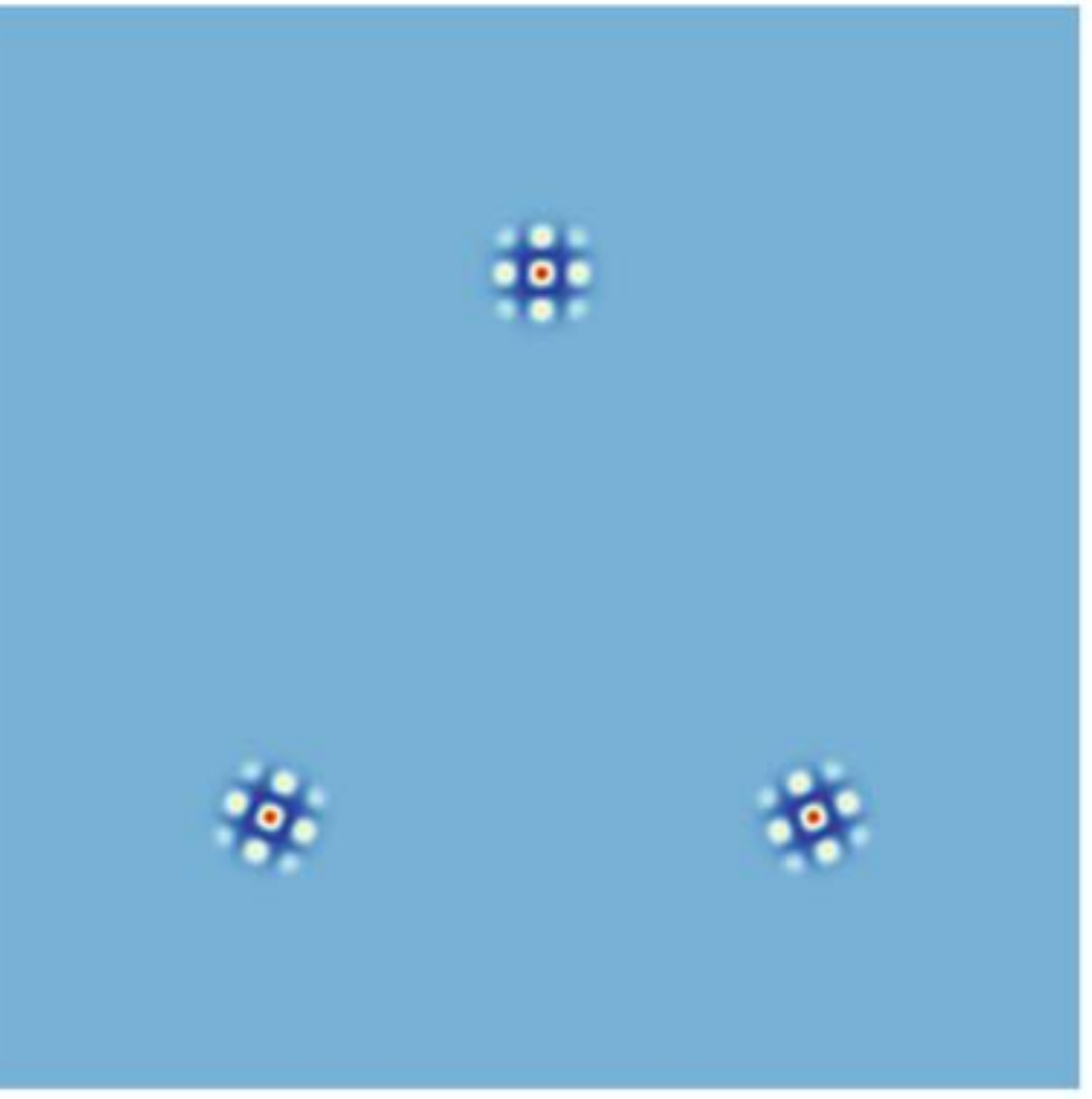}}
\\~~\\
\scriptsize{(b) $t = 100$}\\
{\includegraphics[width=0.4\textwidth]{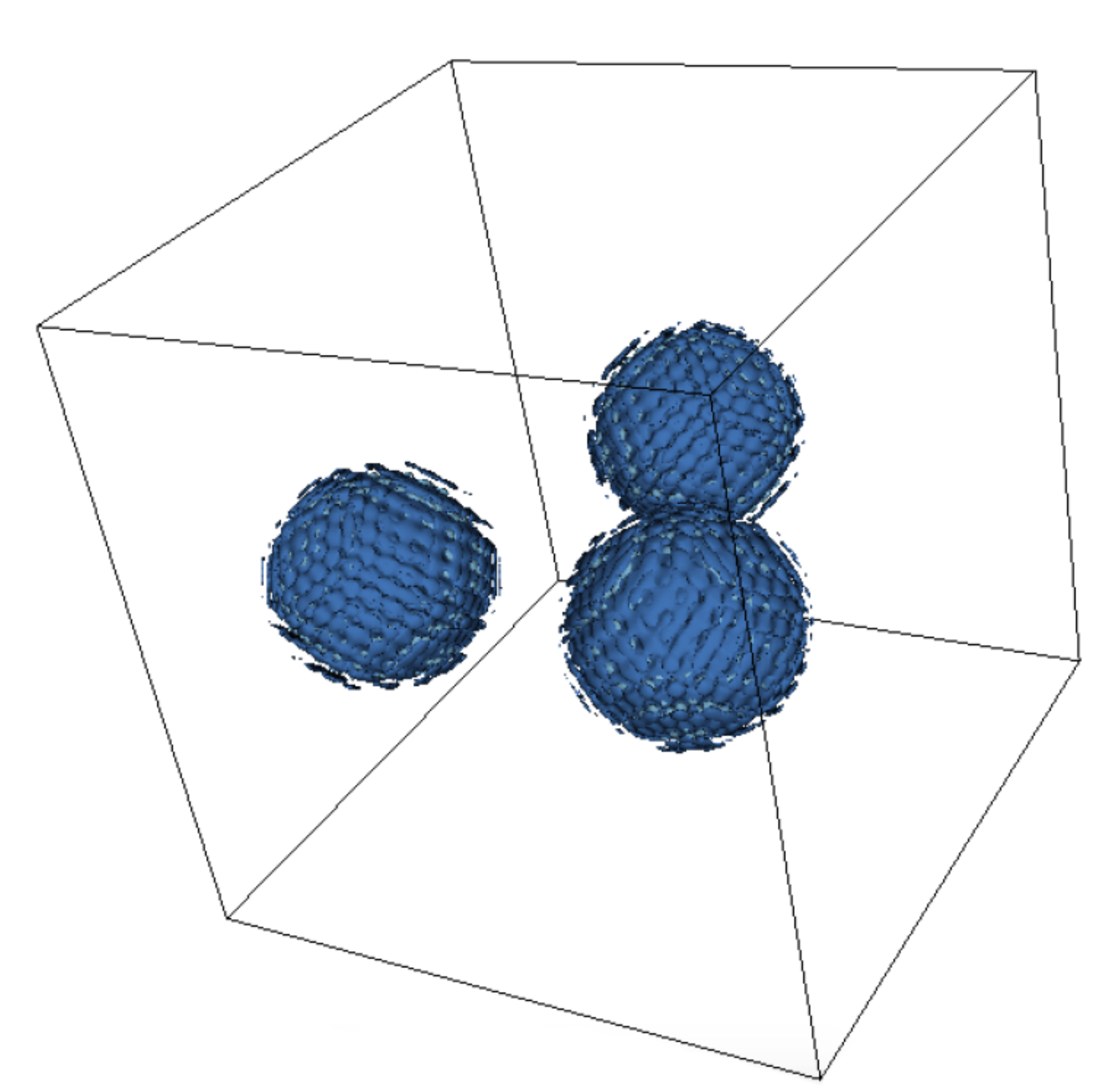}}\qquad
{\includegraphics[width=0.4\textwidth]{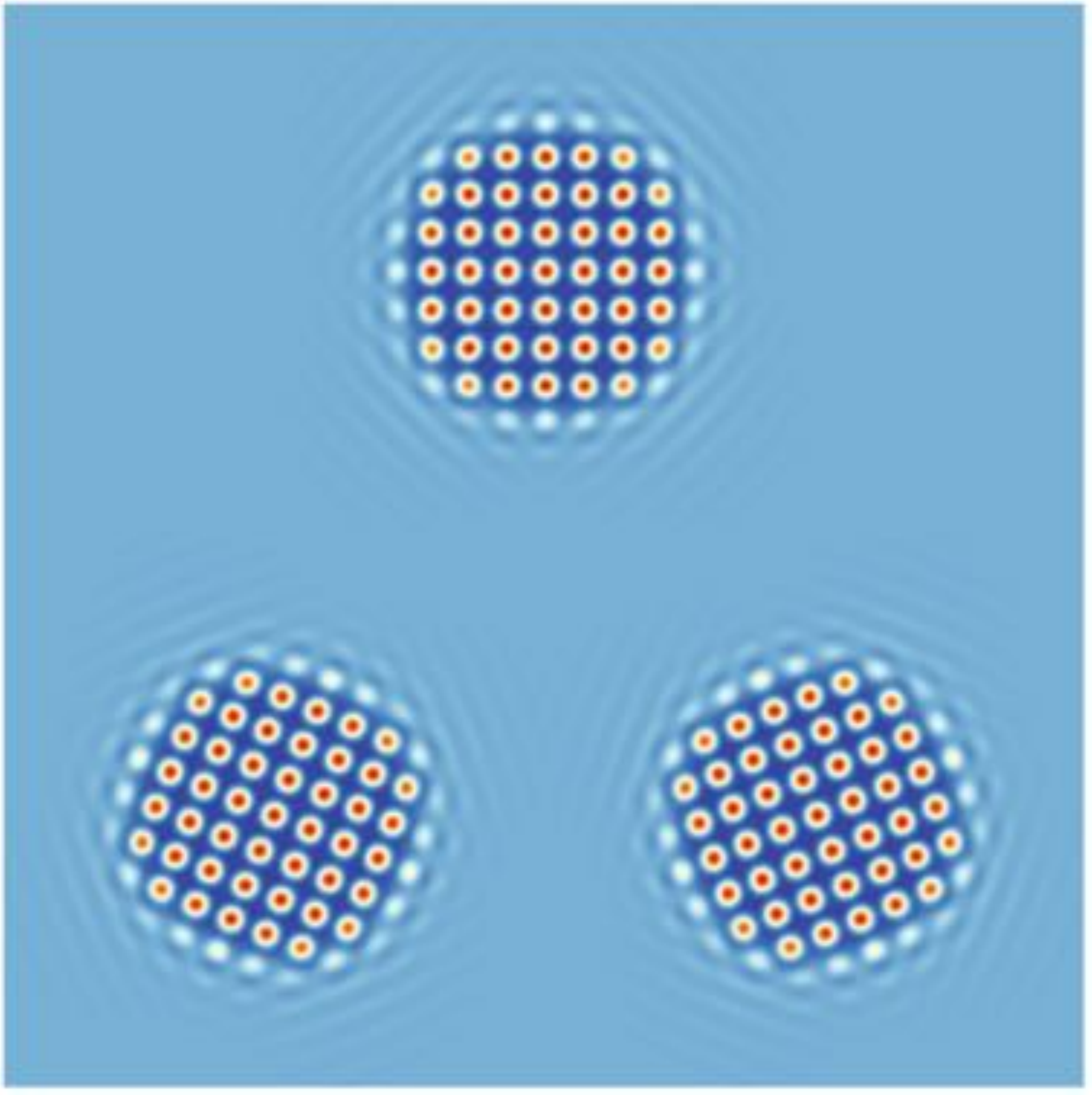}}
\end{center}
\end{figure}

\begin{figure}[!t]
\begin{center}
\scriptsize{(c) $t = 300$}\\
{\includegraphics[width=0.4\textwidth]{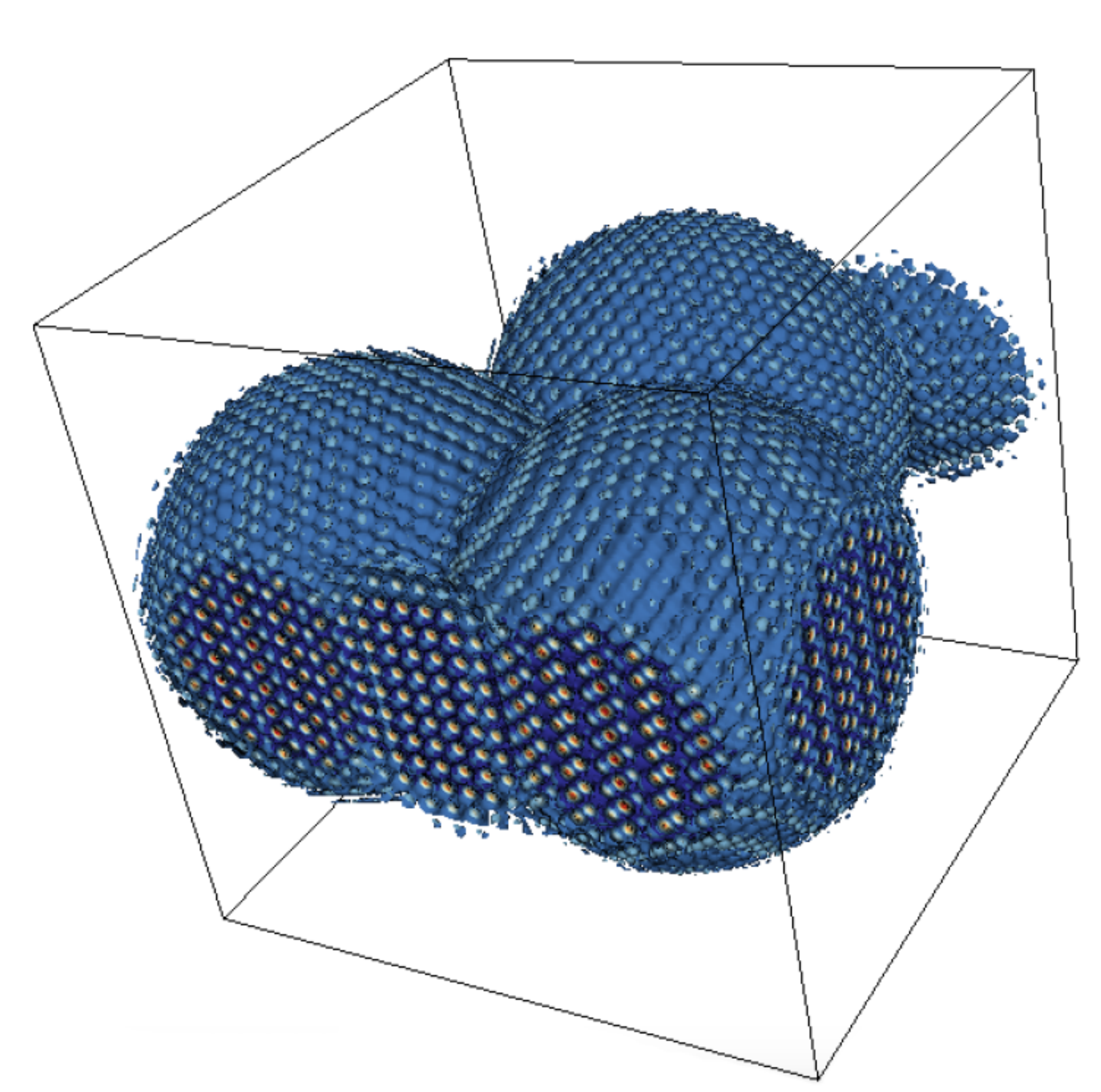}}\qquad
{\includegraphics[width=0.4\textwidth]{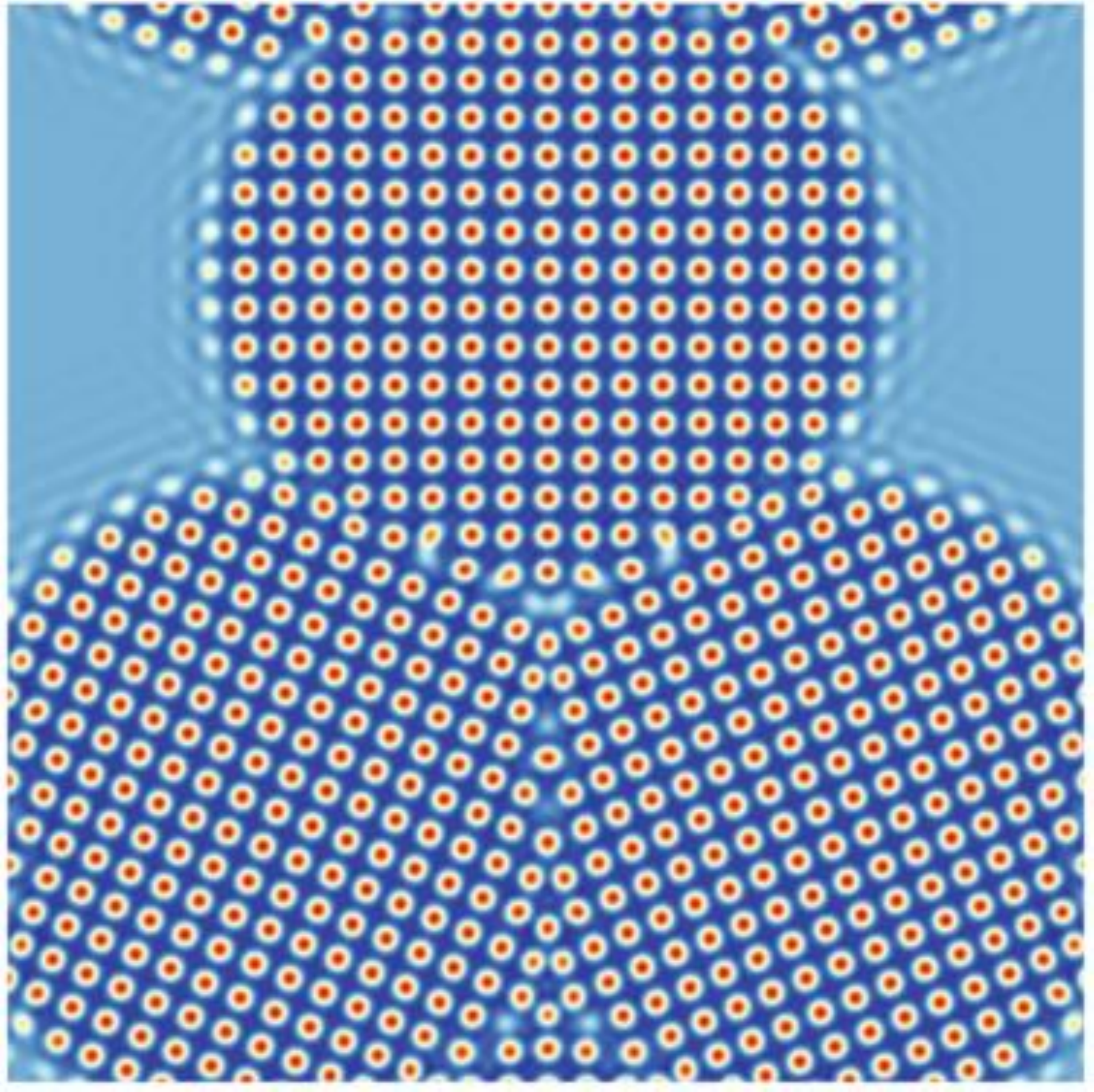}}
\\~~\\
\scriptsize{(d) $t = 650$}\\
{\includegraphics[width=0.4\textwidth]{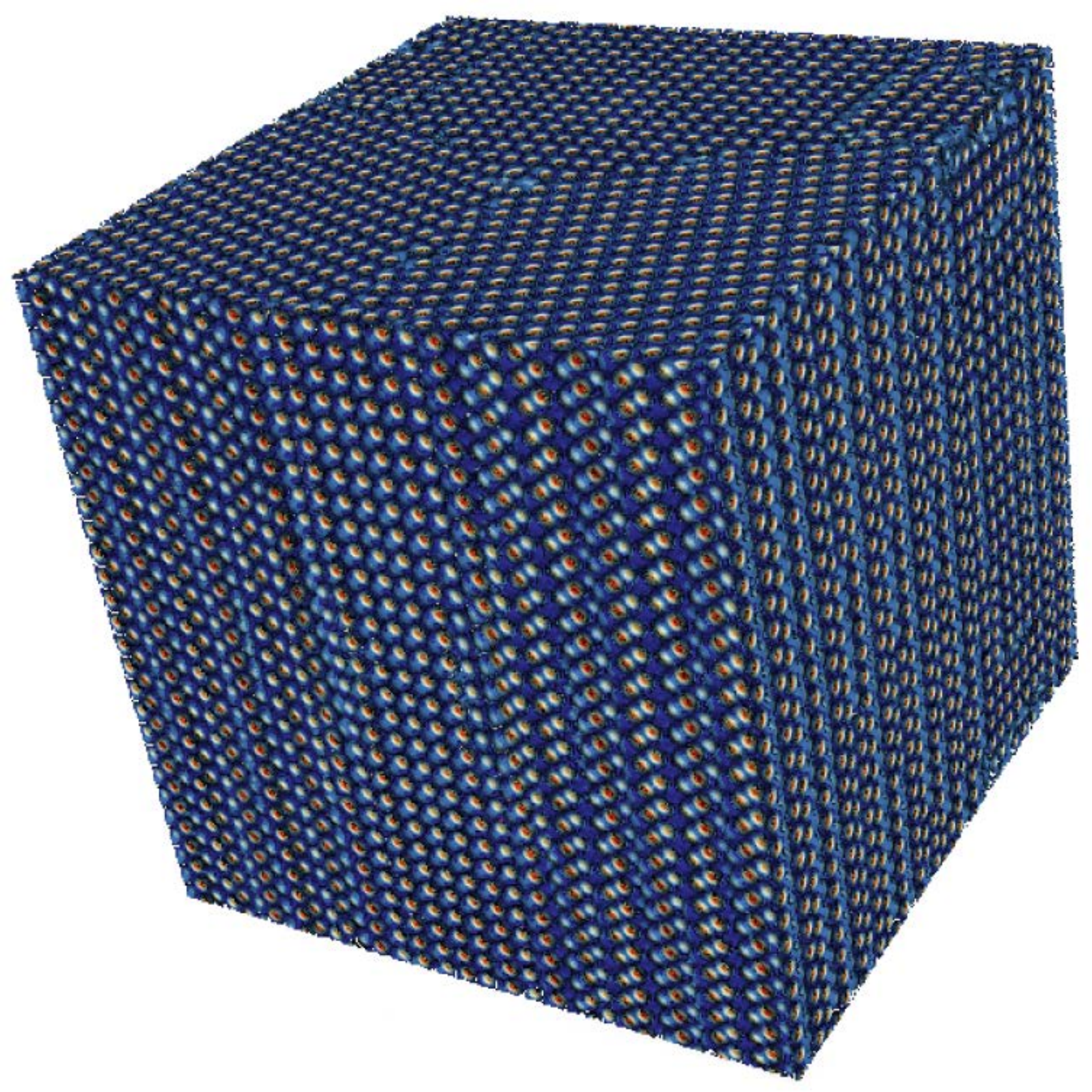}}\qquad
{\includegraphics[width=0.4\textwidth]{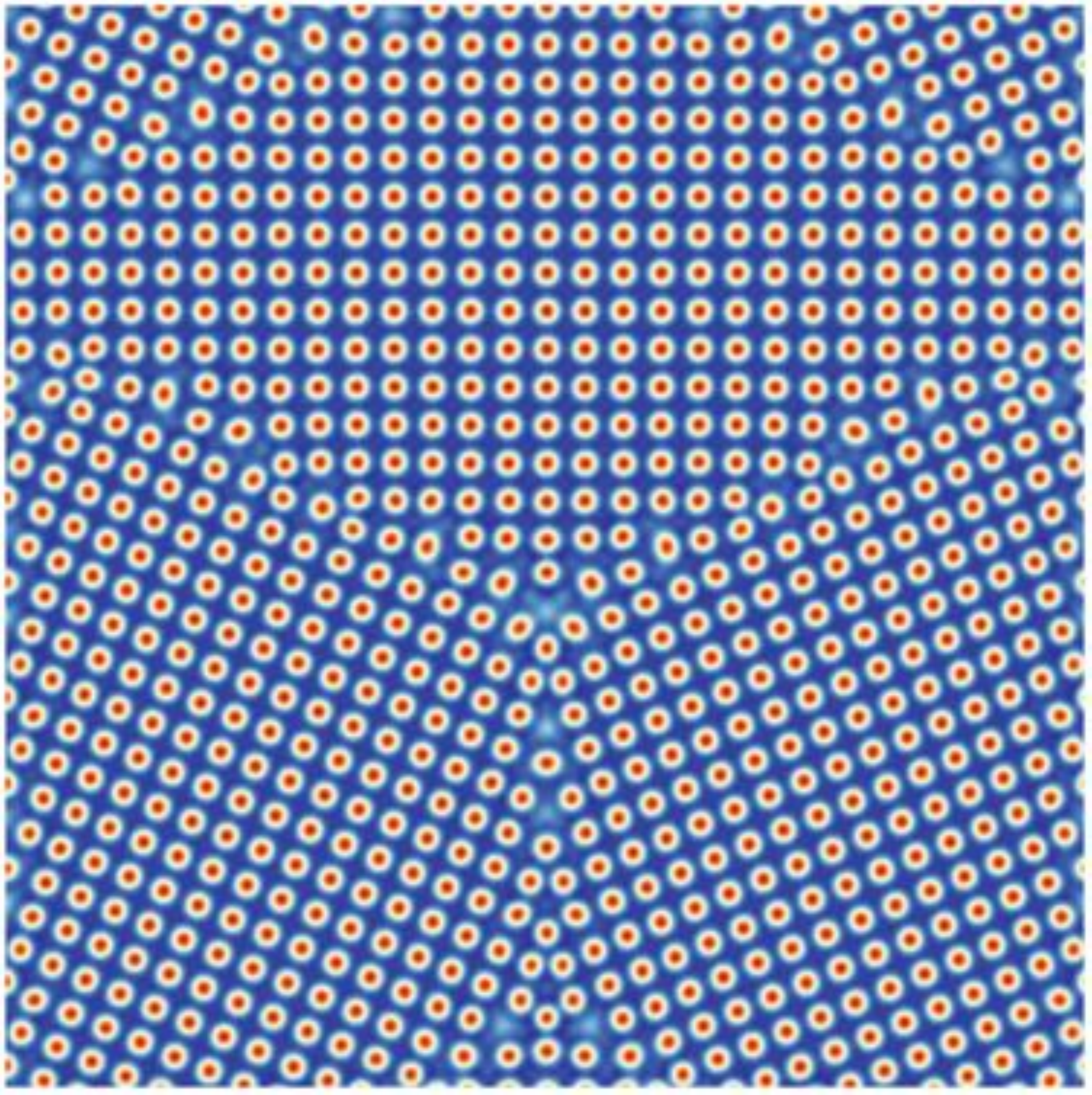}}
\end{center}
\caption{Results of polycrystalline growth in a 3D supercooled liquid. Shown in the pictures are the isosurface (left panel) and the $z = 40\pi$ plane (right panel) of the atomistic density field.}
\label{ex4:sol2}
\end{figure}

Analytically, the BCC configuration is defined as \cite{pfm, epd}
\begin{equation}
\phi_{\mathrm{BBC}}(\bx) = \cos(xq)\cos(yq) + \cos(xq)\cos(zq) + \cos(yq)\cos(zq),
\end{equation}
where $\bx =(x, y, z)$ represents the point in the three-dimensional Cartesian coordinate system and $q$ represents a 
wavelength related to the BCC crystalline structure. 
In our simulation, a regular crystallite has the form as follows 
\begin{equation}
\phi(\bx) = \bar{\phi} + A\rho(\bx)\phi_{\mathrm{BBC}}(\bx),
\end{equation}
where $\bar{\phi}$ represents the average density of the liquid-crystal system, and $A$ represents 
an amplitude of the fluctuations in density. The scaling function $\rho(\bx)$ is defined as
\begin{equation}
\rho(\bx) = \left\{
\begin{aligned}
&~\left(1-\left(\frac{\|\bx-\bx_0\|}{d_0}\right)^2\right)^2 \ \ \ \ \ \textnormal{if}\ \  \|\bx-\bx_0\|\leq d_0,\\
&~\ \ \ \ \ \ \ \ \ \ \ \  \  0 \ \ \ \ \ \ \ \ \ \ \ \  \ \ \ \ \ \ \ \ \ \ \ \ \  \ \textnormal{otherwise},
   \end{aligned}\right.\nonumber
\end{equation}
where $\bx_0=(x_0, y_0, z_0)$ is the center of the crystallite sphere, and $d_0$ is the radius of the crystallite sphere. 
In order to define the crystallites oriented in different direction, we replace $\bx$ with $\tilde{\mathbf{\bx}}$, 
in which a system of local Cartesian coordinates $\tilde{x}, \tilde{y}, \tilde{z}$ is used to generate the 
crystallites in different directions. 
We make an affine transformation of the global coordinates to produce a rotation with an angle 
$\omega$ along the $z$-axis, in which the local Cartesian coordinates are given as follows
\begin{equation}
\tilde{\mathbf{\bx}} =
\left(\begin{array}{c}
\tilde{x} \\
\tilde{y} \\
\tilde{z} \\
\end{array}
\right)=
\left(\begin{array}{c}
\cos(\omega)(x-x_0) - \sin(\omega)(y-y_0)\\
\sin(\omega)(x-x_0) + \cos(\omega)(y-y_0)\\
z-z_0\\
\end{array}
\right).
\end{equation}
In the experiment, we situate the three crystallite spheres on the $z=L/2$ plane with 
the same radius $d_0=5\pi$, where $L=80\pi$. The centers of the three crystallite spheres are 
$(L/2, 3L/4, L/2)$, $(L/4,L/4,L/2)$, $(3L/4,L/4,L/2)$, respectively.
The rotation angles $\omega$ for the three crystallite spheres are $0, -\pi/8, \pi/8$, respectively.
In the simulation, $\bar{\phi}=-0.35$, $q=1/\sqrt{2}$, $A=1$, $\gamma=0.35$ and $M(\phi)=1$  are used.

The numerical solutions at different times are shown in Fig.~\ref{ex4:sol2}. 
Similarly to the two dimensional case, 
the crystallite spheres grow in the liquid, and the grain boundaries appear when the crystals meet due to 
the orientation mismatch.
\begin{figure*}[!h]
\centering
\includegraphics[width=0.72\textwidth]{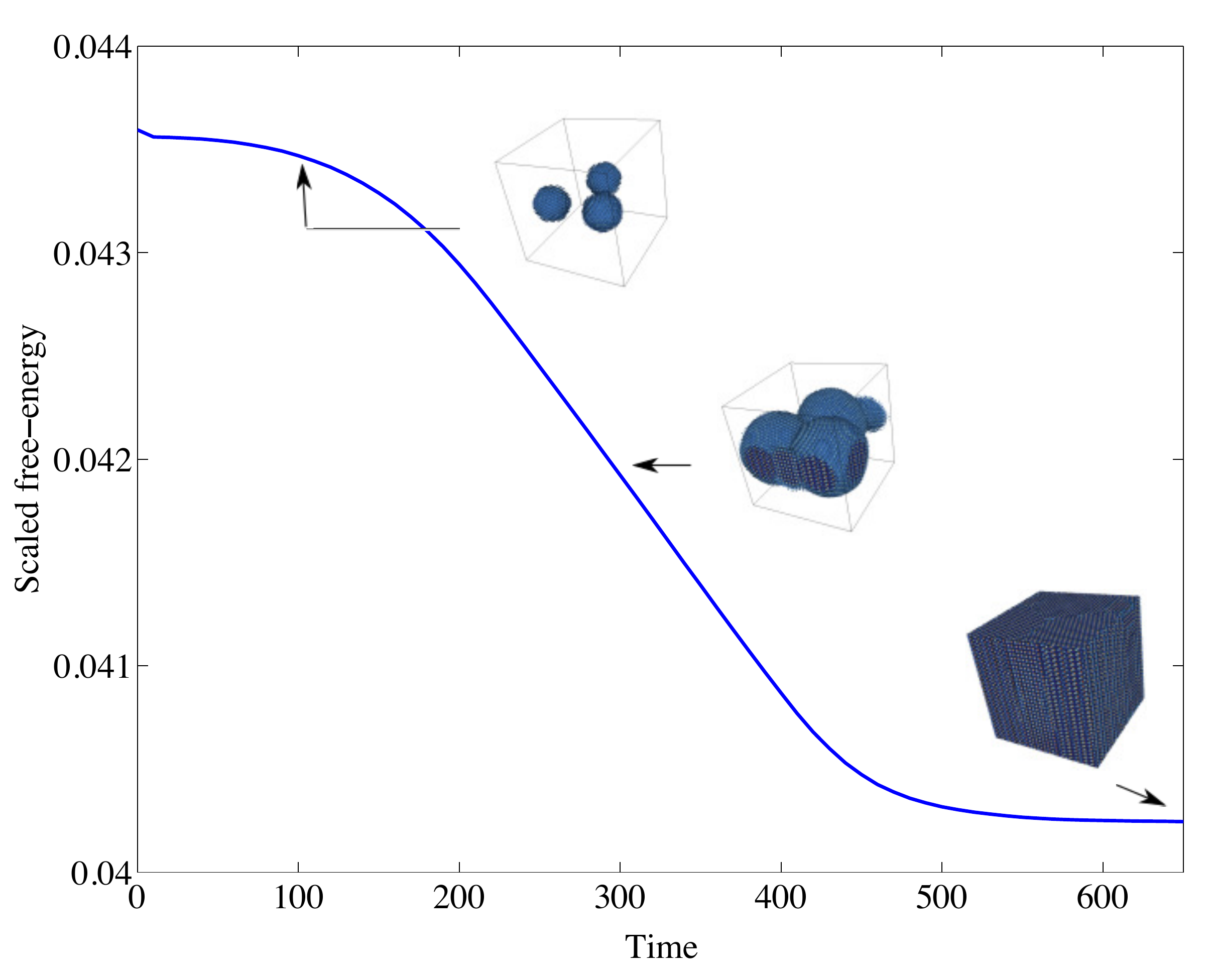}
\caption{The scaled total free-energy of polycrystalline growth in a 3D supercooled liquid eps.}
\label{ex4:energy}       % Give a unique label
\end{figure*}
The free-energy evolution for the simulation is shown in Fig.~\ref{ex4:energy}, in which we can see that 
the free-energy has no increase. 
The time step size keeps the lower bound $\Delta t_{min}$ because the change of free-energy is rapid in 
the whole time interval $[0,650]$.
To observe the grain boundaries clearly, our initial crystallites are fixed on the same plane and only have rotations along 
the $z$-axis, if the readers are interested in other situations, the positions and the rotation angles 
both can be changed freely and similar simulations can be done easily.

\subsection{Performance tuning}
%In this subsection, we focus on the parallel performance of the method. Several issues are carefully discussed including the influence of different overlaps and subdomain solvers, the strong scalability study

It is well known that the performance of the Schwarz preconditioner depends on the choice of 
the subdomain solver. To investigate it, we run the above four test cases for the first $10$ time 
steps, respectively, 
in which the periodic boundary conditions are imposed on computational, and the mobility $M(\phi)$ 
is fixed to be $1$ for convenience. 
To check the influence of the subdomain solvers, we limit the 
test to the classical AS preconditioner and fix the overlapping size to $\delta = 1$. 
The ILU factorizations with $0$, $1$, and $2$ levels of fill-in 
and LU factorizations are considered. 
The number of processor cores, the mesh size, and the time step size for the four tests are listed as follows.
\begin{itemize}
\setlength{\itemsep}{0pt}
\item In the test $A$, $24$ processor cores with a $256\times 256$ mesh and a fixed 
time step size $\Delta t =0.01$ are used.
\item In the test $B$, $288$ processor cores with a $1,024\times 1,024$ mesh and a fixed 
time step size $\Delta t =1$ are applied.
\item In the test $C$, $576$ processor cores with a $4,096\times 1,024$ mesh and a fixed 
time step size $\Delta t =1$ are used.
\item In the test $D$, $4,096$ processor cores with a $128\times 128\times 128$ mesh and a fixed 
time step size $\Delta t =0.1$ are applied. And the computational domain is scaled down to $[0,40\pi]^3.$
\end{itemize}
The numbers of Newton and GMRES iterations together with the total compute time are provided in 
Table~\ref{tab:subdomain}.

\begin{table}[!htb]
\renewcommand{\arraystretch}{1.2}
\caption{\upshape
Performance of the NKS with different subdomain solvers. Here n/c means the divergence of the GMRES solver.
}
\label{tab:subdomain} 
\centering
\begin{tabular}%{|c|c|ccccc|}
{|p{1.2cm}|p{3.2cm}|p{1.1cm}p{1.1cm}p{1.1cm}p{1.1cm}p{1.9cm}|}
\noalign{\smallskip}
\hline
\multicolumn{2}{|c|}{Subdomain solver} &ILU(0) &ILU(1) &ILU(2) &LU &ILU-reuse \\
\hline\hline
\multirow{3}{*}{Test A} & Total Newton        &30       &30      &30      &30       &30  \\
                                    & GMRES/Newton  &6.0      &3.93   &3.87   &3.87    &6.0 \\
                                    &Total Time (s)        &3.17    &3.83   &5.39   &40.43    &2.01 \\
\hline
\multirow{3}{*}{Test B} & Total Newton          &n/c       &30           &30      &30         &30  \\
                                    & GMRES/Newton    &n/c       &101.03    &8.17   &7.87      &8.4 \\
                                    & Total Time (s)         &n/c       &30.47      &8.29   &56.75    &5.12 \\
\hline
\multirow{3}{*}{Test C} & Total Newton        &34       &33         &33       &33        &34  \\
                                    & GMRES/Newton  &5.88     &3.94      &3.91    &3.91     &6.56 \\
                                    & Total Time (s)       &9.99     &11.28     &14.86  &191.75   &6.02 \\
\hline
\multirow{3}{*}{Test D} & Total Newton        &31          &31           &31           &31          &31  \\
                                    & GMRES/Newton  &7.29       &7.23        &7.23        &7.23       & 7.42 \\
                                    &Total Time (s)        &24.01     &284.68    &982.22    &683.94   & 10.41\\
\hline
\end{tabular}
\end{table}

For all test cases, the number of Newton 
iterations is insensitive to the subdomain solver.
It is clear that by increasing the fill-in level, the number of GMRES iterations decreases, 
but the compute time does not necessarily reduce due to the increased cost of the subdomain 
solver. In summary, we find that the optimal choice in terms of the total compute time is the ILU(0) 
or ILU(2) subdomain solver.
In the Newton method, the Jacobian matrices of the each Newton iteration 
have very similar structures, so it is possible to save the compute time by 
only performing the factorization once and reusing the preconditioner matrices within the all Newton iteration. 
Here, we apply the reuse strategy to the optimum subdomain solver which is ILU(0) for test cases $A$, $C$, $D$ 
and is ILU(2) for test case $B$. The results are listed in the last column of Table~\ref{tab:subdomain}, which 
indicates that the reuse strategy can save nearly 50\% of the compute time.

We then investigate the performance of the NKS solver by changing the type of the AS 
preconditioner and the overlapping factor $\delta$. The number of processor cores, 
the mesh size and the time step size for the four test cases 
are the same with the previous simulation. Based on the previous report, we take the optimal 
choice of subdomain solver with the reuse strategy throughout the test cases. 
The classical-AS \eqref{invM}, the left-RAS \eqref{eq:ras}, 
and the right-RAS \eqref{eq:ash} preconditioners with overlapping size $\delta$ gradually 
increasing from 0 to 2 are considered. 
The numbers of Newton and GMRES iterations
together with the total compute time are listed in Table \ref{tab:overlap}.
\begin{table*}[!hbt]
\renewcommand{\arraystretch}{1.2}
\caption{\upshape
Performance of NKS with different types of preconditioner and different overlapping size. }
\label{tab:overlap} 
\centering
\begin{tabular}%{|c|c|ccc|cc|cc|}
{|p{1.2cm}|p{3.2cm}|p{0.9cm}p{0.9cm}p{0.9cm}|p{0.9cm}p{0.9cm}|p{0.9cm}p{0.9cm}|}
\hline
\multicolumn{2}{|c|}{Preconditioner}
&\multicolumn{3}{|c|}{classical-AS}
&\multicolumn{2}{|c|}{left-RAS}
&\multicolumn{2}{|c|}{right-RAS}\\
\hline
\multicolumn{2}{|c|}{$\delta$}  &$0$ &$1$ &$2$ &$1$ &$2$   &$1$ &$2$   \\
\hline\hline
\multirow{3}{*}{Test A} & Total Newton        &30        &30      &30      &30     &30       &30       &30       \\
                                    & GMRES/Newton  &10.0     &6.0     &6.1     &2.67  &3.0      &2.67    &3.0     \\
                                    &Total Time (s)        &2.20     &2.01   &2.28   &1.57  &1.78    &1.55    &1.79   \\
\hline
\multirow{3}{*}{Test B} & Total Newton            &30          &30        &30         &30       &30        &30       &30      \\
                                    & GMRES/Newton      &113.23    &8.4       &84.73    &8.7      &55.13   &7.53    &37.7     \\
                                    & Total Time (s)           &31.31      &5.12    &33.18     &5.21   &22.13    &4.86    &16.07   \\
\hline
\multirow{3}{*}{Test C} & Total Newton        &34           &34      &34     &34        &34        &34       &34     \\
                                    & GMRES/Newton  &13.82      &6.56    &6.56  &2.91     &2.88    &2.91    &2.88    \\
                                    & Total Time (s)       &8.25        &6.02   &6.66    &4.62    &5.11    &4.63    &5.10   \\
\hline
\multirow{3}{*}{Test D} &Total Newton         &32        &31        &32        &31        &31      &31        &31     \\
                                     &GMRES/Newton  &12.88   &7.42     &5.0        &2.06    &2.06   &2.06     &2.06     \\
                                     &Total Time (s)       &3.69     &10.41   &35.06    &8.27    &30.49  &8.24    &30.51   \\
\hline
\end{tabular}
\end{table*}
From the Table, we conclude that the left-RAS and right-RAS preconditioner are superior to 
the classical-AS preconditioner. For the test cases $A$ and $B$, the minimal compute time 
and the least number of GMRES iterations are obtained when the overlapping size $\delta=1$.  
For the test case $C$, the minimal compute time is achieved when the overlapping size $\delta=1$ 
but the least number of GMRES iterations is obtained at $\delta=2$.  
For the test $D$, the minimal compute time 
and the greatest number of GMRES iterations are obtained when the overlapping size $\delta=0$. 
The observations reflect that the number of GMRES iterations does not necessarily reduce as the overlapping size increase.

\subsection{Large-scale scalability}

\begin{figure}[!th]
\begin{center}
\qquad\scriptsize{(a)}\qquad\qquad\qquad\qquad\qquad\qquad\qquad\qquad\qquad\qquad\scriptsize{(b)}\\
{\includegraphics[width=0.48\textwidth]{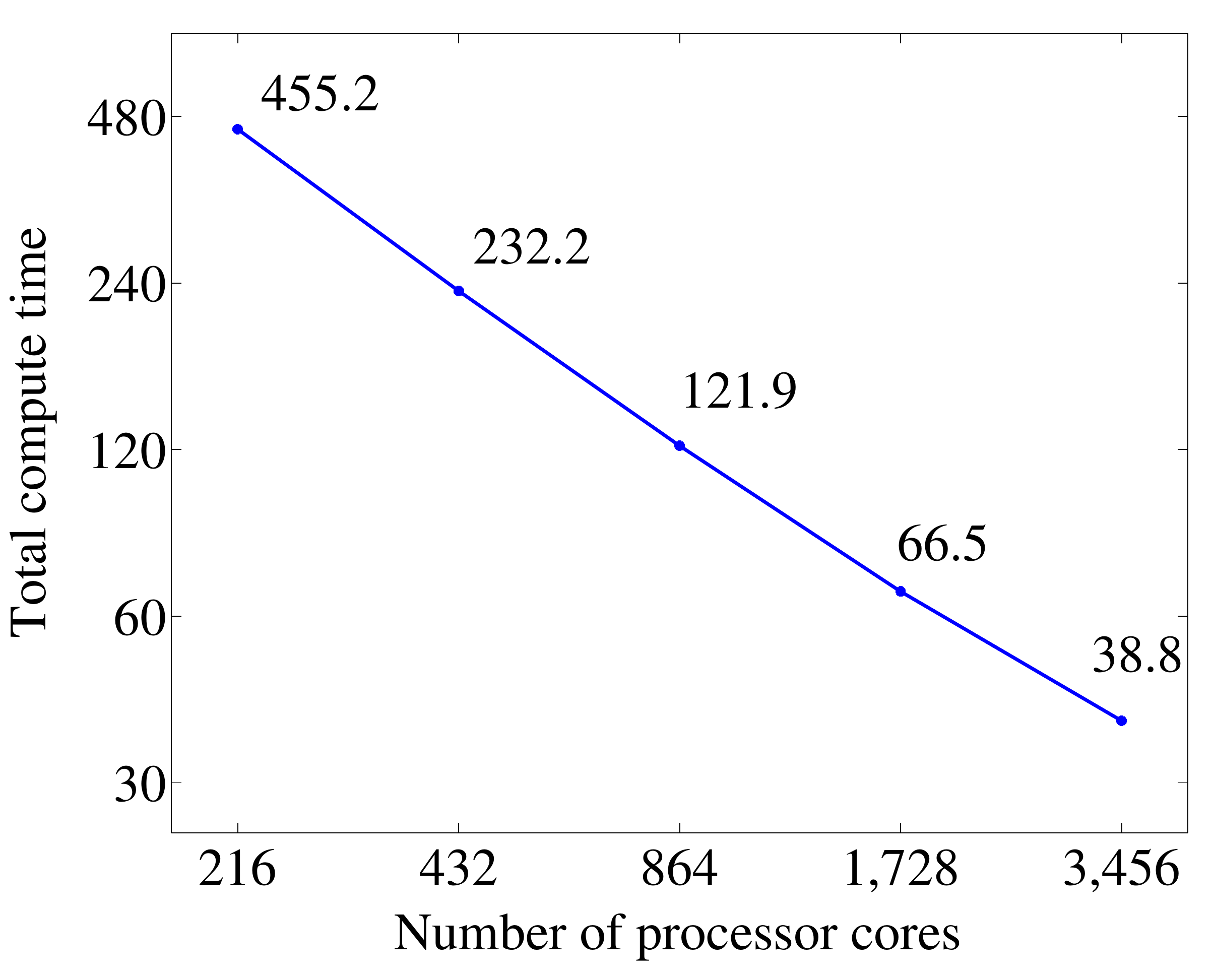}}
{\includegraphics[width=0.48\textwidth]{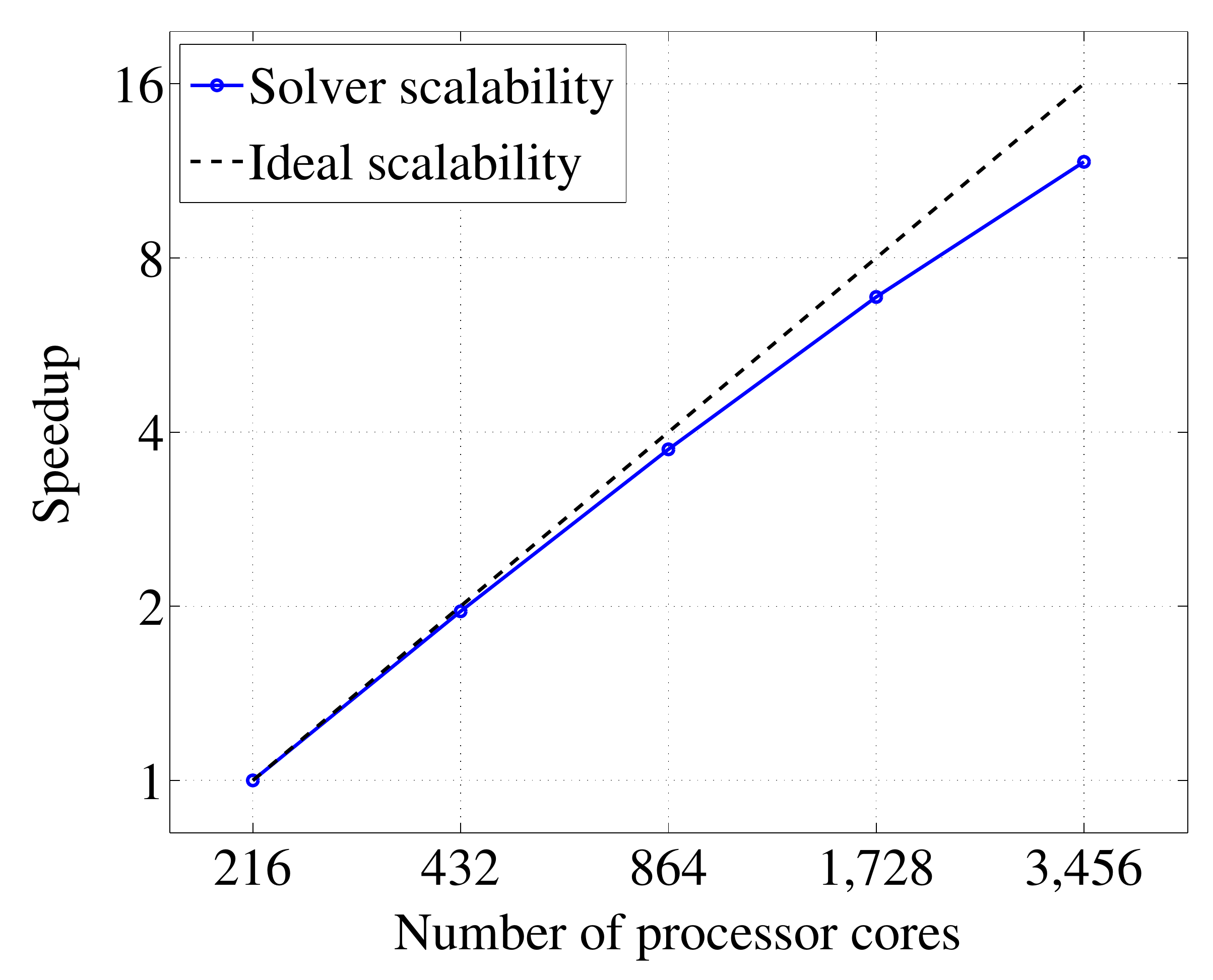}}
\end{center}
\caption{
The total compute time and the strong scalability for test case C. }
\label{scalability1}
\end{figure}

To study the parallel scalability, we run the test case $C$ and $D$ for the first $10$ time steps with 
different number of processor cores, respectively. In the test case $C$, a $24,576 \times 6,144$ mesh is 
considered and time step size $\Delta t$ is fixed to be $1$. Based on the observations from the above 
subsection, we use the left-RAS preconditioner with the overlapping $\delta=1$ and employ the 
sparse ILU(0) factorization with the reuse strategy as the subdomain solver. The numbers of nonlinear 
and linear iterations are reported in Table~\ref{tab:scalability1}, 
\begin{table}[!htb]
\renewcommand{\arraystretch}{1.2}
\caption{\upshape
Performance results with different number of processor cores, here NP means the number of processor cores. 
}
\label{tab:scalability1} 
\centering
\begin{tabular}%{|c|ccccc|}
{|p{3.2cm}|p{1.4cm}p{1.4cm}p{1.4cm}p{1.4cm}p{1.4cm}|}
\noalign{\smallskip}
\hline
 NP &216 &432 &864 &1,728 &3,456 \\
\hline\hline
Total Newton        &34     &34    &34      &34     &34  \\
GMRES/Newton  &2.88  &2.88  &2.88  &2.88  &2.88 \\
\hline
\noalign{\smallskip}
\end{tabular}
\end{table}
which displays that the total number of nonlinear iterations 
and the average number of linear iterations keep constant during the increase process of the number 
of processor cores.
Fig.~\ref{scalability1} shows the results on the total compute time and the parallel scalability.
It can be seen from Fig.~\ref{scalability1} that the total compute time decreases almost linearly,
as the number of processor cores increases.
The overall speedup from $216$ to $3,456$ cores is around $11.7$,
which indicates that the proposed algorithm has a good parallel efficiency for the 2D test case.

\begin{figure}[!th]
\begin{center}
\qquad\scriptsize{(a)}\qquad\qquad\qquad\qquad\qquad\qquad\qquad\qquad\qquad\qquad\scriptsize{(b)}\\
{\includegraphics[width=0.48\textwidth]{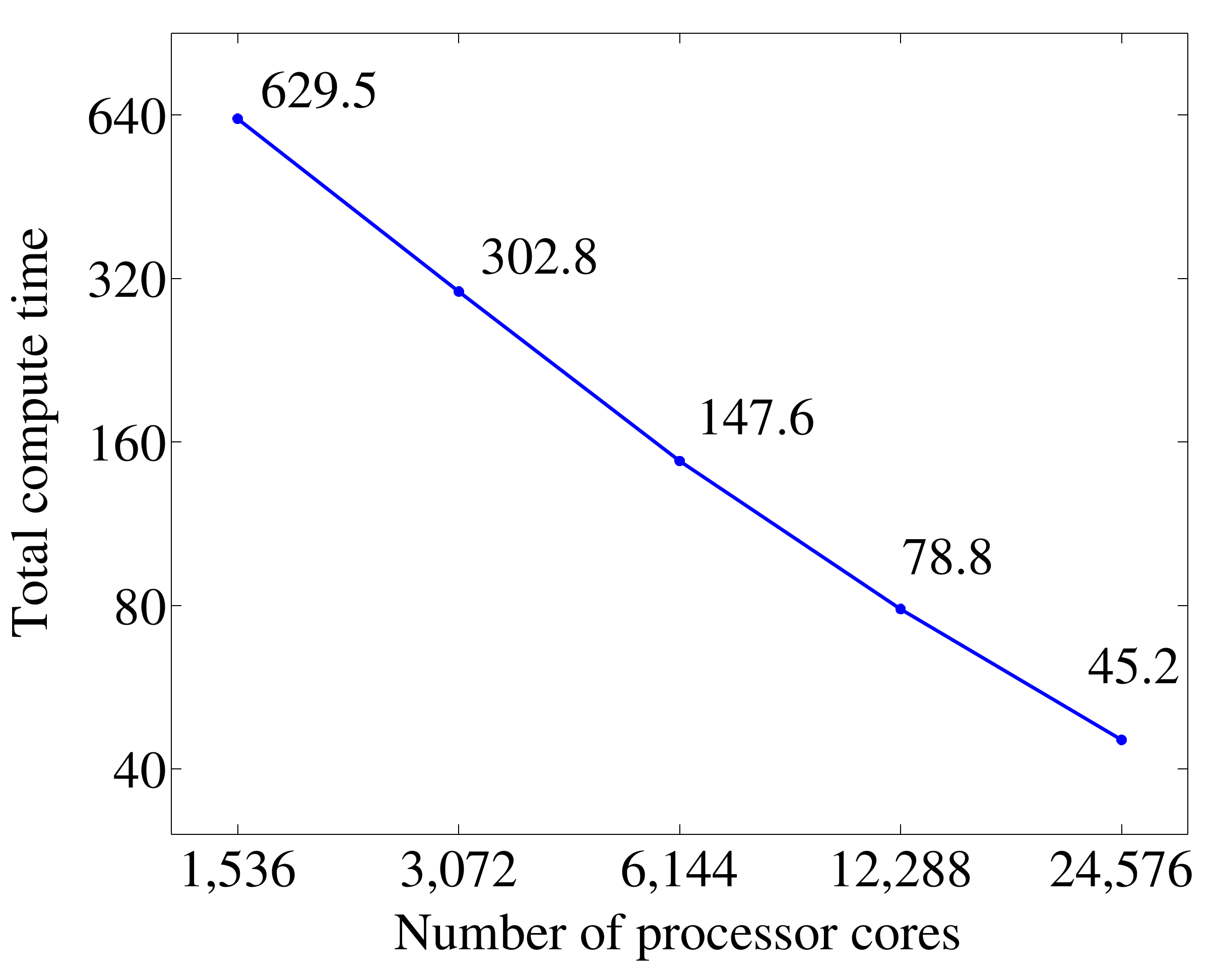}}
{\includegraphics[width=0.48\textwidth]{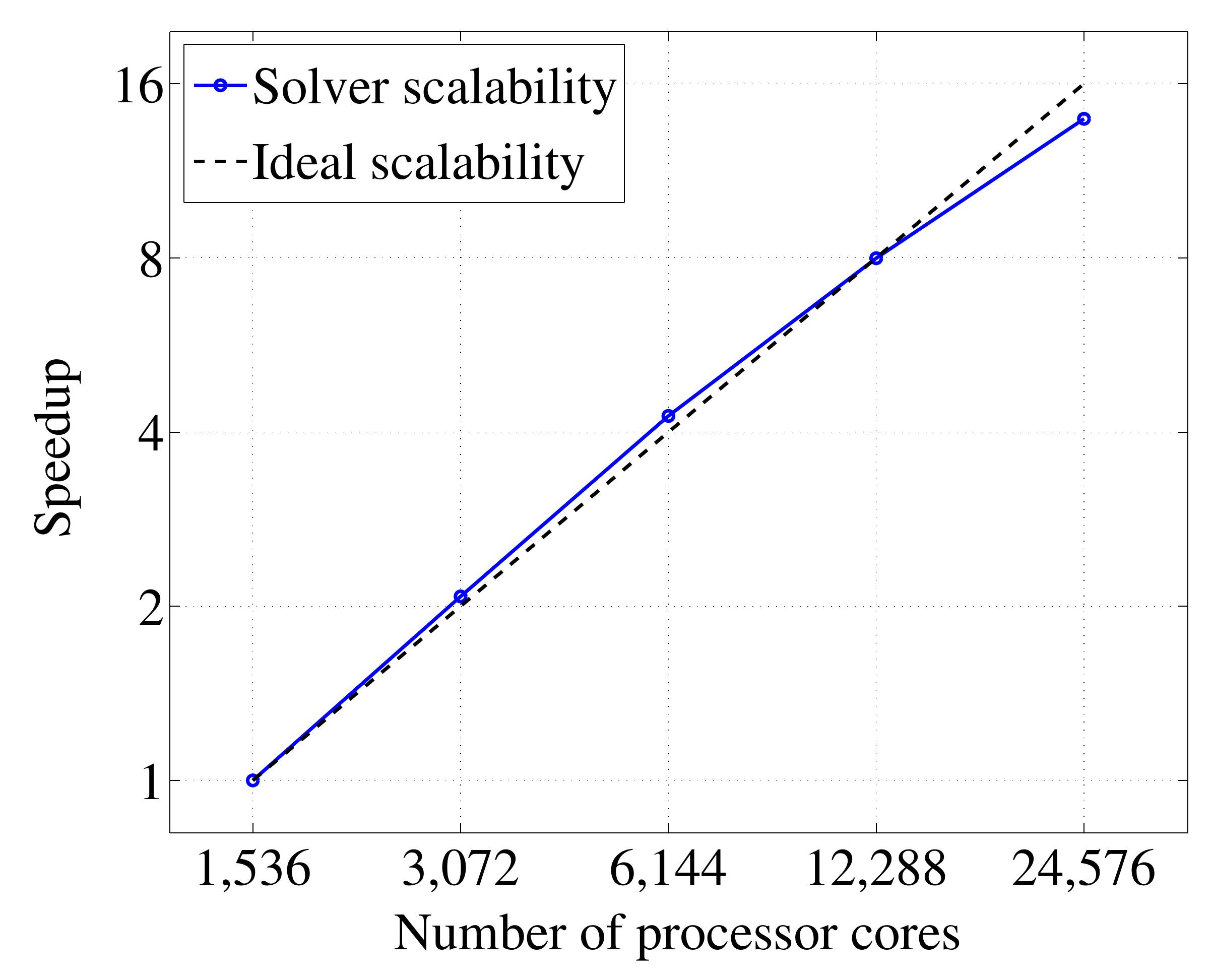}}
\end{center}
\caption{
The total compute time and the strong scalability for the test case D. }
\label{scalability2}
\end{figure}

We run the test case $D$ on $512 \times 512 \times 512$ mesh with a fixed time step size $\Delta t = 0.1$. 
We use the classical-AS preconditioner with overlapping size $\delta=0$
and employ the sparse ILU(0) factorization with reuse strategy as the subdomain solver. 
The numbers of nonlinear and linear iterations are reported in Table~\ref{tab:scalability2}. 
\begin{table}[!htb]
\renewcommand{\arraystretch}{1.2}
\caption{\upshape
Performance results with different number of processor cores. 
}
\label{tab:scalability2} 
\centering
\begin{tabular}%{|c|ccccc|}
{|p{3.2cm}|p{1.4cm}p{1.4cm}p{1.4cm}p{1.4cm}p{1.4cm}|}
\noalign{\smallskip}
\hline
 NP &1,536 &3,072 &6,144 &12,288 &24,576 \\
\hline\hline
Total Newton        &32       &32        &31        &31           &31  \\
GMRES/Newton  &12.63  &12.66   &11.97   &12.65      &12.65 \\
\hline
\noalign{\smallskip}
\end{tabular}
\end{table}
We notices that the number of nonlinear iterations 
and the average number of linear iterations are almost unchanged during the increase of the number 
of processors, which implies that the total number of nonlinear iterations and the average number of linear iterations are insensitive to the number of processor cores. Fig.~\ref{scalability2} shows the results on 
the total compute time and the parallel scalability.
The total compute time decreases almost linearly as the number of processor cores increases. 
The overall speedup from $1,536$ to $24,576$ cores is around $13.9$, 
which indicates an almost ideal parallel efficiency of the proposed algorithm for the 3D test case.

\section{Conclusion}
In this paper, a semi-implicit finite difference scheme and a highly parallel domain decomposition algorithm are 
proposed for solving the PFC equation. The semi-implicit finite difference scheme is derived based on the 
DVD method and is proved to be unconditionally stable and satisfies 
the second order accuracy in time and space. 
For the steady state calculation, an adaptive time stepping strategy is successfully incorporated into 
the semi-implicit time integration scheme so that the time step size is controlled based on the state of solution. 
The nonlinear system constructed by the discretization of PFC equation at each time step is solved by 
the NKS method with modified boundary conditions for the subdomain solves. The accuracy and applicability of the proposed method are validated by several two and 
three dimensional test cases.
The performance of the NKS method is tuned by changing the subdomain solver, 
the type of the Schwarz preconditioner and the overlapping size.
Large scale numerical experiments show that the proposed algorithm can scale well to over ten thousands 
processor cores on the Sunway TaihuLight supercomputer.

\section*{Acknowledgments}

This work was supported in part by Natural Science Foundation of China (grant\# 91530323, 11501554), National Key R\&D Plan of China (grant\# 2016YFB0200603), and Key Research Program of Frontier Sciences from CAS (grant\# QYZDB-SSW-SYS006).

\bibliographystyle{elsarticle-num}
\bibliography{template}

\begin{thebibliography}{10}
\expandafter\ifx\csname url\endcsname\relax
  \def\url#1{\texttt{#1}}\fi
\expandafter\ifx\csname urlprefix\endcsname\relax\def\urlprefix{URL }\fi
\expandafter\ifx\csname href\endcsname\relax
  \def\href#1#2{#2} \def\path#1{#1}\fi

\bibitem{PhysRevLett.88.245701}
K.~R. Elder, M.~Katakowski, M.~Haataja, M.~Grant, Modeling elasticity in
  crystal growth, Phys. Rev. Lett. 88 (2002) 245701.

\bibitem{PhysRevE.70.051605}
K.~R. Elder, M.~Grant, Modeling elastic and plastic deformations in
  nonequilibrium processing using phase field crystals, Phys. Rev. E 70 (2004)
  051605.

\bibitem{zhang2013}
Z.~Zhang, Y.~Ma, Z.~Qiao, An adaptive time-stepping strategy for solving the
  phase field crystal model, J. Comput. Phys. 249 (2013) 204--215.

\bibitem{Yang_ascalable}
C.~Yang, X.-C. Cai, A scalable implicit solver for phase field crystal
  simulations, In Parallel and Distributed Processing Symposium Workshops PhD
  Forum (IPDPSW), 2013 IEEE 27th International (2013) 1409--1416.

\bibitem{Cheng_anefficient}
M.~Cheng, J.~A. Warren, An efficient algorithm for solving the phase field
  crystal model, J. Comput. Phys. 227 (2008) 6241--6248.

\bibitem{Gomez201252}
H.~Gomez, X.~Nogueira, An unconditionally energy-stable method for the phase
  field crystal equation, Comput.~Methods~Appl.~Mech.~Engrg. 249-252 (2012)
  52--61.

\bibitem{ElderPRB}
K.~R. Elder, N.~Provatas, J.~Berry, P.~Stefanovic, M.~Grant, Phase-field
  crystal modeling and classical density functional theory of freezing, Phys.
  Rev. B 75 (2007) 064107.

\bibitem{WuPRB}
K.-A. Wu, P.~W. Voorhees, Stress-induced morphological instabilities at the
  nanoscale examined using the phase field crystal approach, Phys. Rev. B 80
  (2009) 125408.

\bibitem{prb136.864}
P.~Hohenberg, W.~Kohn, Inhomogeneous electron gas, prb 136 (1964) 864--871.

\bibitem{Nucleation}
R.~Backofen, A.~R\"atz, A.~Voigt, Nucleation and growth by a phase field
  crystal {(PFC)} model, Phil. Mag. Lett. 87 (2007) 813.

\bibitem{FASOO}
H.~G. Lee, J.~Shin, J.-Y. Lee, First and second order operator splitting
  methods for the phase field crystal equation, J. Comput. Phys. 299 (2015)
  82--91.

\bibitem{precondition}
S.~Praetorius, A.~Voigt, Development and analysis of a block-preconditioner for
  the phase-field crystal equation, arXiv:1501.06852v1.

\bibitem{Wise2009}
S.~Wise, C.~Wang, J.~Lowengrub, An energy-stable and convergent
  finite-difference scheme for the phase field crystal equation, SIAM J. Numer.
  Anal. 47 (2009) 2269--2288.

\bibitem{wise20092}
Z.~Hu, S.~M. Wise, C.~Wang, J.~S. Lowengrub, Stable and efficient
  finite-difference nonlinear-multigrid schemes for the phase field crystal
  equation, J. Comput. Phys. 228 (2009) 5323--5339.

\bibitem{Elsey2012}
M.~Elsey, B.~Wirth, A simple and efficient scheme for phase field crystal
  simulation, ESAIM: Math. Mod. Num. Anal.

\bibitem{vignal2015}
P.~Vignal, L.~Dalcin, D.~L. Brown, N.~Collier, V.~M. Calo, An energy-stable
  convex splitting for the phase-field crystal equation, Comput. Struct. 158
  (2015) 355--368.

\bibitem{guo2016}
R.~Guo, Y.~Xu, Local discontinuous galerkin method and high order semi-implicit
  scheme for the phase field crystal equation, SIAM~J.~Sci.~Comput. 38 (2016)
  105--127.

\bibitem{DVDM}
D.~Furihata, T.~Matsuo, {Discrete variational derivative method : a
  structure-preserving numerical method for partial differential equations},
  Chapman and Hall/CRC, 2011.

\bibitem{NKS}
X.-C. Cai, W.~D. Gropp, D.~E. Keyes, M.~D. Tidriri, {Newton-Krylov-Schwarz}
  methods in {CFD}, in: R.~Rannacher (Ed.), Proceedings of the International
  Workshop on the Navier-Stokes Equations, Notes in Numerical Fluid Mechanics,
  Vieweg Verlag, Braunschweig, 1994, pp. 123--135.

\bibitem{PhysRevA.15.319}
J.~Swift, P.~C. Hohenberg, Hydrodynamic fluctuations at the convective
  instability, Phys. Rev. A 15 (1977) 319--328.

\bibitem{NMFP}
Z.~Li, {Numerical methods for partial differential equations}, Peking
  University Press, 2010.

\bibitem{NMFU}
J.~E. Dennis, R.~B. Schnabel, Numerical Methods for Unconstrained Optimization
  and Nonlinear Equations, Society for Industrial and Applied Mathematics,
  Philadelphia, 1996.

\bibitem{gmres}
Y.~Saad, M.~H. Schultz, Gmres: A generalized minimal residual algorithm for
  solving nonsymmetric linear systems, SIAM J. Sci. Stat. Comput. 7 (1986)
  856--869.

\bibitem{DDA}
M.~Dryja, O.~B. Widlund, Domain decomposition algorithms with small overlap,
  SIAM~J.~Sci.~Comput. 15 (1994) 604--620.

\bibitem{cai99sisc}
X.-C. Cai, M.~Sarkis, A restricted additive {Schwarz} preconditioner for
  general sparse linear systems, SIAM~J.~Sci.~Comput. 21 (1999) 792--797.

\bibitem{cai03sinum_ash}
X.-C. Cai, M.~Dryja, M.~Sarkis, Restricted additive {S}chwarz preconditioners
  with harmonic overlap for symmetric positive definite linear systems,
  SIAM~J.~Numer.~Anal. 41 (2003) 1209--1231.

\bibitem{Fu2016}
H.~Fu, J.~Liao, J.~Yang, L.~Wang, Z.~Song, X.~Huang, C.~Yang, W.~Xue, F.~Liu,
  F.~Qiao, W.~Zhao, X.~Yin, C.~Hou, C.~Zhang, W.~Ge, J.~Zhang, Y.~Wang,
  C.~Zhou, G.~Yang, The {S}unway {T}aihulight supercomputer: system and
  applications, Science China Information Sciences 59 (2016) 1--16.

\bibitem{petsc}
S.~Balay, S.~Abhyankar, M.~Adams, J.~Brown, P.~Brune, K.~Buschelman, L.~Dalcin,
  V.~Eijkhout, W.~Gropp, D.~Kaushik, M.~Knepley, L.~C. McInnes, K.~Rupp,
  B.~Smith, S.~Zampini, H.~Zhang, {PETS}c users manual, Tech. Rep. ANL-95/11 -
  Revision 3.6, Argonne National Laboratory (2015).

\bibitem{pfm}
N.~Provatas, K.~R. Elder, {Phase-field methods in materials science and
  engineering. 1st ed.}, Wiley-VCH, 2010.

\bibitem{epd}
A.~Jaatinen, T.~Ala-Nissila, Extended phase diagram of the three-dimensional
  phase field crystal model, J Phys: Condensed Matter 22 (2010) 205402.

\end{thebibliography}

\end{document}